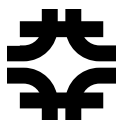

# Fermi National Accelerator Laboratory

# MEASURING MULTIJET STRUCTURE OF HADRONIC ENERGY FLOW

## OR

# WHAT *IS* A JET 

**Fyodor V. Tkachov**

Fermi National Accelerator Laboratory
P.O.Box 500, Batavia, IL 60510, USA

and

Institute for Nuclear Research of the Russian Academy of Sciences
Moscow 117312, RUSSIA[†]

**Abstract** Ambiguities of jet algorithms are reinterpreted as instability wrt small variations of input. Optimal stability occurs for observables possessing property of calorimetric continuity ($C$-continuity) predetermined by kinematical structure of calorimetric detectors. The so-called $C$-correlators form a basic class of such observables and fit naturally into QFT framework, allowing systematic theoretical studies. A few rules generate other $C$-continuous observables. The resulting $C$-algebra correctly quantifies any feature of multijet structure such as the "number of jets" and mass spectra of "multijet substates". The new observables are physically equivalent to traditional ones but can be computed from final states bypassing jet algorithms which reemerge as a tool of approximate computation of $C$-observables from data with all ambiguities under analytical control and an optimal recombination criterion minimizing approximation errors. ☞ *Quick preview on next page*

---

[†] Permanent address.

## Quick preview

• This work gives an answer to the question "what is a jet?" The answer is a purely "kinematical" one (which most experts seem to be ill-prepared for), and is based on an analysis of "kinematics" of calorimetric detectors, "kinematical" aspect of QFT observables, and error enhancement properties of data processing algorithms. The answer is as follows.

"Jets" and "jet finding algorithms" are simply tools of approximate description of data (in a concrete sense explained below). Their physical significance does not go deeper than the leading order QCD predictions (which in many cases is deep enough, of course). As any approximation, it is only useful as long as one is satisfied with its precision.

• Improving upon that approximation does not mean finding a "perfect jet algorithm". The notion of a perfect jet algorithm is physically ill-conceived because of the non-deterministic quantum nature of hadronization.

Instead, the "ideal solution" is based on the formalism of $C$-algebra. The $C$-algebra is a family of "well-behaved" observables together with a translation rules that allow one to reproduce any conventional observable defined via jet algorithms. For instance, take a conventional distribution of events with respect to, say, invariant masses of 3-jet substates, $\rho_{3j}^{\text{old}}(m)$ (e.g. with a peak due to a new particle at some value of mass). One can write down an observable from $C$-algebra — call it $\rho_{3j}^{C}(m)$ — that is an "ideal solution" which $\rho_{3j}^{\text{old}}(m)$ approximates in the following sense. One can compute the new observable from the jet pattern of an event (using any reasonable jet algorithm) and the result will be, by construction, similar to $\rho_{3j}^{\text{old}}(m)$:

$$\rho_{3j}^{C}(m)\big|_{\text{jet pattern}} \approx \rho_{3j}^{\text{old}}(m). \qquad 0.1$$

Note that the r.h.s. cannot be computed otherwise than from the jet pattern. The new observable, on the contrary, is defined in such a way that it can also be computed *directly from the raw event* (i.e. the event prior to application of a jet algorithm) and the result will be as follows:

$$\rho_{3j}^{C}(m)\big|_{\text{raw event}} = \rho_{3j}^{C}(m)\big|_{\text{jet pattern}} + O(y_{\text{cut}}). \qquad 0.2$$

(In the case of cone-type algorithms the error is $O(R)$ where $R$ is the jet cone radius.)

To emphasize: the distribution $\rho_{3j}^{C}(m)$ is physically equivalent to the conventional 3-jet mass distribution and can be computed *without* actually identifying individual jets. A similar translation can be accomplished for any other jet-related quantity.

• The jet algorithms are a useful approximation tool because computing, say, $\rho_{3j}^{C}(m)$ (= the ideal 3-jet mass spectrum) from the jet pattern of an event is easier than from the raw event because there are many fewer jets than particles. It is also clear how the approximation can be improved: one should simply compute the new observable $\rho_{3j}^{C}(m)$ from jet patterns for as small $y_{\text{cut}}$ as possible — or directly from raw data if one has enough computer resources to do so.

• The notorious "ambiguities of jet definition" are then, essentially, the inevitable approximation errors in Eq. 0.2. They are gotten rid of by letting $y_{\text{cut}} \to 0$.

• The formalism of $C$-algebra is a purely kinematical one: it cannot help one e.g. to determine the observable with which a new particle is best seen against background. But it can help one to squeeze the most out of data once such an observable is found. For instance, $\rho_{3j}^{C}(m)$ is, by construction, less sensitive to data errors and statistical fluctuations than $\rho_{3j}^{\text{old}}(m)$.

• Although the perfect jet algorithm is a fiction, an "optimal" one is not: it is an algorithm that minimizes the error in 0.2 for a given $y_{\text{cut}}$ for the observables such as $\rho_{3j}^{C}(m)$. Criteria for such an "optimal preclustering" can be found analytically in an explicit form (see Sec.7).

• The results of this work should *not* be interpreted in the sense that "one should not count jets", nor that one should throw out the old code and start implementing the ideal solution of $C$-algebra. What *is* being suggested is that various features of the ideal solution should be incrementally built into the existing procedures depending on: the computer resources available, the required precision of results, and the level of understanding achieved. The ways to improve the conventional data processing are:

— Regularization of (all) hard cuts (Sec.2.5; perhaps including spline-based schemes described in Sec.15 instead of the bin-type ones).

— One optimal preclustering (Sec.7) instead of all conventional jet algorithms (both recombination and cone-type).

— Replacing conventional observables with their $C$-analogues (for this one has to learn to translate physics into the language of $C$-algebra, which should be possible with the examples in Secs.10–13).

— Exploiting the fully explicit and essentially simple analytical form of $C$-observables to develop more flexible and precise computational schemes than what is possible with the conventional rather rigid algorithms.

## How to read this paper

The subject of jet-related measurements concerns both theorists and experimentalists (the latter more than the former), and there are different obstacles that may make reading this paper difficult for different categories of readers. One is that the philosophy behind the new theory deviates rather considerably from the conventional attitude to jet counting. Moreover, the "philosophical" arguments are mingled with a reasoning that uses unusual logical patterns of general topology (convergences and continuities) that seem to be unfamiliar to a majority of physicists (perhaps because they are so fundamental that are usually taken for granted). This could not have been avoided — the essential uniqueness of the presented construction cannot be appreciated otherwise.

Unfamiliar functional formalism of abstract measures etc. used to describe the mathematical nature of energy flow and $C$-continuous observables may also be difficult for the uninitiated. This could not have been avoided either: The mathematical description of stability of data processing algorithms in calorimetric measurements with respect to errors in the language of a $*$-weak convergence in a space of linear functionals is not an invented hypothesis; it is a straightforward formalization of *what is*.[1]

However, a data processing expert not seeking to understand the mathematical subtleties in complete detail but willing to give the new formalism a try in a realistic problem, may concentrate on the discussion of the role of continuity of observables in Sec. 2.5, and the QFT derivation of $C$-correlators in Sec. 6.4. For practical purposes, one needs the definitions and notations of Sec. 4 and the formulas of Sec. 5 (discrete variants only) while the details of the reasoning of Sec. 3 may be ignored. Then one can turn directly to the discussion of applications in Secs. 8–13. The issues of numerical work with spectral discriminators discussed in Secs. 15 and 12 cannot be skipped.

The material of Sec. 7, Secs. 8–9, and Secs. 10–12 can be studied in any order.

Finally, it should be noted that the validity of the prescription to regularize one's cuts as a matter of routine (Sec. 2.5) extends beyond the measurements in high energy physics. Similarly, Secs. 12 and 15 describe a rather universal scheme based on linear splines for computation of various distributions obtained from irregular "stochastic" approximations such as typically obtained from a sample of events generated by a quantum process. The scheme is hardly more complex, but less sensitive to data errors and statistical fluctuations, than the conventional bin-type algorithms.

## Numbering and cross references

I hope the reader will find it convenient that Fig. 10.24 is located between items 10.23 and 10.25 — irrespective of what the latter are, equations or subsection headings. The universal two-level numbering scheme (with right-aligned numbers conveniently not distracting one's attention) offers an easy way of finding cross references (as with page number references) while remaining coupled to the logical structure of the text. Sub- (and subsub-) section headings are treated along with proposition headings rather like labels in the text (with a liberating effect that one no longer needs to conform to pedantic conventions, e.g., about whether or not a section may be split in the middle by a single subsection heading). For that reason they do not have a separate numbering from equations, figures and tables — it is not clear what such a separate numbering might be good for except a convenience of uncomputerized typographer.

---

[1] If one needs a proof of this fact other than the arguments of the present paper one may note that the definition of calorimetric continuity underwent a noticeable evolution [16], [17], [15] with the original version [16] rather different in form from, but essentially equivalent to, the final one [15].

## Acknowledgments

This work in its final form owes so much to a continuous intensive interaction (which I enjoyed hugely) with the world high energy physics community that justice must be paid to those whose inputs influenced this work:

W. T. Giele introduced me into the theory of jets in 1991 at FERMILAB when my first doubts concerning jet algorithms were raised in connection with the problem of higher order calculations in the QCD theory of jets. The review talks of S. D. Ellis impressed upon me the fundamental importance of the problem ("What is a jet?" [6]). A. V. Radyushkin offered a crucial encouragement after hearing me out at CEBAF in 1992. I thank Z. Kunszt and ETH for hospitality during the Workshop on New Techniques for Calculating Higher Order QCD Corrections (ETH, Zürich, December 1992) — its intensely informative atmosphere catalyzed this work (a key formula for the jet-number discriminators materialized on my way home from ETH; I am indebted to a Swiss train attendant who rescued my notes). S. Catani supplied a useful list of bibliography (any omissions are my fault though), as did R. K. Ellis who also provided a stimulating criticism at an early stage of this work.

Very numerous and lively conversations took place with the participants — both theorists and experimentalists — of: a seminar of the Division of Experimental Physics of INR RAS (Moscow, 1993), the 1993 ITEP Winter School, the VIII and IX Workshops on High Energy Physics (INP MSU, Zvenigorod, 1993 and 1994), the XVI and XVII Seminars on High Energy Physics (IHEP, Protvino, 1993 and 1994), a seminar of the Moscow group of the H1 collaboration (1993), a theory seminar at CERN (1993), a theory seminar at Cornell University (1994), a seminar at NORDITA (1995), the Moscow session of AINHEP-95 (INP MSU, 1995), a theory seminar at FERMILAB (1995).

Stimulating and/or useful were interactions with B. A. Arbuzov, Yu. Arestov, W. A. Bardeen, Yu. Bashmakov, A. V. Batunin, E. E. Boos, F. Cuypers, L. Dudko, F. Dydak, R. K. Ellis, L. Frankfurt, I. F. Ginzburg, B. L. Ioffe, K. Kato, A. B. Kaydalov, V. D. Khovansky, A. Klatchko, V. Kopeliovich, M. Krawczyk, Z. Kunszt, Y. Kurihara, E. Kushnirenko, A. I. Lebedev, D. Lehner, G. P. Lepage, L. Lönnblad, D. Nesic, Kh. S. Nirov, T. A. Osipova, D. Perret-Gallix, Referees A, B and C of Physical Review Letters, M. A. Samuel, E. K. Shabalina, R. Shanidze, D. V. Shirkov, A. Shtannikov, T. Sjöstrand, I. K. Sobolev, N. Sotnikova, B. Straub, M. Strikman, I. I. Tkachev, A. A. Tyapkin, E. Verlinde, B. R. Webber, S. Wolfram and S. Youssef. I thank B. B. Levtchenko for some numerical checks, A. S. Barabash for providing information on how calorimetric detectors actually work [72], and N. A. Sveshnikov for many discussions of the QFT aspect of the $C$-algebra.

Parts of this work were performed during my visits to ETH (1993); Penn State University (1994); and NORDITA (1995) — I thank for hospitality, respectively, Z. Kunszt; J. C. Collins and H. Grotch; L. Brink, P. Hoyer and P. Osland. At various space-time points this work was supported in parts by (chronologically) the International Science Foundation (the George Soros Fund; grants MP9000 and MP9300), the U.S. Department of Energy (grant DE-FG02-90ER-40577), the Russian Fund for Basic Research (grant 95-02-05794).

This work was completed at FERMILAB: I thank its Theory Group for their kind hospitality, and members of the D0 all-jets top group (S. Ahn, H. Prosper, and D. A. Stewart Jr.) for a thorough discussion of the $C$-algebra in a practical context; I owe special thanks to D. A. Stewart Jr. for explanations of the data acquisition system of the D0 detector and many comments, and to N. Sotnikova for much help with preparation of the paper.

## TABLE OF CONTENT

## Introduction and overview 1

### The jet paradigm 1.1

The jet paradigm is the foundation of the modern high-energy physics. It is based on the experimental evidence for hadron jets [1] and a QCD-based theoretical picture [2] of the hadronic energy flow in the final state inheriting the shape of partonic energy flow in the underlying hard process. The association of each jet with a hard parton [3], [4] is the qualitative basis for comparison of data with theory. Its straightforward quantitative implementation leads to the so-called jet finding algorithms. For each final state (event) **P**, such an algorithm computes what may be called its *jet pattern* **Q** — i.e. the total number of jets in **P** and the 4-momentum of each jet:

$$\mathbf{P} = \left\{ p_1, p_2, \dots p_{N_{\text{part}}} \right\} \xrightarrow{\text{jet finding algorithm}} \mathbf{Q} = \left\{ Q_1, Q_2, \dots Q_{N_{\text{jets}}} \right\}, \qquad 1.2$$

where $p_i$ are the 4-momenta of the particles in the final state while $Q_j$ are the 4-momenta of the resulting jets. While the event may have $O(100)$ particles, a majority of events have only a few jets. Therefore, the jet pattern **Q** has typically much fewer 4-momenta than the original event **P** and can be studied much easier. Moreover, such a reduction of information brings out the physics of the process being studied: For instance, if an unstable virtual particle decayed into quarks and gluons that hadronized into jets, the invariant mass of the corresponding group of jets is close to that of the decayed particle.

On the use of such algorithms the current practice of measurements in high-energy physics is founded.[i]

### Ambiguities of jet algorithms 1.3

Unfortunately, hadron jets have a finite angular width and irregular shape. So when their angular separation is not large enough they are hard or impossible to resolve. Then the answer (the jet pattern **Q** on the r.h.s. of 1.2) depends on inessential details of jet finding algorithms. As a result, the ambiguities of jet definition remain a subject of the ongoing discussion ("What is a jet?" [6], [8]). Moreover, as the research moves on to physical problems with more stringent precision requirements, the ambiguities of jet definition manifest themselves quite tangibly: Recent results of Monte Carlo modeling indicate that the dominant uncertainty in the determination of the top quark mass at the LHC is expected to be due to the ambiguities of jet definition [9].

### Purpose of this work 1.4

The purpose of this work is to reexamine the problem of ambiguities of jet algorithms in a systematic fashion from the point of view of first principles of physical measurements. I maintain that a clarification of the issues of kinematics and measurements must be achieved before one will be in a position to discuss dynamics in a completely meaningful fashion. For instance, the qualitative physical notion of length (of, say, sticks) is known to be quantified by real numbers, and the continuity of the latter plays a crucial role in the analysis of measurement errors.

What is the precise mathematical analogue of the qualitative physical notion of energy flow of collision events? And what is the corresponding continuity that would allow one to control effects of measurement errors in a systematic fashion? (Cf. Table 4.28 below.)

The focus on the aspect of measurements and kinematics constitutes the key difference between this work and the conventional approach that emphasizes the dynamics of jets.[ii]

---

[i] For an introduction to the uses of hadronic jets in experimental studies of the Standard Model see e.g. [5].

[ii] This fact might ignite an ideological discussion. As a preemptive measure, I've made an attempt to clarify certain implicit assumptions behind the conventional point of view on jet counting. This resulted in

## Understanding the problem　　1.5

The physical problems where jets are involved can be divided into two classes:

• In the problems of exploratory type one may be satisfied with observing an effect at a minimal sufficient level of statistical significance. This is usually the first stage in studies of any new class of physical phenomena. For instance, in what may be called descriptive theory of hadronic jets one studies the dynamics of jets as such (a good review mostly devoted to this sort of physics is [10]). Here one is mostly interested in qualitative effects that occur already in the leading order of perturbation theory (cf. the effects of QCD coherence [11]). Another example is the search for top quark [12], [13].

• _Precision measurements_ comprise quantitative studies of the Standard Model (for a review see e.g. [5]) such as precision measurements of $\overline{\alpha}_S(Q^2)$, the mass and width of the top quark etc. In such problems jets are only an auxiliary intermediate notion.

Clearly, the two classes of problems differ with respect to the requirements on the data processing algorithms employed. In the first case, one would like to have simple and fast algorithms that allow one to represent the corresponding qualitative physical features in a most direct and visual manner even, perhaps, at the expense of numerical precision.[i] On the other hand, in applications of the precision measurement class the overriding concern is the highest precision attainable for a given event sample.

Note that even in the exploratory problems one may encounter a situation with a low signal/background ratio so that even to establish existence of an effect one may need to squeeze every bit of physical information out of data, which may be impossible with crude methods.

## Measurements or a modeling of dynamics?　　1.6

One may notice that jet algorithms were invented in the context of the problems of exploratory type[ii] and their systematic use for data processing in the precision measurement applications may not be accepted uncritically.

Indeed, theoretical research into jet definition algorithms has traditionally been performed in the context of studies of hadronization. The latter is a non-perturbative phenomenon, is ill-understood, and presents a fascinating field of study. As a result, the problem of adequate numerical description of mutlijet structure has come to be regarded as a "physics issue" that concerns the "underlying jet dynamics" rather than "kinematics" and "geometry"[iii], and that instead "the main point is that one would like to identify a class of <…> observables that are in closer correspondence with the underlying jet dynamics".[iv]

The importance of understanding hadronization and jet dynamics is indisputable. But in the context of precision measurements such a point of view makes hardly more sense than requiring that rulers and clocks reflect the "underlying dynamics" of gravity when one studies the laws governing trajectories of falling bodies. The rulers and clocks do reflect properties of the external world — but only in the kinematical aspect: first and foremost, we want our rulers to be straight and clocks, precise.

Of course, there is a valid point in the "dynamical" argument, too. Indeed, some a priori information on the underlying dynamics helps one to perform measurements where they are most informative (e.g.

---

"philosophical" arguments that are somewhat lengthy and which a pragmatist may not approve of. The subject of hadron jets, however, is fraught with prejudice (as I discovered in many conversations with both theorists and experimentalists), and it is hard to discuss unarticulated attitudes in brief. I may add that all "philosophical" arguments of this paper are more or less direct answers to the questions raised and attitudes exhibited, by the many physicists with whom I discussed these matters.

[i] For instance, one may need to find — very much by trial and error — an observable with respect to which to perform cuts so as to suppress background [12], [13].

[ii] Cf. [2] as well as the fact that the recombination algorithms of a modern type were invented in the context of Monte Carlo event generators [14].

[iii] which are, no doubt, lowly subjects that have nothing to do with The Profound Physics.

[iv] The quotations are from the report of Referee A on the first version of [15].

choosing objects of convenient weight to drop from a tower of convenient height and inclination; or by restricting a search for a new particle to events satisfying certain cuts).

But the kinematical requirement of correctness of measurements (including the data processing stage) is a matter of firstest concern when the precision of results is at stake. Therefore, it makes sense to examine the problem of studying hadronic energy flow as a problem of measurement, and to try — instead of attempting to solve the mathematically ill-posed problem of inverting hadronization — to find "straight rulers" for measuring the multijet structure of multihadron final states.

One may further notice that defining individual jets is only an auxiliary intermediate step in any precision measurement application. A good deal of its attractiveness is due to simplicity of the jet pattern one has to handle at the final stage (when the parameters one measures are actually obtained) and the simplicity of the conventional jet algorithms. Simplicity and speed, however, are not the matters of primary concern in precision measurements — the precision is. It makes sense, therefore, to ask which algorithms may be best for that particular purpose. Having found such algorithms, one can then proceed to optimize them while being fully aware of all the tradeoffs involved.

Then the crucial first step is to subject to a scrutiny the key implicit assumption already present in the question "what is a jet?" — the assumption that the data processing of the form 1.2 is the only way to quantify the multijet structure. The theory of calorimetric observables for measuring multijet structure outlined in [15], [16], [17], [18], [19], [20], [21] and systematically developed in this paper proves the assumption to be wrong.

## C-correlators and C-algebra    1.7

I argue that the fundamental means (the "straight rulers") to measure the multijet structure are not the jet algorithms 1.2 but the so-called C-correlators[i] — i.e. the observables of the following general form:

$$\mathbf{F}_m(\mathbf{P}) = \sum_{i_1 \dots i_m} E_{i_1} \dots E_{i_m} f_m(\hat{\boldsymbol{p}}_{i_1}, \dots, \hat{\boldsymbol{p}}_{i_m}),    1.8$$

where $E_i$ and $\hat{\boldsymbol{p}}_i$ are the energy and direction of the $i$-th particle (or calorimeter cell), each summation runs over all particles in the event (all cells that lit up), $m$ is any positive integer, and $f_m$ is any sufficiently smooth symmetric function of its $m$ arguments.

Special cases of such correlators have been well-known for a long time, e.g. the special sequences of observables studied in [22], [23]. The energy-energy correlators of [24] are also closely related to the C-correlators but correspond to discontinuous $f_m$ of a special form. However, it has never been realized that the following two properties of the functions 1.8 ensure them a very special role in studies of multijet structure:

First, the observables 1.8 have a form of multiparticle correlators that is natural from the point of view of Quantum Field Theory. This is further discussed in Secs. 2.2 and 6 (see also [54]).

Second, they possess special stability properties with respect to data errors — a stability that can be briefly described as follows.

## C-continuity    1.9

Consider any observable 1.8, any two events and any calorimetric detector installation. If the two events $\mathbf{P}'$ and $\mathbf{P}''$ (whatever their numbers of particles etc.) are seen by the detector installation as the same, i.e. $\mathbf{P}' \approx \mathbf{P}''$ within data errors, then the corresponding two values of the observable are guaranteed to stay close, $\mathbf{F}_m(\mathbf{P}') \approx \mathbf{F}_m(\mathbf{P}'')$, with the difference depending on the observable and vanishing together with data errors. Note that with jet algorithms this is not always the case: two such events may yield different numbers of jets — and a difference of two unequal integer numbers cannot be made less than 1.

_______________

[i] C from "calorimetric". The prefix will be widely used below.

The described stability is a form of continuity (we call it *C-continuity*; its precise description is given below in Secs. 4–5). Its importance is due to the fact that, given the same measurement errors in the input data the error of an observable that is continuous in the above sense is, in general, less than the error for the corresponding cut-type observables (Secs. 2.5 and 3; see also [41]).

It is also important to understand that the role of *C*-continuity is not limited to the issues of data errors as such but concerns other types of small variations as well, such as higher order corrections, computational approximations, minor programming variations, etc.

The above two properties of *C*-correlators allow one to say that the collection of values of all *C*-correlators constitutes the entire *physical information* about an event. Once one has understood that, one realizes that any aspect of multijet structure that can be meaningfully quantified and measured, allows an expression in the language of *C*-correlators.

## Expressing physics in the language of *C*-correlators. *C*-algebra                    1.10

The next question is, how to translate the qualitative phenomenon of jets into numbers using only *C*-correlators? The answer is not obvious because their energy dependence is rigidly fixed (cf. 1.8). But once the central role of *C*-correlators is understood (in a sense, *C*-correlators is all one can "see" about events using calorimetric detectors from the point of view of QFT), it becomes clear that such a translation must be possible.

The translation is based on a few rules that allow one to organize values of various *C*-correlators into new observables that (i) inherit the property of *C*-continuity and (ii) describe various qualitative features of multijet structure one happens to be interested in. The expressive power of the resulting *C*-algebra of observables proves to be sufficient for the purposes of precision measurement applications.

There are two main applications for jet finding algorithms that can be classified as precision measurements. Both usually involve a classification of the events with respect to the number of jets ($N_{\text{jets}}$ in 1.2).

The first application may be described as "jet counting". The idea here is to count the fraction of events for each number of jets in the final state. One then obtains what is called $n$-jet fractions (usually denoted as $\sigma^{n\,\text{jets}}$). They provide a more detailed information than the total cross section and prove to be greatly useful both for studying QCD and the Standard Model[i]. This is due to their simplicity and their more or less direct connection with hard subprocesses (each additional jet is associated with an additional radiated parton). For instance, the 3-jet fraction in $e^+e^-$ annihilation in the lowest order of perturbation theory receives contributions from radiation of a gluon from the quark-antiquark pair. Therefore, it is directly sensitive to the coupling constant $\overline{\alpha}_S$ and can be used to extract the latter from experimental data. It turns out that the physical information about the "number of jets" in the final state can be equivalently represented in terms of a special sequence of *C*-correlators 1.8 (the so-called *jet-number discriminators* $\mathbf{J}_m$ introduced in [16], [15]; see also Sec. 8 below).

The other class of applications involves studies of particles that decay into jets, which requires investigation of mass spectra of multijet substates. Here one normally selects events with a given number of jets corresponding to the partonic subprocess where the particle is expected to be seen and then studies invariant masses of groups of jets/partons in a more or less exhaustive fashion (i.e. considering all combinations of two jets if the particle decays into two partons). An example is the top search in the purely hadronic channel [25] where one would have to select 6-jet events and then identify 3-jet substates that may be decay products of the top quark. The observables that can be used in such a context are the so-called *spectral discriminators* ([18], [21] and Sec. 10 below). These are somewhat more complex than the *C*-correlators 1.8 but use the latter as elementary building blocks. Spectral discriminators contain more information about each event than multijet mass distributions obtained via jet algo-rithms, and allow one, in principle, to measure the signal from a virtual particle that decays into jets *even in situations where the jets are too wide to be resolved unambiguously* by the conventional algorithms. If the higher computer requirements of the new formalism prove not to be prohibitive, this may allow one to bypass the limitations of the conventional jet algorithms (cf. Sec.1.3).

---

[i] For instance, a major part of the proceedings [7] is devoted to such applications.

## *C*-algebra and jet algorithms 1.11

It is remarkable that the relation between the two approaches to jet measurements — the one based on jet algorithms 1.2 and the other on the *C*-correlators 1.8 — is precisely that between an approximation trick and an exact answer. Indeed, it is quite natural to regard the jet pattern **Q** on the r.h.s. of 1.2 as an approximate representation of the final state **P** on the l.h.s. But what is an exact interpretation of the adjective "approximate"? The answer is, in the sense that the values of any *C*-correlator 1.8 for the r.h.s. and the l.h.s. are numerically close, and what is known as the jet resolution parameter $y_{cut}$ in the context of jet algorithms is simply a parameter that controls the approximation error:

$$\mathbf{F}_m(\mathbf{P}) \approx \mathbf{F}_m(\mathbf{Q}) + O(y_{cut}) \ . \qquad\qquad 1.12$$

If one requires that the approximation error be minimal one obtains a criterion for an "optimal" jet algorithm ([18], [19], [20] and Sec.7 below). Such a criterion turns out to combine features of the jet definitions currently in use. Then the ambiguities of the conventional jet algorithms are naturally interpreted as approximation errors which it becomes possible to control (and, in principle, eliminate) if one systematically employs the *C*-correlators 1.8 (and other *C*-continuous observables correctly constructed from *C*-correlators) to quantify the multijet structure.

## "Linguistic" restrictions of *C*-algebra 1.13

It should be emphasized that there are questions about multijet structure of an event that can not be asked in the precise language of *C*-algebra. By construction, the linguistic restrictions of this sort reflect the intrinsic limitations of the measurement procedures based on the use of finite precision calorimetric detectors. For instance, it is incorrect to ask how many jets a given event has, and what their 4-momenta are. Nevertheless, as was already mentioned, it is possible to define observables that are physical equivalents of the conventional *n*-jet fractions and of the mass spectra of "multijet substates". Moreover, since jet algorithms retain the role of an approximation trick for computation of *C*-continuous observables, the concepts of "jets" and their "number" retain their value within the precision of such an approximation. (Recall e.g. quantum mechanics where it is, strictly speaking, forbidden to ask what was the exact trajectory of an electron. However, if the electron's motion is quasiclassical, the question becomes meaningful to the same degree to which the quasiclassical approximation is acceptable.)

## Plan of the paper 1.14

The theory of jet measurements described in this paper consists — like any other theory — of four parts: *philosophy* (a critique of jet algorithms and a discussion of a special role of continuity of observables in precision measurements, Secs.1–3); *formalism* (elucidation of the precise mathematical nature of energy flow and the construction of *C*-algebra, Secs.4–6); *computational methods* (esp. the optimal jet clustering criterion, Sec.7; also Secs.12, 15 and 6.16); *applications* (a description of various quantities of physical interest in the language of *C*-algebra, Secs.8–13).

In Sec.2 the ambiguities of the conventional jet algorithms are subjected to a scrutiny. It is then argued that instead of ambiguities one should talk about instabilities of the corresponding numerical procedures, and "instability" is, essentially, just another term for "discontinuity". We consider a simple one-dimensional example and demonstrate how a use of continuous weights instead of hard cuts suppresses statistical data errors. (This is a key ideological point of the present work.) Then we consider the geometry of discontinuities and the associated instability regions in the continuum of multiparticle final states. We investigate the potential numerical effects of instabilities taking into account specifics of QCD. A conclusion is that instabilities are a fundamental feature of any algorithm that attempts to describe a final state in terms of an integer number of jets.

Once the role of continuity of observables for minimization of errors is understood, it should be clarified which particular form of continuity is the right one. The point here is that in infinitely dimensional spaces (such as the continuum of events/final states/energy flows), many radically different continuities are possible. This issue is addressed in Sec. 3. It is pointed out that a prerequisite to defining continuity of observables is a definition of a convergence (topology) in the space of final states. We show that the appropriate convergence is determined uniquely by the structure of data errors of a particular class of measurement installations.

In Sec. 4 the general scenario is specialized to the case of multimodule calorimetric detectors, and the so-called $C$-convergence is described. This completes a precise mathematical interpretation for the physical notion of "energy flow" which is central in the discussion of jet-related measurements. In Sec. 5 the observables possessing the corresponding property of $C$-continuity are described. They are shown to constitute a sort of algebra (the $C$-algebra). The basis[i] of the latter consists of $C$-correlators. Other physically useful $C$-continuous observables can be constructed using a few rules. In Sec. 6 the $C$-correlators are discussed from the point of view of Quantum Field Theory. It is argued that they fit naturally into the general framework of QFT. In particular, an alternative derivation of $C$-correlators is presented. The derivation starts with a QFT-compatible correlator-type observable and makes use of the condition of fragmentation invariance extended to take into account finite energy and angular resolution of calorimeters in the spirit of the preceding discussion of the role of continuity. Prospects for theoretical studies and calculations of $C$-correlators are briefly discussed.

Since $C$-correlators play such a central role in the new theory, Sec. 7 addresses the issue of their optimal computation from data. First of all, expansions in masses of soft particles are possible — an option that has no analog in the conventional data processing (except in the form of the problem of the missing energy, which is discussed in Sec. 7.14). Furthermore, a stability with respect to almost collinear fragmentations allows one to employ an approximation trick ("preclustering") that is similar to conventional jet algorithms (cf. the discussion in Sec. 1.11). An optimal preclustering criterion is derived in a purely analytical fashion from the well-defined requirement of minimizing approximation errors. The new criterion is shown to possess features of the popular jet definitions but is more general in that it allows a simultaneous clustering of more than 2 particles into one.

Secs. 8–13 are devoted to phenomenological applications, and we show how to express physics in the language of $C$-continuous observables ($C$-correlators and their derivatives). In Sec. 8 a special sequence of $C$-correlators is derived (the so-called jet-number discriminators $\mathbf{J}_m$) that are physically equivalent to the conventional $n$-jet fractions. The properties of the jet-number discriminators are considered in Sec. 9.

Sec. 10 introduces and studies properties of a spectral discriminator. The latter are $C$-continuous observables for measuring mass spectra of multijet substates without identifying individual jets, which allows one to avoid instabilities of the conventional algorithms. It is demonstrated that a presence of isolated clusters of particles results in the so-called δ-spikes — δ-functional contributions to spectral discriminators. δ-spikes are a universal feature of spectral discriminators and, after averaging over all final states, result in enhancements that signal presence of new particles. Sec. 11 defines more complex spectral discriminators that allow one to better focus on multijet substates. Sec. 12 summarizes an algorithm of computation of spectral discriminators from data. Sec. 13 describes the options available in the formalism to enhance signal/background ratio by taking into account a priori dynamical information. It also describes modifications needed for the case of hadronic collisions.

Sec. 14 contains a summary and concluding remarks. The two appendices (Secs. 15 and 16) provide background information on abstract measures.

---

[i] A non-linear one.

## "Ambiguities" of jet finding algorithms    2

The jet algorithms currently in use come in two flavors: cone-type and recombination (alternatively called, respectively, clustering [10] and successive combination [26]). The cone-type algorithms define a jet as a group of particles that belong to a cone of certain angular radius [2], [27]. They involve a free parameter (the radius of jets $R$) and an optimization [2] or an iterative procedure [27] to determine jet axes, as well as a rather arbitrary prescription to deal with overlapping cones. In the case of recombination algorithms[i], the underlying idea was, apparently, to invert the hadronization. They also involve a free parameter (the so-called jet resolution parameter $y_{cut}$, $0 < y_{cut} \ll 1$) and proceed in an iterative fashion, taking at each step two particles and deciding (using a criterion that involves $y_{cut}$ as a cut) whether or not to combine them into a jet.

For the purposes of our discussion, the details of jet algorithms are less important than the structure of the results they yield. Such an algorithm is applied to a multiparticle state and computes its jet pattern (cf. 1.2). Note that the jet pattern contains an implicit dependence on the control parameter of the algorithm ($y_{cut}$ or $R$). There is no criterion to fix the control parameter; the conventional practice is to perform computations for an interval of its values. Note that the control parameter has a physical significance in the conventional schemes that becomes manifest when one studies mass spectra of substates consisting of a fixed number of jets. This is because the number of such substates and their invariant masses can be completely different for different values of the control parameter.

### Jet finding algorithms and precision measurements    2.1

Once one has realized what were the context in which the conventional jet algorithms were invented and the implicit assumptions behind them, it is no longer self-evident that they should be an adequate tool for the needs of precision measurement applications.

### Theoretical aspects: jet algorithms vs. QFT    2.2

Precision measurements imply a comparison with high-quality theoretical calculations. For systematic theoretical calculations to be possible (including various kinds of corrections), they should be performed within the formalism of Quantum Field Theory. The first objection to jet finding algorithms is that they do not fit well into the QFT framework.[ii]

A practical manifestation of this circumstance is that jet algorithms are hard to study theoretically: For instance, analytical calculations of theoretical predictions are impossible even in the simplest case of the cross section for $e^+e^- \to 3$ jets in the leading order of perturbation theory, which is to be contrasted with the purely analytical next-to-next-to-leading order calculations of the total cross section [33] made possible by the algorithms [34], [35], [36] that exploit the structure of multiloop Feynman diagrams — the structure predetermined by QFT — to the fullest extent. Moreover, a theoretical study of such algorithms seems to be rather involved (Sudakov resummations etc.; cf. [31]). Also, it is not clear how to approach the issue of power corrections that have been argued to be numerically important in jet physics [37].

On the other hand, one can notice that the ensemble of multihadron final states is a system with a

---

[i] Luclus [14] and its derivatives JADE [26], Durham [27] and Geneva [28]; for a review and comparison see [28], [29].

[ii] This argument is sometimes underrated. Recall, however, that QFT is a very tightly knit construction; it combines quantum mechanics, special relativity, and the experimental fact that particle interactions occur via exchange of quanta; and it does so in a practically unique way — which is expressed most clearly in the construction of perturbative QFT in [30]. It thus summarizes a huge body of experimental knowledge about the Universe. Therefore, such a "purely theoretical" objection may not be dismissed lightly.

varying number of particles. The appropriate theoretical language to describe such systems is in terms of multiparticle correlators (see any systematic textbook on QFT, e.g. [38], or statistical mechanics, e.g. [39]). Next one notices that "jettiness" is but a special sort of geometrical correlations between particles' momenta. Then one is driven towards a logical conclusion that it should be possible to describe the multijet structure in terms of multiparticle correlators, which would provide a direct connection to the formalism of QFT. We will see that such a description is indeed possible.

## Instabilities of jet algorithms                                                                         2.3

Let us now turn to the key idea behind jet algorithms, which is to reconstruct the pattern of underlying hard partons' momenta by, in effect, inverting the hadronization. The first observation is that the evolution of a hard parton state into hadrons is neither classical nor deterministic. Its inversion, therefore, is bound to be a mathematically ill-posed problem. This means that such an inversion may not be always stable with respect to small changes of input data.[i]

Instability of an algorithm means that there are large changes in the output when the input is changed a little. The simplest reason for that is a discontinuity of the algorithm with respect to input data. For instance, the number of jets (whatever algorithm is used to define it) is an integer-valued function of events, and such a function cannot be continuous in any non-pathological sense on a connected continuum, i.e. a continuum in which any two points can be connected by a continuous curve (as is the case with multiparticle final states where any such state can be continuously deformed into any other). In such a situation the maximal output error due to input errors is of order of the discontinuity — irrespective of the size of the input errors (cf. Fig. 2.11 below). Although the size of the region in the continuum of final states where the instability occurs does get smaller as the errors decrease so that the cumulative integral error also decreases, nevertheless the numerical estimates of Sec. 2.5 demonstrate that a well-behaved continuous algorithm may have a considerable advantage over (and is never worse than) discontinuous algorithms in this respect.

Another form of instability of jet algorithms — also connected with their discontinuity — affects jets' 4-momenta. Recall that masses of particles that decay into jets are reconstructed from the invariant masses of the corresponding jets. Now a jet algorithm enforces a representation of the final state by a few jets' 4-momenta in spite of the jets' non-zero angular width and overlaps. In the latter case different jet algorithms will never assign particles to jets in exactly the same way, and the resulting jet pattern will be different. Accordingly, invariant mass of, say, pairs of jets will be measured with an uncertainty due to the ambiguity of jet definition. This phenomenon is behind the fact already mentioned that the error of the top mass determination at the LHC is expected to be dominated by uncertainties due to ambiguities of jet algorithms [9].

The above effect takes place even in the absence of data errors. Moreover, since (non-negligible) data errors are always present, they would influence the results of different algorithms in a different way (sometimes causing, say, a recombination algorithm to combine a pair of particles when it would not do so in the absence of data errors, or vice versa), and the resulting ambiguity increases. In fact, there takes place an enhancement of data errors.

To understand this point recall that some sort of optimization (maximization, minimization) procedure is explicit in some algorithms (e.g. the scheme described in [2]), and the iterative procedures of the popular algorithms of both recombination and cone type, may also be formally thought of as implement-ing some such implicit optimization. Therefore, consider the following situation as a toy model[i]. Suppose one deals with a function $f(x)$ of a real argument $x$, represented by a collection of pairs $[f(x_i), x_i]$ both elements of which are known with errors. Suppose a theory predicts that the function is a $\delta$-function smeared by unknown small corrections, but one nevertheless expects that the function retains a form of a peak, and the shape of the function is controlled by a parameter $A$ which one wishes to determine (this is analogous to determining some Standard Model parameter from hadronic final states). Then one may attempt to extract $A$ by finding the position of the maximum of the function.

---

[i] A non-iterative — and presumably more stable — version of recombination algorithms is described in [38]. However, its clustering criterion is rather ad hoc, and it also does not eliminate the instability with respect to jets' 4-momenta that leads to ambiguities in mass measurements. It should best be considered as a shell in which various clustering criteria can be used (cf. the analytical criterion for $n \to 1$ clustering of the "optimal" algorithm described in Sec. 7).

Further suppose that one does so in a straightforward manner — e.g. by choosing one value of the argument that gives a maximal function value. (This is similar to trying to represent a final state by a finite number of jets via some iterative or optimization procedure.) Then the resulting error is determined not only by data errors but is also enhanced by the width of the peak near its top within the margin of data errors as shown in the following figure:

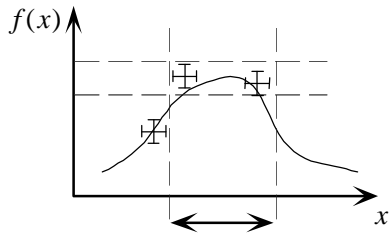

2.4

where the crosses are the measured points $[f(x_i), x_i]$ with error bars and the horizontal double arrow represents the interval of uncertainty in the position of the maximum — an enhanced error inevitably inherited by the resulting value of $A$.

It is clear that the uncertainty is a result of one's using a too crude representation for $f(x, A)$ with a single $\delta$-peak. Now suppose one performs some more calculations and computes not just one number (the position of the peak) but the entire shape, and fits the theoretical curve against experimental data. Then the above enhancement of errors would not occur because the shape would be held better in place by the slopes. Correspondingly, the determination of $A$ would be more precise.

The above model illustrates the mechanism of how the instabilities of jet algorithms are avoided when one studies mass spectra of multijet substates using spectral discriminators of Secs. 10–11.

## Discontinuities and data errors                                                           2.5

Let us study numerical effects of discontinuities on the integral error of the result. The mathematical mechanism considered here is very general and not specific to jet physics. For clarity, we consider a simple example with a 1-dimensional continuum of final states but the resulting estimate remains valid in a general situation.

### A simple 1-dimensional model                                                             2.6

One deals with a continuum of final states $\mathbf{P}$. We take each $\mathbf{P}$ to be a point from the interval $[0, 1]$. ($\mathbf{P}$ can be regarded as a parameter with respect to which a cut is being imposed on a sample of events. A more realistic description of hadronic final states is given in Sec. 4.) The events are generated according to some probability distribution $\pi(\mathbf{P})$ (determined by the $S$-matrix). Its form is controlled by the parameters one usually wants to extract from experimental data, while various physical phenomena manifest themselves through its qualitative behavior (bumps etc.).

There are two typical examples of such phenomena: (i) Virtual production of a particle which is manifested via an enhancement of the probability distribution $\pi(\mathbf{P})$, say, near $\mathbf{P} \sim 1$ but not near $\mathbf{P} \sim 0$.
(ii)  A dynamical mechanism of the underlying theory that, say, suppresses the production of events near $\mathbf{P} \sim 1$. (For instance, $\mathbf{P} \sim 0$ could correspond to 2-jet events while the region $\mathbf{P} \sim 1$, to 3-jet events.) In either case, the events $\mathbf{P} \sim 1$ correspond to a physical "feature" one wishes to quantify. The conventional way of describing such a feature is in terms of a "region" in phase space. So one introduces a cut in a more or less arbitrary fashion (say, $\mathbf{P} = \frac{1}{2}$) to describe the region, and counts the fraction of events that fall within the region defined by the cut. But the position of the cut is ill-defined in either case — in the first example, because the virtual particle has a finite width; in the second example, because any 3-jet event can be continuously deformed into a 2-jet one and there is no unambiguous criterion to separate 2-jet events from the 3-jet ones. It is clear, therefore, that a hard cut does not correspond to any physical reality — a fact made manifest by the ambiguity of its placement.

---

[i] motivated by the comparison in Sec. 11.25 of our spectral discriminators with the conventional procedures.

We have come to an important point. The above argument offers a physical reason for why a continuous weight varying between 0 at $\mathbf{P} = 0$ and 1 at $\mathbf{P} = 1$ should be a better tool to quantify that feature than a hard cut. Such a weight can be thought of as representing the degree to which a final state resembles those that exhibit the feature most clearly[i] — the "weight" of the feature carried by the final state.[ii] If one computes and adds up the values of the weight for all events (normalizing by the total number of events) then the result represents the integral "weight" of the physical feature one studies in a given sample of events.

Of course there is an ambiguity of choosing a weight. But it is often possible to invoke additional considerations to reduce the ambiguity to a minimum. Such considerations can be theoretical — cf. the construction of jet-number discriminators in Sec. 8 where the weight is built of "natural" scalar products analogous to those from which matrix elements consist. Or such considerations could be empirical — one could e.g. choose a linear "regularization" of the weight to minimize effects of data errors etc. (cf. the analysis of errors below and Sec. 2.21; also cf. the fact that control parameters of jet algorithms are chosen to minimize distortions of mass spectra due to hadronization).

Let us return to the example. A quantum observable yields a number for a given reaction. It is defined by specifying a function that returns a number $\mathbf{F}(\mathbf{P})$ for each final state $\mathbf{P}$. Then the value of the observable on the statistical ensemble of final states is the mean value

$$\langle \mathbf{F} \rangle = \int d\mathbf{P}\, \pi(\mathbf{P})\, \mathbf{F}(\mathbf{P}). \qquad\qquad 2.7$$

Choosing $\mathbf{F}(\mathbf{P})$ to take values 1 and 0 within and outside the chosen region of phase space is equivalent to counting the fraction of events that fall into the region described by a hard cut.

The case of a continuous weight discussed above corresponds to $\mathbf{F}(\mathbf{P})$ that is continuous, bounded (which assumption does not incur much loss of physical generality), and is such that $0 \le \mathbf{F}(\mathbf{P}) \le 1$ (which can be always achieved by a normalization).

From the point of view of numerical mathematics (recall e.g. various schemes of numerical integration), it is perfectly obvious that different properties of continuity should result in a different numerical sensitivity of the two schemes to errors and approximations.

Taking into account measurement errors    2.8

In practice, the ideal formula 2.7 is distorted in several ways (e.g. the integration is replaced with a summation etc.). But here we are interested only in the data errors. The latter can be taken into account as follows. For each ideal final state $\mathbf{P}$, the measurement device "sees" another final state $\tilde{\mathbf{P}}$ distorted by data errors. The state $\tilde{\mathbf{P}}$ is seen with the probability $h_\varepsilon(\tilde{\mathbf{P}}, \mathbf{P})$, where $\varepsilon$ describes the error intervals and is determined by the detector hardware. The same data processing algorithm $\mathbf{F}$ is then applied to $\tilde{\mathbf{P}}$, and one obtains the following result instead of 2.7:

$$\langle \mathbf{F} \rangle_{\mathrm{exp}} = \iint d\mathbf{P} \, d\tilde{\mathbf{P}} \, \pi(\mathbf{P}) \, h_\varepsilon(\tilde{\mathbf{P}}, \mathbf{P}) \, \mathbf{F}(\tilde{\mathbf{P}}) \equiv \int d\mathbf{P} \, \pi(\mathbf{P}) \, \tilde{\mathbf{F}}_\varepsilon(\mathbf{P}), \qquad 2.9$$

where

$$\tilde{\mathbf{F}}_\varepsilon(\mathbf{P}) = \int d\tilde{\mathbf{P}} \, h_\varepsilon(\tilde{\mathbf{P}}, \mathbf{P}) \times \mathbf{F}(\tilde{\mathbf{P}}). \qquad 2.10$$

---

[i] or those from which it differs most.
[ii] This interpretation was influenced by a probabilistic interpretation of the jet-number discriminators suggested by F. Dydak.

It is natural to assume that the measurement errors are distributed continuously, then Eq. 2.10 defines a function that is continuous everywhere — even if $\mathbf{F}$ has discontinuities. Fig. 2.11 illustrates the difference between how $\tilde{\mathbf{F}}_\varepsilon(\mathbf{P})$ fluctuates around $\mathbf{F}(\mathbf{P})$ in the cases when the latter represents a hard cut and a continuous weight.

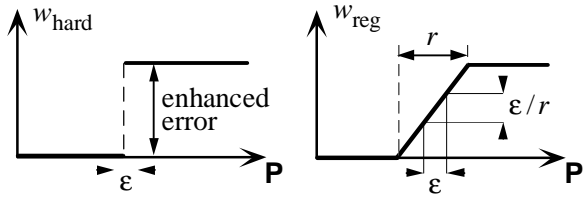

2.11

One sees that however small $\varepsilon$, the fluctuations of the values of $\mathbf{F}$ around a discontinuity are of the same order as the discontinuity itself. Therefore, one can talk about an *instability region* of the observable with respect to data errors around the points where the discontinuity occurs. In such instability regions small data errors cause the values of the observable to fluctuate within an interval that does not vanish as $\varepsilon \to 0$.

The qualitative difference of the two pictures 2.11 translates into a difference between how the measurement errors from many events accumulate when the integration is performed. The key point here is that measurement errors for different events are independent, so it is the corresponding $\sigma^2$ that are added rather than $\sigma$ (should the latter be true, the discontinuity would have played no role). As a result, the integral statistical error is quite different in the two cases.

(It may be pointed out here that the notion of "continuity" is unambiguous only in our simple one-dimensional example. Defining an appropriate continuity in the realistic case of multihadron final states is less straightforward; cf. Sec. 3.10.)

### Integral error for a hard cut                                                      2.12

Suppose one decides to count the number of the events above the cut. The exact result is given by

$$\int_0^1 d\mathbf{P} \, \pi(\mathbf{P}) \, w_{\mathrm{hard}}(\mathbf{P}), \qquad 2.13$$

where the weight $w_{\mathrm{hard}}(\mathbf{P})$ corresponds to the hard cut and is 1 or 0 depending on whether $\mathbf{P}$ is above the cut or not.

We neglect statistical fluctuations of the density of generated events but assume that their positions within [0,1] are measured with an error of order $O(\varepsilon)$. For simplicity suppose $\pi(\mathbf{P}) \sim \mathrm{const}$, so that the generated events are distributed more or less uniformly. Then the measurement errors for events that occur close to the cut (their fraction is $O(\varepsilon)$) will induce fluctuations of order $O(1)$ for the value of $w_{\mathrm{hard}}(\mathbf{P})$ because of the discontinuity. Then the induced variance for 2.13 is

$$\sigma_{\mathrm{hard}}^2 = O(\varepsilon) \times [O(1)]^2 = O(\varepsilon). \qquad 2.14$$

Integral error for continuous weight                                                        2.15

Now suppose one quantifies the same physical feature using a continuous weight (= *regularized cut*), e.g. one that linearly interpolates between 0 and 1 within a subinterval of length $r$:

$$w_{\text{reg}}(\mathbf{P}) = \begin{cases} 1, & \text{for } \mathbf{P} - 1/2 > r/2, \\ \frac{r-1}{2} + \frac{1}{r}\mathbf{P}, & \text{for } -r/2 \leq \mathbf{P} - 1/2 \leq r/2, \\ 0, & \text{for } \mathbf{P} - 1/2 < -r/2. \end{cases}$$                2.16

Then one computes

$$\int_0^1 d\mathbf{P}\, \pi(\mathbf{P})\, w_{\text{reg}}(\mathbf{P}).$$                2.17

We take $r > \varepsilon$; otherwise the situation is equivalent to the hard cut.

The integral error for 2.17 is induced by the events whose weight is computed with errors due to errors in their position on the horizontal axis. The fraction of such events is $O(r)$. The variance of the value of $w_{\text{reg}}$ for each such event is

$$\left[O(\varepsilon) \times \tfrac{1}{r}\right]^2 = O\!\left(\tfrac{\varepsilon^2}{r^2}\right).$$                2.18

Since measurement errors for different events are independent, the total variance for 2.17 is estimated by Eq. 2.18 times the fraction of such events which is $O(r)$, i.e.

$$\sigma_{\text{reg}}^2 = O\!\left(\tfrac{\varepsilon^2}{r}\right).$$                2.19

For $r = O(\varepsilon)$, this degenerates into 2.14, as expected.

The net effect is that the resulting error interval in the regularized case is suppressed as compared with the hard cut case by

$$\boxed{\frac{\sigma_{\text{hard}}}{\sigma_{\text{reg}}} = O\!\left(\sqrt{\frac{r}{\varepsilon}}\right)}.$$                2.20

Comparison and discussion                                                                   2.21

(i)   From 2.20 it is clear that if the ambiguity in defining the physical feature is larger than measurement errors then it is generally advantageous to quantify that feature using a continuous weight instead of a hard cut. Of course, both measurements and theoretical calculations should be done with the same weights.

(ii)   Note that one may choose the position of the cut so as to optimize the signal/background ratio. With a regularized cut of the above type, one has an extra parameter with respect to which to optimize. In the worst case the optimum would be reached for the hard cut ($r = 0$), but it is clear that such situations are exceptional. This means that with a regularized cut one is never at a disadvantage as compared with a hard cut as far as the resulting errors are concerned. The simplest prescription is to make the weight linearly interpolate between 0 and 1 over the interval of uncertainty of cut's placement; the end points of such an interval presumably correspond to some physical reasons why one would not like to place the cut above and below certain points. The exact size of the interval should not usually be very important but it should not be less than the size of errors.

(iii)   Strictly speaking, the above suppression of errors is due to not just simple continuity but a somewhat stronger regularity. In our example it was existence of a bounded (even if discontinuous) first de-rivative. A Hölder's condition would also be sufficient as well as some other types of regularity. In practice such a stronger regularity is usually ensured automatically (if one does not aim to construct a counterexample). On the other hand, in more complex cases there are many continuities possible (because the space of final states is infinitely dimensional), each one determining the corresponding differentiability etc. The real difficulty is to choose the correct form of continuity. This is why in subsequent sections we will be talking only about continuity without mentioning differentiability etc.

(iv)   It is easy to see that the suppression factor 2.20 for integral errors that is induced by a regularization remains operative in a general case (not necessarily one-dimensional). It is sufficient to interpret $\varepsilon$ as the fraction of events that fall within the instability region of the hard cut and $r$, as the fraction of events that fall within the interpolation region of the regularized cut provided the variation of the weight in the region is more or less uniform.

(v)   Take into account a small additional computational cost of implementing a regularized cut. Then it may be suggested that one should *__regularize one's cuts as a matter of routine__* — unless the computational overhead proves to be prohibitive or other sources of errors clearly dominate. This conclusion remains valid irrespective of the physical nature of the problem.

(vi)   From the standard formulas of the theory of Monte Carlo integration, one can see that regularizing the cut also reduces the variance of the purely statistical error (assuming the regularization is such that the mean value remains roughly the same). This point is discussed in more detail in [41].

## Instability regions of jet algorithms    2.22

The above toy model ignored two effects: the geometry of cuts in a multidimensional case and the shape of the probability distribution $\pi(\mathbf{P})$. The examples presented below — although not constituting a proof — indicate that in the case of QCD the two effects tend to conspire in an unfavorable fashion. We consider the two aspects in turn.

### Geometry of instability regions in the continuum of final states    2.23

The geometry of the cut plays a role in the enhancement of errors because the more compact the boundary corresponding to the cut, the less its volume (for a fixed magnitude of data errors), and the less the error due to instabilities. Unfortunately, the situation with the boundaries between $n$-jet regions for different $n$ seems to be quite opposite. A few examples will illustrate this point.

A jet finding algorithm splits the continuum of final states into "$n$-jet regions":

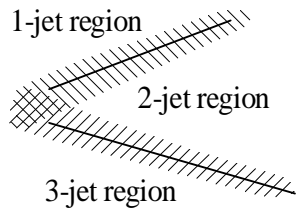

1-jet region

2-jet region

3-jet region

2.24

Consider a 1-jet state $\mathbf{P}_1$ that consists of one particle with the total energy $E$ going in the direction $\theta = 0$. Also consider a continuous family of final states $\mathbf{P}_{\Theta,n}$, each consisting of $n$ particles carrying equal shares of the total energy $E$ and going in the directions described by $\varphi_i = (i-1)\frac{2\pi}{n}$, $\theta_i = \Theta$ for all $i$. As $\Theta$ varies, $\mathbf{P}_{\Theta,n}$ describes a continuous curve in the continuum $\{\mathbf{P}\}$.

Let $n = 2$. For $\Theta = 0$, $\mathbf{P}_{\Theta,2}$ is equivalent to the 1-jet state $\mathbf{P}_1$. For $\Theta = \frac{\pi}{2}$ one has two pure jets. Obviously, there is an intermediate value $\Theta = \Theta'$ when the curve $\mathbf{P}_{\Theta,n}$ crosses the boundary between 1- and 2-jet regions. Because of measurement errors there is an interval of $\Theta$ around $\Theta'$ where the jet al-gorithm cannot count jets stably — a region of instability that separates 1- and 2-jet regions (cf. Fig. 2.24). The thickness of this region is determined by experimental errors and the jet algorithm used.

Let $n = 3$. For $\Theta = 0$, $\mathbf{P}_{\Theta,3}$ is again equivalent to the 1-jet state $\mathbf{P}_1$ while for $\Theta = \frac{\pi}{2}$ one now has *three* pure jets. But what happens in between? It is not difficult to understand that even in the absence of experimental errors, there is an interval $\Theta' < \Theta < \Theta''$ where the algorithm would see 2 jets. But which of the three particles are combined into a jet — and therefore the resulting 4-momenta — depends on the order in which the algorithm treats the particles and on minor programming details. It is also clear that to regard the states $\mathbf{P}_{\Theta,3}$ for $\Theta' < \Theta < \Theta''$ as having two jets does not make much sense. Experimental errors make the situation worse. Thus there exists an instability region that is adjacent simultaneously to 1-, 2- and 3-jet regions, in which the conventional jet counting is rather meaningless. This is also reflected in the fact that even if one forces the algorithm to "count" jets, the resulting jets' 4-momenta cannot be determined in stably. Such an instability region is cross hatched in Fig. 2.24.

Similar instability regions occur for higher $n$. Consider the following configuration:

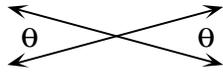

        2.25

As $\theta$ varies from 0 to $\frac{\pi}{2}$ the 2-jet region directly connects to the 4-jet region (unlike the connection between 1-jet and 3-jet regions considered above). However, the connection point is also adjacent to the 3-jet region so that there is an instability region around that point in which the number of jets fluctuates between 2 and 4. A similar direct connection exists e.g. between 3-jet and 6-jet regions.

So, the geometry of $n$-jet regions and boundaries between them as well as the instability regions, become rather intricate: The $n$-particle subspace of $\{\mathbf{P}\}$ consists of interlaced $m$-jet regions for all $m \leq n$, with many boundary points adjacent to more than two such regions.

## Effects of the shape of $\pi(\mathbf{P})$        2.26

Consider the $n$-jet fractions defined using any conventional jet finding algorithm:

$$\sigma^{n\,\text{jets}} = \int d\mathbf{P}\, \pi(\mathbf{P})\, f_n(\mathbf{P}),        2.27$$

where $f_n(\mathbf{P}) \equiv \theta(\mathbf{P}$ has $n$ jets) is 1 or 0 depending on whether or not the condition in the argument of $\theta$ is satisfied. In particular,

$$\sum_{n=1,2\ldots} \sigma^{n\,\text{jets}} = 1 \ .        2.28$$

The number of jets is determined incorrectly for events from instability regions, which results in a redistribution of events between $n$-jet fractions for different $n$, i.e. a smearing between them. The smearing affects not only pairs of adjacent $n$-jet fractions, say 2- and 3-jet fractions, but also 2- and 4-jet ones (cf. Fig. 2.25), etc.

Furthermore, recall that within QCD radiation of each additional parton involves a factor $\alpha_S$ so that

$$\sigma^{n+1\,\text{jets}} = O(\alpha_S) \times \sigma^{n\,\text{jets}} \ .        2.29$$

This means that the smearing affects quantities that may differ by an order of magnitude: 1% of 2-jet events incorrectly identified as having 3 jets may mean an $O(10\%)$ error in the 3-jet fraction which would affect the determination of $\bar{\alpha}_S$ from $\sigma(e^+e^- \to 3\,\text{jets})$. There is also a smearing between, say, $\sigma_{3\,\text{jets}}$ and $\sigma_{6\,\text{jets}}$ in which case the two quantities may differ by $O(\alpha_s^3) \sim O(1000)$ although it is hard to estimate the contribution of the corresponding instability region.

Another feature of QCD is that the probability of additional parton radiation is enhanced in collinear regions by collinear singularities. Therefore radiation of an additional jet is enhanced in configurations where it is closer to, and harder to differentiate from, the other jets.

It is clear that the smearing affects most significantly the $n$-jet fractions for higher $n$. The obvious geometric reason is that the surface of the unit sphere remains $4\pi$ so that angular distances between jets must decrease while jets on average tend to be softer (less energy per jet), therefore, wider. On the other hand, the collinear singularities make it more likely that for a larger number of jets at least one pair of jets has a smaller angular distance.

Numerical importance of such effects in a realistic situation has already been discussed (cf. 1.3).

## Summary 2.30

(i)    Mathematically, instead of "ambiguities" of jet algorithms one should talk about their instabilities resulting from their discontinuity.

(ii)    The discontinuity of jet finding algorithms manifests itself through various instabilities. One is the enhancement (as compared to the case of continuous weights) of data errors near the boundaries separating regions corresponding to events with different jet numbers. This results in a smearing between such regions. Its numerical effect is enhanced by the intricate geometry of the $n$-jet regions in the continuum of all final states and by the specific form of the probability distribution in QCD. It can be eliminated neither by varying the jet resolution parameter $y_{cut}$, nor by increasing statistics (for a given precision of detectors). It is more important for smaller $y_{cut}$, lower energies and larger numbers of jets. Another manifestation of instabilities is a systematic error in determining jets' 4-momenta. In the latter case the numerical effects of instabilities are expected to remain significant in physical problems of much interest even at the energies of LHC [9].

(iii)    The enhancement of statistical measurement errors by discontinuities can be eliminated by using continuous weights (regularized cuts) instead of hard cuts in order to quantify the physical features one studies. Moreover, the use of continuous weights here conforms better to the physical reality of absence of boundaries between events with different numbers of jets.

(iv)    It can be suggested that in high precision/low signal situations one should regularize cuts as a matter of routine — irrespective of the physical nature of the problem (jets or not; see Sec. 2.21).

(v)    The term "ambiguities" used in connection with the problems of jet algorithms, conveys a wrong impression that the ambiguities may perhaps be fixed by invoking additional considerations. It should be emphasized that this is not possible because one actually deals with instabilities that are due to discontinuities that cannot be eliminated as a matter of principle: A mapping of any connected continuum (such as the space of final states ) into integer numbers (such as classification of events according to their "number of jets") cannot be continuous in any non-pathological sense. The discontinuity, therefore, is a fundamental intrinsic property of any jet finding algorithm. It is impossible to eliminate the effects of "ambiguities" of such algorithms without abandoning them altogether as a primary data processing tool in jet-related measurements.

## Mathematics of measurements and continuity     3

Everything that happens to experimental data prior to the moment when they are confronted with theoretical numbers constitutes a *physical measurement*. In high energy physics this involves two stages:

- First one records the raw data from a detector installation. In the context of precision measurement problems, the data thus obtained constitute a result of a physically meaningful measurement to no higher degree than, say, a digital photograph used as an intermediate representation to measure the length of a stick.[i]

- At the second stage a reduction of information is performed, i.e. extraction of a "physical meaning" from the raw data. This involves a rather complex manipulation of numbers with a computer.[ii]

The above definition of physical measurement is an expression of an attitude which is appropriate for the problems of precision measurement class: It makes one focus on what is the primary concern of any measurement, namely, the *precision* of the final results.

### Data processing and errors     3.1

The problem, therefore, boils down to studying how the properties of the algorithms employed for reduction of information interact with data errors etc.[iii] As follows from the discussion of a one-dimensional example in Sec. 2.5, the integral output errors for a quantum observable are minimized if the function that defines the observable is chosen sufficiently regular — as a minimum, continuous.[iv]

The crux of the matter is that continuity is a simple notion only for numeric functions of a finite number of numeric arguments. But the number of particles in final states in high energy physics experiments cannot be meaningfully restricted (the actual numbers are $O(100)$), so the continuum of multihadron final states should be regarded as infinitely dimensional. In an infinitely dimensional space many radically nonequivalent continuities are possible (see any textbook on functional analysis, e.g. [43], [44]).

For instance, consider the infinitely dimensional space of ordinary continuous functions; the Fourier series for such a function converges to it in the $L^2$ topology but need not converge to anything at all in the uniform sense (i.e. in the so-called $C^0$ topology; cf. Sec. 15). An appropriate convergence in each case is determined by a particular application and cannot be postulated a priori.

So, the problem is to choose an appropriate continuity for functions defined on final states. There is a concrete criterion to make the choice: the desired continuity should ensure a minimization of integral errors in accordance with the conclusions of Sec. 2.21. Obviously, the structure of data errors of a particular class of detectors should play a crucial role here.

Consider the difference between the ideal expression of an observable, Eq. 2.7, and its version that takes into account measurement errors, Eq. 2.9 (we consider here the most general case making no assumptions about the dimensionality of the continuum of **P**):

$$\langle \mathbf{F} \rangle_{\mathrm{exp}} - \langle \mathbf{F} \rangle = \int d\mathbf{P} \, \pi(\mathbf{P}) \times \left[ \widetilde{\mathbf{F}}_{\varepsilon}(\mathbf{P}) - \mathbf{F}(\mathbf{P}) \right]. \tag{3.2}$$

---

[i] Of course, this depends on the concrete problem: if one aims at modeling data in full detail (as is the case, e.g., with weather forecasts) then such digitized data are immediately compared with theoretical predictions, and are to be regarded as the final results of measurement.

[ii] Inclusion of a mathematical algorithm into measuring system is a familiar concept, e.g., in spectroscopy and other related fields; cf. the textbook [40].

[iii] Although we are speaking about data errors, similar arguments apply to other sources of uncertainties such as unknown corrections (logarithmic and power) in theoretical studies as well as minor variations of program implementations.

[iv] More stringent types of regularity — differentiability etc. — require that a continuity be defined first.

The integration over the events $\mathbf{P}$ with the weight $\pi$ can be simply interpreted as a summation over a sample of events. Then the first term in the square brackets can be thought of as the value of the observable $\mathbf{F}$ for the event $\mathbf{P}$ distorted by the detector errors ($\varepsilon$ describes their magnitude). To make things more transparent consider the following analogue of 3.2:

$$\langle \mathbf{F} \rangle_{\text{exp}} - \langle \mathbf{F} \rangle = \int d\mathbf{P}\, \pi(\mathbf{P}) \times \left[ \mathbf{F}(\mathbf{P}_\varepsilon) - \mathbf{F}(\mathbf{P}) \right], \qquad\qquad 3.3$$

where $\mathbf{P}_\varepsilon$ is the distorted event. We have seen in Sec. 2.5 that the integral error is minimized if the fluctuations of the values of $\mathbf{F}$ induced by errors in the argument vanish together with the magnitude of errors, which can be expressed as follows:

$$\left[ \mathbf{F}(\mathbf{P}_\varepsilon) - \mathbf{F}(\mathbf{P}) \right] \xrightarrow[\varepsilon \to 0]{} 0 \quad \text{for any } \mathbf{P}. \qquad\qquad 3.4$$

This is nothing but a form of continuity, and it will be discussed later on in more detail.

## A mathematical subtlety                                                     3.5

Strictly speaking one should require that $\mathbf{F}$ satisfy a somewhat more stringent condition:

$$\max_{\mathbf{P}} \left[ \mathbf{F}(\mathbf{P}_\varepsilon) - \mathbf{F}(\mathbf{P}) \right] \xrightarrow[\varepsilon \to 0]{} 0. \qquad\qquad 3.6$$

If $\mathbf{P}$ were real numbers (or finite dimensional vectors) then Eq. 3.4 would simply describe continuity of the function $\mathbf{F}(\mathbf{P})$ while Eq. 3.6, its uniform continuity. The latter does not, in general, follow automatically from the former — but it does if the region in which the function is defined is compact (e.g. a sphere). In finite dimensional situations compactness of a region is equivalent to its being bounded and closed [45]. In infinitely dimensional spaces the issues of compactness and uniform continuity are more tricky.

A mathematician will note that the required compactness in our case follows from the standard Banach-Alaoglu theorem (see e.g. [44]). Indeed, we will see (Sec. 4) that $\mathbf{P}$ in our case are interpreted as measures on the unit sphere while the continuity expressed by 3.4 will be seen to be the continuity with respect to the ∗-weak topology in the space of $\mathbf{P}$. Furthermore, the integral of any $\mathbf{P}$ over the entire unit sphere with unit weight is limited by a constant (the total energy of the colliding particles). According to the Banach-Alaoglu theorem the set of such $\mathbf{P}$ is compact. Then Eq. 3.4 automatically implies 3.6.

On the other hand, a reader to whom ∗-weak topologies mean nothing need not worry much about the reasoning in the preceding paragraph. Indeed, in the final respect our objective is to find a class of observables that could serve as an alternative to jet algorithms. The condition 3.4 is restrictive enough for that purpose — it will allow us in Sec. 5 to derive the so-called $C$-correlators that: form a rather narrow class; allow one to express practically any physical feature; allow an alternative derivation (Sec. 6).

Whether or not the resulting formalism can become a viable alternative to jet algorithms is a complex issue that can only be decided by practice, and we simply accept 3.4 as a basic requirement which our observables should satisfy.

## Data errors and convergence                                                 3.7

Recall that to define continuity of a function one first has to define convergence of its arguments[i]. In our case, the arguments are final states $\mathbf{P}$ that form a space that we denote as $\{\mathbf{P}\}$. Suppose we have defined which sequences of final states $\mathbf{P}_n \in \{\mathbf{P}\}$ are considered as "convergent". Then the observable $\mathbf{F}(\mathbf{P})$ is continuous (with respect to the specified convergence in $\{\mathbf{P}\}$) if its values $\mathbf{F}(\mathbf{P}_n)$ form a conver-gent numerical sequence for any sequence $\mathbf{P}_n$ that converges in the specified sense.

---

[i] The standard mathematical term to denote what we call convergence is "topology" [44]. Note that experimentalists speak about "topology of an event" (= geometry of the event) which is a completely different thing.

## Reminder: length and real numbers                                                 3.8

It may be helpful to begin the discussion of connection of measurement devices, errors, and convergences by briefly reviewing the issue in the simple case of length measurements.

Recall the formal construction of real numbers. One begins with rational numbers and defines the so-called Cauchy sequences of rational numbers that have the following property: If $r_n$ is any such sequence then for any $N$ one has $|r_{n'} - r_{n''}| < \frac{1}{N}$ for all sufficiently large $n'$ and $n''$. (The l.h.s. of the criterion is the usual distance between two numbers, i.e. the modulus of difference.) Such sequences are declared to be "convergent" (with respect to the chosen criterion). The real numbers are then defined as ideal[i] objects that correspond to such sequences: One says that each Cauchy sequence converges to a real number. Two different convergent sequences $a_n$ and $b_n$ are said to be equivalent and represent the same real number if the sequence obtained by combining the terms of the two, $a_1, b_1, a_2, b_2, a_3, b_3 \ldots$, is again a convergent sequence. The construction is called _completion_ of the space of rational numbers with respect to the defined convergence. Arithmetic operations are extended to the ideal numbers via continuity.

One can see that the construction of real numbers is connected with length measurements as follows.[ii] Imagine a stick whose length $L$ one wants to measure. One takes a ruler graduated every $1/N_1$ part of, say, meter. One aligns its zero end with one end of the stick and finds the two adjacent marks on the ruler between which the other end happens to be. One counts the marks and obtains a pair of rational numbers $r_1'$, $r_1''$. One chooses either of the two or any rational number in between depending on a convention. Denote the chosen number as $r_1$. One says that $r_1$ represents $L$ within the precision $\varepsilon_1 = 1/N_1 = r_1'' - r_1'$. Next one takes a more precise ruler, i.e. graduated every $1/N_2$-th part of the meter with $N_2 > N_1$ and repeats the procedure. One obtains another interval $r_2'$, $r_2''$ and a rational number $r_2$ which is said to represent $L$ within precision $\varepsilon_2 = 1/N_2$, etc. It is an experimental (sic!) fact that all the intervals $r_n'$, $r_n''$ overlap.[iii] This is reflected in that whatever the convention for choosing $r_n$, all such sequences happen to be convergent and equivalent to one another in the sense of the preceding paragraph.

This should be compared with the construction of the so-called $p$-adic numbers (for each prime $p$ there exists a continuum of such numbers). The latter are also defined as ideal numbers represented as "limits" of Cauchy sequences of rational numbers except for one difference: the "distance" — and therefore the notion of convergence — is chosen differently from the case of real numbers (for details see [47]). Despite the difference, the $p$-adic numbers share many properties with the real numbers, e.g. arithmetic operations remain continuous in the space of $p$-adic numbers. The $p$-adic numbers are by no means a pathology: They emerged naturally in connection with certain fundamental problems of the theory of numbers (the latter has vital applications to cryptography and digital signature systems on which, e.g., modern electronic banking is increasingly dependent [48]).

Lastly, it is only rational numbers that are actually used in practice, and one might wonder whether choosing a particular idealization makes any difference. In particular, what makes us stick to the real numbers in theoretical constructions? The answer should by now be obvious: it is the fact that their continuity properties directly formalize the structure of measurement errors of the concrete measurement procedures that prove to be useful in a particular kind of applications. Similarly, we will see that there is just one convergence in the space of final states/energy flows that formalizes the structure of errors of calorimetric detectors.

---

[i] The ideal nature of real numbers is to be contrasted with the fact that rational numbers exist in a very tangible form as finite sequences of digits on paper or bits in computer memory.

[ii] One would normally discuss such measurements while taking for granted that length is represented by a real number. We wish to make no such assumption, which explains a somewhat pedantic exposition.

[iii] The complication of statistical errors is ignored here for simplicity.

## Convergence in terms of resolution of measurement device        3.9

In what follows it will be useful to reformulate the definition of Cauchy sequences in terms of measurement devices in such a way as to make it easily generalizable to more complex situations. The new definition is as follows: A sequence of measurements $l_n$ is convergent in the sense determined by measurement devices (e.g. rulers) if whatever the resolution $\varepsilon$ of a ruler, $l_n$ cannot be distinguished using that ruler for all sufficiently large $n$.

The definition remains valid after the completion that results in the space of ideal lengths/real numbers: A sequence of lengths/real numbers $l_n$ converges if whatever the resolution $\varepsilon$ of the ruler, $l_n$ cannot be distinguished using that ruler for all sufficiently large $n$.

## Complex measurement installations        3.10

In general, a complex measurement installation — e.g. those used in high energy physics — consists of more than one elementary detector module each yielding a (rational) number. The total number of modules need not be fixed, i.e. it may vary depending on a particular installation. But it is always finite.

An example is a typical high-energy physics detector consisting of many detector modules. Each measurement from such an installation (corresponding to one event) yields a data record — a finite array of numbers organized in a certain way. Denote such a data record as $\mathbf{P}$. Consider a particular class $C$ of measurement installations (e.g. calorimetric detectors). The detector records $\mathbf{P}$ from all possible installations of the class $C$ can be regarded as elements of a mathematical space (denoted $\{\mathbf{P}\}$). The structure of installations of the class $C$ reflected in the structure of errors of the records $\mathbf{P}$ determines a convergence in $\{\mathbf{P}\}$.

## Definition of convergence        3.11

First define the convergence in $\{\mathbf{P}\}$ following the pattern of Sec. 3.9: A sequence of data records $\mathbf{P}_n$ is called _convergent in the sense of C_ if whatever detector installation from this class is used, the elements of the sequence cannot be distinguished by that installation for all sufficiently large $n$. (We will also use the term _C-convergent_ and _C-convergence_ in such cases.)

Let us now take into account that the number of elementary detector modules in any such installation is finite. Denote the modules of the installation as $m_i$. Consider any one such module $m_i$; let $m_i(\mathbf{P}_n)$ be the number it measured for $\mathbf{P}_n$. The module $m_i$ has a finite precision, and the numbers $m_i(\mathbf{P}_n)$ become indistinguishable within the precision of $m_i$ starting with, say, $n = N_i$. Then the entire installation (the collection of all $m_i$) will not "see" the difference between $\mathbf{P}_n$ starting from $n = \max_i N_i$. One notices that because the number of elementary modules is always finite, the convergence in the sense of $C$ may be equivalently defined using only elementary modules: One says that the sequence $\mathbf{P}_n$ converges in the sense of $C$ if for any elementary module $m$ from measurement installations from the class $C$ the numbers $m(\mathbf{P}_n)$ become indistinguishable for all sufficiently large $n$ — whatever the precision of $m$.

Finally, one can reformulate the definition in such a way that the precision of modules is not mentioned: The sequence $\mathbf{P}_n$ converges in the sense of $C$ if for any elementary module $m$ there exists the limit

$$\lim_{n \to \infty} m(\mathbf{P}_n) < +\infty.$$        3.12

This is the definition we adopt in what follows.

It remains to note that the above definition is completely general, for the very nature of measurement is such that only a finite number of "detector modules" can be involved in any instance of measurement process.

A concretization of this scheme to the case of multihadron final states and calorimetric detectors together with precise descriptions of $\mathbf{P}$, $m$, $\{\mathbf{P}\}$ and the resulting "$C$-convergence" is presented in Sec. 4.

Convergence in terms of open neighborhoods                                      3.13

A mathematician may note that the above scheme is easily transformed into the usual mathematical definition of a topology. Indeed, the "$C$-topology" in the space $\{\mathbf{P}\}$ that is equivalent to the above $C$-convergence is exactly the weakest topology in which all numeric functions $m(\mathbf{P})$ corresponding to all allowed elementary detectors are continuous. Note also that in applications $\{\mathbf{P}\}$ is always separable. In the concrete situation considered below in Sec. 4, $\{\mathbf{P}\}$ consists of linear functionals on a linear space whose elements correspond to elementary detector modules $m$. Then the $C$-topology is a variant of the well-known $*$-weak topology [44]. In our special case we deal with a linear space of measures, and the convergence is also known as the weak convergence of measures [49].

Summary                                                                          3.14

(i)   Integral errors of data processing in complex measurements — the errors that are due to measurement errors in the raw data — are minimized if one uses sufficiently regular continuous observables. (This is also true of the purely statistical fluctuations; cf. [41].)

(ii)   To define the required continuity of observables one must specify a notion of convergence in the continuum of mathematical images of physical objects one deals with.

(iii)   The required convergence is determined by the structure of measurement errors in the raw data, eventually, by the kinematic structure of the measurement device.

(iv)   A natural completion construction results in an ideal representation of the physical objects one measures as some sort of ideal mathematical objects (e.g. length as a real number).

(v)   The notion of convergence that is specific to multimodule detectors can be characterized in terms of elementary detector modules (Eq. 3.12). (The resulting convergence happens to be a variant of the so-called $*$-weak topologies that are well-known in mathematics.)

The plan of our construction is now clear: First, we should precisely describe the objects we wish to measure (i.e. events/multiparticle states/energy flows) as well as the elementary calorimetric detector modules. Then we will be in a position to describe the convergence determined by the calorimetric detectors following the above scheme. This will give us a complete formalization of what "energy flow" is from mathematical point of view.

**Multiparticle states, energy flows and calorimetric detectors**    **4**

In this section we apply the scenario described in the preceding section to the special case of multi-module calorimetric detectors in order to arrive at a precise mathematical description of energy flow including the convergence defined by such detectors.

As was explained e.g. in [26], there are two different geometric frameworks in which jets are studied that correspond to two different physical situations. The first case is the $e^+e^-$ annihilation experiments where the hadronic system as a whole is at rest in the laboratory reference frame and one aims to emphasize rotational symmetry. Then it is natural to regard the detector modules as covering a sphere. The other case corresponds to hadronic collisions where one emphasizes invariance with respect to boosts along the beam direction. One then deals with a cylindrical geometry. In what follows we consider the spherically symmetric case. Our notations are mostly independent of a particular parametrization, so only a few formulas are affected by modifications needed for the case of hadronic collisions discussed in Sec. 13.

## Multiparticle states and their energy flows    4.1

One deals with multiparticle final states produced in collisions at a fixed point within a detector where we place the origin of the coordinate system. For each particle in such a state, a detector installation measures its characteristics. Calorimetric detectors measure its energy $E \geq 0$ and the direction in which it goes (denoted as $\hat{\boldsymbol{p}}$).

## Parametrization of directions $\hat{\boldsymbol{p}}$    4.2

A simple way is to represent the direction $\hat{\boldsymbol{p}}$ as a unit 3-vector $\hat{\boldsymbol{p}}^2 = 1$, which is a point of $S_2$, the unit sphere in 3-dimensional Euclidean space. One can parametrize it, e.g., with the two standard angles, $\theta$ and $\varphi$. But what really fixes the spherical geometry in our context is how one defines integration over $\hat{\boldsymbol{p}}$. If $\Omega$ is a set of directions then for the spherical geometry one defines:

$$\int_{\Omega} d\hat{\boldsymbol{p}} = \text{surface of } \Omega. \qquad 4.3$$

In terms of the angles $\theta$ and $\varphi$:

$$\int d\hat{\boldsymbol{p}} \, \psi(\hat{\boldsymbol{p}}) = \int_0^{\pi} \sin\theta \, d\theta \int_0^{2\pi} d\varphi \, \psi(\theta, \varphi). \qquad 4.4$$

## Angular distance between directions: "angular separation"    4.5

Another quantity that depends on whether one uses spherical or cylindrical geometry is the measure of distance between two directions corresponding to two particles. One such measure used in the context of spherical geometry is the angle between the two directions:

$$\theta_{ij} = \arccos \hat{\boldsymbol{p}}_i \hat{\boldsymbol{p}}_j. \qquad 4.6$$

It turns out, however, that another quantity appears in the expressions on a regular basis — the quantity we call *angular separation* and denote as $\Delta_{ij}$ systematically throughout this work. Its specific expression in terms of angles etc. depends on a particular kinematic situation. For the spherically symmetric case of $e^+e^- \to$ hadrons the angular separation is defined as follows:

$$\Delta_{ij} = 1 - \cos\theta_{ij} = 1 - \hat{\boldsymbol{p}}_i \hat{\boldsymbol{p}}_j \quad \left(\Delta_{ij} \sim \tfrac{1}{2}\theta_{ij}^2 \text{ for small } \theta_{ij}\right). \qquad 4.7$$

The considerations that went into the definition 4.7 are as follows:

(i)    For small angular distances between the two directions $\hat{\boldsymbol{p}}_i$ and $\hat{\boldsymbol{p}}_j$ the factor $\Delta_{ij}$ should be useable as the angular factor in the optimal preclustering criterion 7.39.

(ii)    For large angular distances, it should be useable as a natural building block for the jet-number discriminators $\mathbf{J}_m$ defined in Sec. 8 (cf. Eq. 8.8).

(iii)   It should be as geometrically natural as possible for a given kinematical framework. In particular, it should conform to the structure of typical factors in matrix elements; this facilitates theoretical study of, e.g., $\mathbf{J}_m$ that play a central role in the $C$-algebra of observables described in subsequent sections.

The formulas expressed in terms of the angular separation remain valid in the case of cylindrical kinematics of hadron-hadron collisions — all one has to do is to redefine $\Delta_{ij}$ appropriately; cf. Sec. 13.

Calorimetric information and energy flow                                                4.8

We will distinguish _calorimetric information_ (i.e. $E$ and $\hat{\boldsymbol{p}}$ for all particles constituting a given final state) from other types of information. If one retains only calorimetric information about the final state, then one obtains what is called _energy flow_ of that state. It is the ways to measure the geometric shape of the energy flow that is our object of study. All the non-calorimetric information is then treated as additional parameters (e.g. momenta of "high-$p_\mathrm{T}$ muons" etc.).

There is some arbitrariness in separating what one will treat as calorimetric information: If one believes one can identify an intermediate particle from purely calorimetric data with enough confidence and precision, one may be willing to treat it as a non-calorimetric information. We are interested in situations where not all data can be treated as non-calorimetric, and it is the calorimetric information that we focus upon.

Thus, a final state $\mathbf{P}$ is characterized by its energy flow (calorimetric information) and perhaps some other parameters which we will treat as implicit. For simplicity of notation, we will not distinguish the multiparticle state from its energy flow.[i]

A final state as seen by an ideal calorimetric detector (with perfect energy and angular resolution) is represented as

$$\mathbf{P} = \{E_i, \hat{\boldsymbol{p}}_i\}_{i=1,\dots,N_{\mathrm{part}} \leq \infty} = \{p_i\}_i \,.$$                4.9

Calorimetric detectors do not distinguish types of particles, so the states are not affected by permutations of $p_i$.

It is convenient to have a notation for the state $\mathbf{P}$ consisting of all particles from other states, e.g. $\mathbf{P}_a$ and $\mathbf{P}_b$. Then we write

$$\mathbf{P} = \mathbf{P}_a \oplus \mathbf{P}_b \,.$$                4.10

Lorentz invariance                                                                      4.11

The concrete formulas below are motivated by the idealization that is usual in high energy experiments, namely, that all particles are massless. If such an idealization is valid, then the 3-momentum of a particle is $\boldsymbol{p} = E\hat{\boldsymbol{p}}$, and the 4-momentum is $p = (E, \boldsymbol{p})$, $p^2 = 0$. A Lorentz-covariant expression for the energy is

$$E = pP, \quad P^2 = 1,$$                4.12

---

[i] Strictly speaking, energy flows are elements of the factor space of the space of final states with respect to the equivalence relation of fragmentation invariance. Drawing such a distinction explicitly would be too cumbersome.

where $P$ is a 4-vector whose rest frame defines the coordinate system. Then a Lorentz covariant expression in the $P$'s rest frame for the angles $0 \le \theta_{ij} \le \pi$ between pairs of particles can be obtained from the following relation:

$$p_i p_j = E_i E_j - \boldsymbol{p}_i \boldsymbol{p}_j = E_i E_j (1 - \hat{\boldsymbol{p}}_i \hat{\boldsymbol{p}}_j) = E_i E_j (1 - \cos\theta_{ij}) \,. \qquad 4.13$$

The concept of jet is not Lorentz-invariant, and the calorimetric data (energies and directions of particles) are specific to the detector rest frame. In some cases one may wish to look at an event in a reference frame other than the detector rest frame [50]. This entails performing a Lorentz transformation that makes use of the masslessness assumption. But although Lorentz invariance provides motivations for fixing details in concrete formulas for observables, it is an extrinsic concept for the issues we address in the theory of calorimetric observables. Thus, the 4-momentum introduced above is just a convenient label, while the basic objects are the energy and direction, both immediately measurable and subject to errors. This means that studying corrections due to non-zero masses is relegated to the theory department where it properly belongs. Note that such corrections can be studied in the general context of power corrections. This is further discussed in Sec. 6.16.

## Detectors                                                          4.14

A realistic detector installation consists of a large but finite number of elementary calorimeter modules. Each such module may be represented by a continuous function $\psi(\hat{\boldsymbol{p}})$ that takes the values between 0 and 1 and describes the local efficiency of the module, and the energy it measures for a state $\boldsymbol{P}$ is

$$\langle \boldsymbol{P}, \psi \rangle \equiv \sum_{i=1}^{N_{\text{part}}} E_i \, \psi(\hat{\boldsymbol{p}}_i) \,. \qquad 4.15$$

## Data records                                                       4.16

Let the $c$-th module of the detector be described by $\psi_c(\hat{\boldsymbol{p}})$. Then what the entire detector sees about the final state $\boldsymbol{P}$ is the collection of numbers

$$E_c = \langle \boldsymbol{P}, \psi_c \rangle, \quad c = 1, \dots \,. \qquad 4.17$$

The modules are chosen to have small angular sizes, and one assumes that all the information one needs is the approximate position $\hat{\boldsymbol{q}}_c$ of the calorimeter module. Then a typical data record is a finite array of the form

$$\{E_c, \hat{\boldsymbol{q}}_c\}_c \,. \qquad 4.18$$

Given sufficient energy and angular resolutions of individual detector modules, a physicist regards such a data record as an adequate representation of the exact energy flow 4.9.

All possible different detector records from all imaginable detector installations form a mathematical infinitely dimensional space (= the space of data records). Formally, multiparticle states 4.9 with finite numbers of particles are elements of that space.

I emphasize that when translated into the language of mathematics, the implicit physical convention that a detector record is "close within detector resolution" to the corresponding physical multiparticle state, becomes an axiom that determines one of many mathematically possible forms of convergence in the space of data records — a convergence predetermined by the structure of measuring devices (calorimetric detectors in the present case). This is a central point to which we will soon return.

## Energy flows as "abstract measures"          4.19

The structure of 4.15 suggest the following mathematical interpretation: Each multiparticle final state **P** as seen by all imaginable calorimetric detectors is represented by a sum of δ-functions:

$$\mathbf{P} = \mathbf{P}(\hat{p}) = \sum_{i=1}^{N_{\text{part}}} E_i \, \delta(\hat{p}, \hat{p}_i) \,. \qquad\qquad 4.20$$

To interpret this expression correctly, recall that, by definition, the δ-function $\delta(\hat{p}, \hat{p}_i)$ acquires a concrete numerical meaning only in integrals with continuous functions:

$$\int_{S_2} d\hat{p} \; \psi(\hat{p}) \, \delta(\hat{p}, \hat{p}_i) \equiv \psi(\hat{p}_i) \,. \qquad\qquad 4.21$$

Correspondingly, Eq. 4.20 represents the collection of all values of integrals with all possible continuous functions ψ (the values of such integrals reduce to 4.15).

In mathematical language, the objects represented by 4.20 are linear functionals defined on all continuous functions on $S_2$ — for each such function they define a number (given by 4.15). Such objects are called _measures_ on $S_2$. To avoid confusion with physical measurements, we will use the term _abstract measure_ to denote such objects. (For a background mathematical information on abstract measures see Secs. 15–16.)

For applications it would be sufficient to have in view the two formulas 4.20 and 4.21.

## Calorimetric convergence (*C*-convergence). When are two energy flows close?          4.22

We have come to the key point of our study, namely, a mathematical description of closeness of energy flows of multiparticle states. This is achieved by introducing a convergence in the space of all energy flows — the convergence determined by the measurement devices (cf. Secs. 3.7–3.14).

Suppose one has a sequence $\mathbf{P}_n$ of multiparticle states (abstract measures, or linear functionals). We say that it converges to a multiparticle state **P** if for any detector module described by a continuous function ψ, the values of energies measured by ψ for $\mathbf{P}_n$ converge to the energy measured by ψ for **P**:

$$\lim_{n\to\infty} \langle \mathbf{P}_n, \psi \rangle = \langle \mathbf{P}, \psi \rangle \quad \text{for any allowed } \psi. \qquad\qquad 4.23$$

The reasoning of Sec. 3.10 shows that $\mathbf{P}_n$ will then be indistinguishable for any detector installation consisting of a finite number of detector modules for all sufficiently large *n*. We call this _calorimetric convergence_ (_C-convergence_ for short).

Mathematically, this is a special case of the general notion of ∗-weak topology in a space of linear functionals (Sec. 3.13). Such topologies cannot, in general, be usefully described by a simple single-valued distance or norm, which makes them seem somewhat amorphous for applications. Nevertheless, the above definition is precise enough to allow one to find constructive ways to deal with the situation.

## *C*-convergence and collinear fragmentations          4.24

Let us explain the meaning of *C*-convergence with a few examples. One easily verifies from the definition that an energy flow changes *C*-continuously under any of the following modifications:

(i)   Continuous variations in the energies of the particles constituting the event (energies are never known precisely).

(ii)   Adding any number of arbitrarily directed particles with the total energy going to zero (and the number and directions of particles may change arbitrarily in the process). Indeed, soft particles may escape undetected and their total number is unknown.

(iii) Continuous variations of the angular parameters of particles in the event. This is because particles' directions are determined up to the size of calorimeter modules.

(iv) *Almost-collinear fragmentations*. This can be regarded as a combination of an exact fragmentation and (iii). An *exact fragmentation* [i] consists in a replacement of a parent particle by any number of collinear fragments,

$$E, \hat{\boldsymbol{p}} \to E_a, \hat{\boldsymbol{p}}_a; \ldots; E_b, \hat{\boldsymbol{p}}_b,$$ 4.25

where

$$E = E_a + \ldots + E_b, \quad \hat{\boldsymbol{p}}_a = \ldots = \hat{\boldsymbol{p}}_b = \hat{\boldsymbol{p}}.$$ 4.26

This of course does not change the energy flow 4.20. If after such a fragmentation one changes the directions of the fragments a little, one obtains an almost-collinear fragmentation.

## Correctness of the scheme 4.27

We have introduced a mathematical model for energy flow via a somewhat loose reasoning using the physical notions of particle etc. that are extrinsic with respect to the measurement process as such. We should verify that no inconsistency was introduced thereby. To this end, it remains to recall the simple mathematical fact that any measure can be approximated in the ∗-weak topology by finite sums of δ-functions, and the space of abstract measures is complete in that topology (cf. Secs. 15–16).

The data record 4.18 can be viewed as a finite sum of δ-functions on the sphere (the energies measured by individual detector cells are the coefficients, and their positions are the points where the δ-functions are localized). The completion of the space of all data records with respect to the *C*-convergence restores abstract measures as ideal representations of energy flow. Another aspect of this is that we never see the final state as such (except in Monte Carlo modeling) but only the energies measured by calorimetric cells; in an operational sense, a final state *is* a collection of the data records produced by all possible calorimetric detectors.

The physical meaning of the *C*-convergence clarifies in a broader context of the discussion of Sec. 3. It may be helpful to represent the analogy between calorimetric measurements and the measurements of length as follows:

| length | energy flow |
|---|---|
| ruler | calorimetric detector |
| rational number | data record |
| usual convergence of rational numbers | *C*-convergence (∗-weak topology) |
| real number | abstract measure |

4.28

---

[i] The notion of fragmentation was introduced in [2] and formalized in [50], where the adjective "exact" was not used because data errors were not considered.

## Comparison with the $L^2$ formalism 4.29

As was already emphasized, the space of all energy flows is infinitely dimensional, and there are many nonequivalent continuities in such spaces. Choosing a wrong one may be rather misleading. To appreciate the subtlety of the problem, recall the large scale study described in [22] where an idealization of energy flows as ordinary functions on $S_2$ was adopted. To measure closeness of energy flows, the following norm was chosen:

$$\| \mathbf{P} \|_{L^2} \overset{\text{def}}{=} \left( \int_{S_2} \mathrm{d}\hat{p} \, | \mathbf{P}(\hat{p}) |^2 \right)^{1/2} . \qquad 4.30$$

The resulting $L^2$ topology is familiar from courses of quantum mechanics, is very similar to norms in finite-dimensional Euclidean spaces, is fairly easy to deal with, and allows a number of sophisticated constructions, e.g. expansions in spherical harmonics, summations involving 3-$j$ symbols, etc.

Unfortunately, such superficial analogies have strictly nothing to do with physics. Indeed, it is well-known (and is rather obvious) that the $L^2$ norm is a poor tool for comparing shapes, which is exactly what one does in classifying events with respect to the geometry of energy flow. In particular, the $L^2$ norm of the energy flow that corresponds to a finite number of particles, Eq. 4.20, is infinite.

One reason why the research of [22] was led astray was an unmotivated transition to "continuous limit" (interpreted there as a replacement of the energy flow with an ordinary function) prior to specifying the notion of convergence with respect to which the limit is understood.

## Correct continuous limit 4.31

Our interpretation allows taking infinite — and even continuous — sums in 4.20, which can be done as long as the total energy remains finite. The limiting procedures implied thereby are to be understood as applied to each value of the functional representing energy flow, on each continuous function. For instance, consider a continuous sum,

$$\mathbf{P}(\hat{p}) = \int \mathrm{d}\alpha \, E_\alpha \, \delta(\hat{p}, \hat{p}_\alpha) , \qquad 4.32$$

where $E_\alpha$ is, e.g., a continuous function of $\alpha$ and the integration may run, e.g., over a non-zero length arc on $S_2$, or over a part of $S_2$ with non-zero surface. Then by definition one has

$$\langle \mathbf{P}, \psi \rangle \equiv \int \mathrm{d}\alpha \, E_\alpha \, \psi(\hat{p}_\alpha) , \qquad 4.33$$

where the integral is defined in the usual sense.

Eq. 4.33 remains well-defined as long as $\psi$ is continuous and the total energy is finite:

$$\int \mathrm{d}\alpha \, E_\alpha < \infty . \qquad 4.34$$

In particular, the definition of $C$-convergence remains valid in the case of more general energy flows that cannot be associated with states with finite number of particles (e.g. in QED).

## Calorimetrically continuous (*C*-continuous) observables                    5

In the preceding Sec. 4 we established the mathematical nature of energy flow as abstract measure on the unit sphere, together with a correct convergence (the *C*-convergence) that corresponds to the structure of errors of multimodule calorimetric detectors. But that was only an intermediate step. What one actually needs is a practical description of functions on final states/energy flows that are continuous with respect to the *C*-convergence of their arguments. In this section we describe a large class of such functions. (The special role of continuous observables was discussed in Sec. 2.)

Recalling that an (ideal) energy flow **P** is an abstract measure (= a linear functional on continuous functions on the unit sphere; Sec. 4.19), one cannot miss the mathematical subtlety of the situation: an observable **F**(**P**) is a numeric-valued *function* defined on linear *functionals* defined on continuous *functions* defined on the unit sphere. There seems to exist no systematic mathematical study of the structure of non-linear functions on functional spaces in the constructive aspect.

Nevertheless, it turns out possible to point out a useful basic class of such functions in our case (*C*-correlators; Sec. 5.1, esp. Sec. 5.15). *C*-correlators also turn out to have a natural QFT interpretation (discussed in Sec. 6.1). The latter property opens a prospect of their systematic theoretical study. Moreover, as described in Sec. 5.22, *C*-correlators can be organized into more sophisticated *C*-continuous observables (e.g. the spectral discriminators introduced in Sec. 13) thus forming an algebra (*C*-algebra) in terms of which any physical feature can be correctly expressed.

### *C*-correlators: basic *C*-continuous observables                    5.1

### Linear functions of energy flows                    5.2

Abstract measures **P** are defined as linear functionals $\langle \mathbf{P}\varphi \rangle$ on continuous functions $\varphi$, formally represented as

$$\langle \mathbf{P}\varphi \rangle = \int_{S_2} d\hat{\boldsymbol{p}}\, \mathbf{P}(\hat{\boldsymbol{p}})\varphi(\hat{\boldsymbol{p}}). \qquad 5.3$$

Fix one such function $\varphi(\hat{\boldsymbol{p}})$ and consider the expression $\langle \mathbf{P}\varphi \rangle$ as a function of **P**:

$$\mathbf{F}_\varphi(\mathbf{P}) = \langle \mathbf{P}\varphi \rangle \ . \qquad 5.4$$

It is a tautology to say that this is a *C*-continuous function of **P**. Indeed, recall that the *C*-convergence was defined in Sec. 4.22 by requiring continuity of functions of exactly this form with special $\varphi$ varying between 0 and 1. However, linearity ensures that the expression 5.4 with arbitrary continuous $\varphi$ are also continuous.

If **P** describes a finite number of particles (4.9 and 4.20) or a detector record 4.18, then Eq. 5.4 becomes

$$\mathbf{F}_\varphi(\mathbf{P}) = \sum_{i=1}^{N_{\text{part}}} E_i\, \varphi(\hat{\boldsymbol{p}}_i). \qquad 5.5$$

Note that this expression is obviously invariant with respect to exact fragmentations 4.25. Equally obviously, small variations of angles and energies as well as almost collinear fragmentations (Sec. 4.24) result in small variations of $\mathbf{F}_\varphi$.

### Examples                    5.6

Let $0 \leq \varphi(\hat{\boldsymbol{p}}) \leq 1$ be a continuous function on the unit sphere. It can be thought of as describing the local acceptance of a detector cell. Then the expression 5.5 gives the energy deposited in that cell. A trivial special case ($\varphi(\hat{\boldsymbol{p}}) = 1$) is the total energy of the event:

$$E_{\text{tot}} = \mathbf{E}(\mathbf{P}) \stackrel{\text{def}}{=} \sum_{i=1}^{N_{\text{part}}} E_i \; .$$    5.7

Less trivial examples are the multipole (non-spherically symmetric) moments used as elementary building blocks in the constructions of [22].

Bilinear scalar functions of energy flows    5.8

A standard mathematical construction is to take a tensor product of two or more measures ([45], sec. IV.8). Consider a direct product $S_2 \times S_2$ of two unit spheres (its points are parametrized by a pair of unit 3-vectors, $(\hat{\mathbf{p}}', \hat{\mathbf{p}}'') \in S_2 \times S_2$ or, equivalently, by four angles). Let $\varphi(\hat{\mathbf{p}}', \hat{\mathbf{p}}'')$ be a continuous function on $S_2 \times S_2$. Then the well-defined expression

$$\overline{\varphi}(\hat{\mathbf{p}}') = \int_S \mathrm{d}\hat{\mathbf{p}}'' \, \mathbf{P}''(\hat{\mathbf{p}}'') \varphi(\hat{\mathbf{p}}', \hat{\mathbf{p}}'')$$    5.9

is a continuous function on the unit sphere and one can perform another integration with $\mathbf{P}'$:

$$\int_S \mathrm{d}\hat{\mathbf{p}}' \, \mathbf{P}'(\hat{\mathbf{p}}') \overline{\varphi}(\hat{\mathbf{p}}') \; .$$    5.10

This defines a measure on $S_2 \times S_2$ that can be denoted as

$$[\mathbf{P}' \times \mathbf{P}''](\hat{\mathbf{p}}', \hat{\mathbf{p}}'') = \mathbf{P}'(\hat{\mathbf{p}}') \times \mathbf{P}''(\hat{\mathbf{p}}'').$$    5.11

Correctness of the construction — the required continuity properties etc. — is verified in detail in Theorem 73 of [45]. In particular, the result is independent of the order of integrations.

In particular, one can take $\mathbf{P}' = \mathbf{P}'' = \mathbf{P}$. Applying the arguments of Sec. 5.2 to the measure $\mathbf{P} \times \mathbf{P}$ thus constructed, one obtains a $C$-continuous function that is bilinear in energy flow:

$$\mathbf{F}_{\varphi}(\mathbf{P}) = \int_{S_2} \mathrm{d}\hat{\mathbf{p}}' \int_{S_2} \mathrm{d}\hat{\mathbf{p}}'' \, \mathbf{P}(\hat{\mathbf{p}}') \mathbf{P}(\hat{\mathbf{p}}'') \varphi(\hat{\mathbf{p}}', \hat{\mathbf{p}}'').$$    5.12

If $\mathbf{P}$ has a discrete form of 4.20 (cf. also 4.18), then one obtains:

$$\mathbf{F}_{\varphi}(\mathbf{P} \times \mathbf{P}) = \sum_{ij} E_i E_j \, \varphi(\hat{\mathbf{p}}_i, \hat{\mathbf{p}}_j).$$    5.13

Without loss of generality, one takes $\varphi$ to be symmetric in its two arguments.

For instance, the total invariant mass[i] of an event $\mathbf{P}$ is given (under the usual assumption of masslessness of all particles) by the following special case of 5.13:

$$S_{\text{tot}} = \mathbf{S}(\mathbf{P}) \stackrel{\text{def}}{=} \sum_{ij} E_i E_j (1 - \hat{\mathbf{p}}_i \hat{\mathbf{p}}_j).$$    5.14

Multilinear $C$-correlators    5.15

More generally, let

$$f_m(\hat{\mathbf{p}}_1, \ldots, \hat{\mathbf{p}}_m) \geq 0$$    5.16

be a continuous symmetric function of $m$ unit vectors (i.e. a function on a direct product of $m$ copies of the unit sphere, $S_2^m$; also, the assumption of positivity results in no loss of generality because one can always write $f_m = f_m^+ - f_m^-$ with $f_m^{\pm} \geq 0$). Then the following expression defines a $C$-continuous observable:

---

[i] We are always talking about invariant mass *squared* unless explicitly stated otherwise.

$$\mathbf{F}_m(\mathbf{P}) = \int_{S_2^m} \mathrm{d}\hat{\boldsymbol{p}}_1 \ldots \mathrm{d}\hat{\boldsymbol{p}}_m \; \mathbf{P}(\hat{\boldsymbol{p}}_1) \ldots \mathbf{P}(\hat{\boldsymbol{p}}_m) \, f_m(\hat{\boldsymbol{p}}_1, \ldots, \hat{\boldsymbol{p}}_m). \tag{5.17}$$

The discrete analogue is

$$\mathbf{F}_m(\mathbf{P}) = \sum_{i_1 \ldots i_m} E_{i_1} \ldots E_{i_m} \, f_m(\hat{\boldsymbol{p}}_{i_1}, \ldots, \hat{\boldsymbol{p}}_{i_m}). \tag{5.18}$$

For example, a sequence of observables of this form with specially chosen $f_m$ quantify the feature known as the "number of jets" (the jet-number discriminators of Sec. 8).

## Regularity restriction from IR safety 5.19

In general one would prefer some sort of a more stringent regularity rather than a mere continuity (cf. (iii), Sec. 2.21). For instance, the requirement of IR safety [2] implies that the observables should be e.g. Hölder continuous [51] (it is sufficient that they have bounded first derivatives; cf. Sec. 6.9 below). For $C$-correlators, it is sufficient to impose appropriate restrictions onto their component functions $f_m$ in 5.18. In the cases of practical interest that we will consider $f_m$ happen to be infinitely differentiable.

## Comparison with EEC observables 5.20

A sequence of multiparticle correlators of a special form similar to 5.18 was introduced in [24] (the antenna pattern, the energy-energy correlators etc.; cf. the discussion of their phenomenological applications in [37]). However, those correlators are defined with discontinuous angular functions in the form of cuts (taking values 0 and 1). Therefore, they are, strictly speaking, not $C$-continuous.

## $C$-correlators and "physical information" 5.21

It is convenient to call the observables of the special form 5.17, 5.18 _C-correlators_. They play a central role in our theory:

- They are directly connected with the underlying QFT formalism (Sec. 6.1).

- One can express any interesting physical feature in terms of them (Secs. 5.22 and 8–13).

It therefore makes sense to say that the collection of values of all $C$-correlators on a given final state constitutes the entire physical information about that state.

Note that any discontinuous observable can be approximated (in an appropriate integral sense with respect to integration over all final states) by $C$-continuous ones. From such a viewpoint, no physical information is lost due to the restriction of $C$-continuity.

## $C$-algebra 5.22

By $C$-algebra we understand a collection of $C$-continuous observables built from $C$-correlators following a few simple operations listed below that preserve $C$-continuity. An example involving integration over a parameter already occurred in the construction of bilinear $C$-continuous functions. The basic operations provide a sufficient flexibility to allow one to express jet-related physics in terms of observables from the $C$-algebra. In particular, the spectral discriminators (Secs. 10–11) are built this way. In view of this, whether or not the $C$-algebra comprises all possible $C$-continuous observables becomes an issue of somewhat academic interest and will not be discussed.

Note that limiting procedures may violate $C$-continuity (cf. the counterexample in Sec. 6.9), so some (minimal) care has to be exercised.

## Compositions of $C$-continuous functions 5.23

Obviously, finite algebraic combinations (linear combinations and products) of $C$-continuous functions yield again $C$-continuous functions.

More generally, if a $C$-continuous $\mathbf{F}(\mathbf{P})$ takes values in a region $D$, and $f(z)$ is a continuous function of $z \in D$, then $f(\mathbf{F}(\mathbf{P}))$ is $C$-continuous. For instance, $\exp\mathbf{F}(\mathbf{P})$ is $C$-continuous.

Division is allowed provided the denominator can not become infinitesimally small for the final states one works with. In particular, one can divide the basic observable 5.18 by $m$-th power of the total energy 5.7. Then one only has to deal with energy fractions instead of absolute energies of the particles. (We will use this option in the definition of jet-number discriminators in Sec. 8. This is said to reduce systematic errors due to undetected soft particles.)

A generalization of the above constructions is as follows. Let $\mathbf{H}(\mathbf{P}, F)$ be a function of two arguments, the energy flow and a numeric parameter; suppose it is continuous in its pair of arguments (in the sense of $C$-convergence in its first argument, and in the numeric sense in the second argument). Let $\mathbf{F}(\mathbf{P})$ be another $C$-continuous function taking values of the same nature as the parameter. Then the composition

$$\mathbf{F}'(\mathbf{P}) = \mathbf{H}(\mathbf{P}, \mathbf{F}(\mathbf{P}))  \qquad 5.24$$

is $C$-continuous. This construction is extended straightforwardly to any finite number of arguments of both types and to the case when the second argument has a non-numeric nature (e.g. is a measure).

## Filters and substates     5.25

One often wishes to study certain substates of a final state $\mathbf{P}$ (e.g. when searching for a particle that decays into jets). It would then be natural to treat the substate as a final state in its own right and compute observables for it (e.g. its invariant mass).

The first question is how to specify a group of particles within our formalism, i.e. using the language of $C$-observables. A simple and natural way to do so is with a continuous function $0 \le \Phi(\hat{p}) \le 1$ which we call *filter*. Then the subset consists of those particles for which $\Phi(\hat{p}_i) > 0$; we say that the resulting substate is "filtered" by $\Phi$. The energies of the particles of the substate are taken to be $E_i \Phi(\hat{p}_i)$ with their directions unaffected. Denote the substate thus obtained as

$$\Phi \circ \mathbf{P} \overset{\text{def}}{=} \{E_i \Phi(\hat{p}_i), \hat{p}_i\}_{i=1,\ldots}  \qquad 5.26$$

(cf. Eq. 4.9). The *filtering*

$$\mathbf{P} \to \Phi \circ \mathbf{P}  \qquad 5.27$$

is a $C$-continuous operation: $C$-convergent sequences of final states become $C$-convergent sequences of filtered substates (which, formally, are also elements of the space of final states).

One can take a composition of the mapping 5.27 with any $C$-continuous observable $\mathbf{F}(\mathbf{P})$. The result, $\mathbf{F}(\Phi \circ \mathbf{P})$, is $C$-continuous.

For instance, the invariant mass of the substate $\Phi \circ \mathbf{P}$ is given by the following expression:

$$\mathbf{S}(\Phi \circ \mathbf{P}) = \sum_{i,j} E_i E_j \left(1 - \hat{p}_i \hat{p}_j\right) \times \Phi(\hat{p}_i) \Phi(\hat{p}_j).  \qquad 5.28$$

Similarly, the expression $\mathbf{J}_3(\Phi \circ \mathbf{P})$ measures how well the substate can be characterized as having no less than 3 jets. (The observables $\mathbf{J}_m$ for measuring the "number of jets" are studied in Secs. 8–9.)

## Integration over a parameter     5.29

Another construction that preserves $C$-continuity is integration of a $C$-continuous observable over a parameter (which may be multidimensional):

$$\int d\gamma\, \mathbf{H}(\mathbf{P},\gamma).$$                                                                       5.30

In practice, it is sufficient to require that the domain of variation of $\gamma$ is compact, and that $|\mathbf{H}(\mathbf{P},\gamma)|$ is bounded by a constant independent of $\mathbf{P}$ (cf. the Lebesgue convergence theorem; see e.g. Theorem 34 of [45]).

## Minimization with respect to a parameter                                       5.31

The integration considered above is the most useful example of a construction that involves a limiting procedure but, nevertheless, yields a $C$-continuous function. Another useful example is a minimization (maximization) with respect to a parameter. If the parameter is discrete and takes values from a finite set then the result,

$$\min\left(\mathbf{F}_1(\mathbf{P}),\mathbf{F}_2(\mathbf{P})\right),$$                                                                       5.32

is again a $C$-continuous function. (Extension to more than two functions is obvious.)

With a parameter taking an infinite number of values (e.g. a continuous one) the situation is more tricky because, in general, limiting procedures may violate continuity even in the case of ordinary continuous functions of a real argument. (An example is discussed in Sec. 6.9.) However, the precautions one has to take to avoid such pathologies prove neither difficult to understand nor too restrictive.

Consider the following expression:

$$\min_{\gamma}\mathbf{H}(\mathbf{P},\gamma),$$                                                                       5.33

where $\gamma$ is a continuous parameter varying in a connected compact region, e.g. an interval with both ends included, a sphere, etc. To functions of this class belong the standard observables known as spherocity, acomplanarity, etc. (see [10] for a complete list and references). Of course, the function $\mathbf{H}$ must be continuous in its pair of arguments but that is not enough. The two useful cases when $C$-continuity of 5.33 is ensured are as follows.

The first (obvious) case is when $\mathbf{H}$ is a composition of an ordinary continuous function of two real arguments, $h(x,\gamma)$, and a $C$-continuous function $\mathbf{F}(\mathbf{P})$,

$$\mathbf{H}(\mathbf{P},\gamma) = h(\mathbf{F}(\mathbf{P}),\gamma),$$                                                                       5.34

and $\min_{\gamma} h(x,\gamma)$ is a continuous function of $x$.

The second case is when the angular function $f_m$ of a $C$-correlator 5.18 depends on $\gamma$. The dependence should be such that, roughly speaking, the rate of variation of $f_m$ with respect to its angular arguments is independent of $\gamma$. For instance, $\gamma$ may describe rotations of $f_m$ as a whole (as is the case with spherocity etc.). But one should avoid the cases when $\gamma$ is, say, the radius of the region within which $f_m$ is localized, and $\gamma$ is allowed to go to zero (cf. Sec. 6.13, esp. Eq. 6.15 and the remarks thereafter).

## Differential $C$-continuous observables                                       5.35

It is a common practice to consider differential distributions of events with respect to an observable. Then one considers expressions of the following form:

$$\int d\mathbf{P}\,\pi(\mathbf{P})\,\delta(z - A(\mathbf{P})),$$                                                                       5.36

where integration is over all final states $\mathbf{P}$ against the probability $\pi(\mathbf{P})$ defined by the $S$-matrix (cf. 2.7). Such a construction is also possible with $C$-continuous observables, in particular, with $C$-correlators.

The expression 5.36, however, is too simple to be really useful. Indeed, even to describe the feature known as "number of jets" one properly needs an infinite sequence of scalar observables (Sec. 8). Consider a $C$-continuous observable $\mathbf{F}(\mathbf{P}, \gamma)$ that depends on a parameter $\gamma$ and takes values from the interval [0,1] (which can be achieved by appropriate normalization). Consider the expression

$$\int d\mathbf{P}\,\pi(\mathbf{P})\int d\gamma\,\delta(z - \mathbf{F}(\mathbf{P}, \gamma))\,. \qquad 5.37$$

One can regard this construction as an "extension of phase space", $\mathbf{P} \to \mathbf{P}, \gamma$ and defining a differential observable on the "extended events". If one introduces a normalized weight depending on $\gamma$ (cf. Sec. 13.30) then the analogy with 5.36 is complete.

Eq. 5.37 is equivalent to measuring the observable defined by the following "function" on final states:

$$f(z) = \int d\gamma\,\delta(z - \mathbf{F}(\mathbf{P}, \gamma))\,. \qquad 5.38$$

$f(z)$ is not necessarily a continuous function of $z$ but may, in general, contain singular $\delta$-functional components (see the examples in Sec. 11.11). Therefore it is essential to understand its mathematical nature (in order e.g. to correctly treat such issues as approximations and data errors):

A $\delta$-function is defined by integrals with continuous functions. The same is true for 5.38: Let $\chi(z)$ be continuous for $z \in [0,1]$. Then $\int d\gamma\,\chi(\mathbf{F}(\mathbf{P}, \gamma))$ is $C$-continuous. It can also be represented as follows:

$$\int d\gamma\,\chi(\mathbf{F}(\mathbf{P}, \gamma)) \equiv \int_0^1 dz\,\chi(z)\,f(z)\,. \qquad 5.39$$

The net effect is that the object $f(z)$ defined by 5.38 acquires a numerical meaning after integration with a continuous function. Mathematically, this means that $f(z)$ is an "abstract measure" on $z \in [0,1]$ for each final state $\mathbf{P}$ (for definitions see Sec. 15). This means that Eq. 5.38 represents a measure-valued $C$-continuous observable.

It may be worthwhile to emphasize that the expression 5.38 is, in fact, a shorthand notation for the collection of 5.39 for all allowed $\chi$. Therefore, the convergence of values of 5.38 is to be understood as numerical convergence of *all* expressions 5.39 — without, however, requiring any correlation of the rate of convergence for different $\chi$.

Measure-valued observables emerge most naturally in the context of studying spectral properties of multijet substates. Recall e.g. that the spectral density of a quantum propagator is a measure (a single $\delta$-function in the case of a free particle). More generally, spectral densities of self-adjoint operators (e.g. quantum mechanical Hamiltonians) are also, in general, measures, etc. Measures are singled out from among all distributions by the fact that it makes sense to talk about their positivity — which is exactly why they occur in spectral problems.

The occurrence of measure-valued observables, however, *has nothing to do whatsoever* with the fact that the energy flow of an event is also interpreted as a measure on the unit sphere. Even if all energy flows were ordinary continuous functions, the spectral observables (e.g. the spectral discriminators introduced in Secs. 10–11) could contain singular $\delta$-functional components. Vice versa, a spectral discriminator for an energy flow corresponding to a few isolated particles (and, consequently, represented as a sum of $\delta$-functions on the unit sphere) may happen to be a continuous function.

The above construction becomes particularly useful in combinations with the filtering of Sec. 5.25 — see Sec. 11.1. Further options for construction of $C$-continuous observables are discussed in Sec. 13.29.

## *C*-correlators and Quantum Field Theory                                      6

To compare results of precision measurements with theoretical predictions, one wishes the latter to be as precise and reliable as the former. A high quality of theoretical calculations can only be achieved within the systematic formalism of Quantum Field Theory. The *C*-correlators 5.18 fit well into the framework of QFT. Moreover, the form 5.18 for jet-related observables can be easily obtained if one follows the very basic guidelines of QFT together with minimal continuity requirements motivated by the conclusions of Sec. 2.21. That fact strengthens the central position of *C*-correlators in the theory of jets and opens a prospect for advances in the study of various kinds of corrections — logarithmic perturbative higher-order terms, power-suppressed non-perturbative contributions, etc.

### *C*-correlators from the point of view of QFT                                 6.1

A fundamental ingredient of QFT is the fact that interactions of elementary particles occur via emission/absorption of quanta. Therefore, one deals here with systems with a varying number of particles. Much thought was given to this, and a general, systematic, and well-studied formal scheme for analysis of such systems is provided by the formalism of secondary quantization. Then one works with two-, three-, … particle correlators, each "particle" of a correlator corresponding to an operator that probes a particular physical feature (e.g. energy density etc.)[i] Moreover, one can make a stronger statement, namely, that if a complex physical feature cannot be expressed in terms of multiparticle correlators, the observables indigenous to QFT, then it cannot be a correct observable at all. Recall in this respect also that vacuum averages of products of field operators — the Wightman functions — carry a complete information on QFT models (the Wightman's reconstruction theorem; see e.g. [52]).

It is not difficult to realize that any *C*-correlator 5.18 has exactly the following form of a correlator of Bose-Einstein type:

$$\sum_{1 \le i_1 < \ldots < i_n \le N_{\text{part}}} \sum_{k_1 + \ldots + k_n = m} \frac{m!}{k_1! \ldots k_n!} \; E_{i_1}^{k_1} \ldots E_{i_n}^{k_n} \; f_m(\underbrace{\hat{p}_{i_1}, \ldots, \hat{p}_{i_1}}_{k_1 \text{ times}}, \; \ldots, \; \underbrace{\hat{p}_{i_n}, \ldots, \hat{p}_{i_n}}_{k_n \text{ times}}). \qquad 6.2$$

Because we deal with quantum averages expressed via matrix elements squared rather than wave functions, the Fermi statistics does not occur. The corresponding expression in operator terms has the following general structure:

$$\left\langle p_1, \ldots, p_{N_{\text{part}}} \left| O \right| p_1, \ldots, p_{N_{\text{part}}} \right\rangle,$$

$$O = \int d\mu(q_1) \ldots \int d\mu(q_m) \times j(q_1) \ldots j(q_m) \times f_m(\hat{q}_1, \ldots \hat{q}_m),$$

$$j(q) = E_q \, a^+(\boldsymbol{q}) a(\boldsymbol{q}). \qquad 6.3$$

(μ is the standard 1-particle phase space measure.) Such an operator interpretation leaves open a possibility of their investigation using methods of QFT. An example of such a study is [53]. Ref. [54] reports a representation of 6.3 in terms of the energy-momentum tensor.

The above QFT form of *C*-correlators (cf. also the results of [54]) is to be appreciated in view of the fact that, as follows from the results of Secs. 8–13, it is theoretically sufficient to study *C*-correlators because other jet-related observables can be expressed in their terms.

---

[i] See e.g. [36], [37] for a detailed discussion of such issues. The treatment of [37] particularly emphasizes the role of correlators in studies of multiparticle systems. Note that here again we are dealing with a purely "kinematical" aspect that has never been properly addressed in the theory of jets.

### An alternative derivation of *C*-correlators                                      6.4

The derivation of *C*-correlators in the form 5.17 did not use assumptions about the underlying theory. Then we showed that they have a natural QFT interpretation (Sec. 6.1). Let us now approach the issue from the opposite end: Let us begin with a multiparticle correlator, impose the restriction of fragmentation invariance 4.25, and obtain an observable of the form 5.18. In fact, we will use a somewhat stronger form of fragmentation invariance including a requirement of continuity inspired by the analysis of Sec. 2.5. This and the obligatory correlator form of observables are the two ingredients that were lacking in [2] and [51].

An *m*-particle correlator has the following form with an arbitrary correlator function $f_m$:

$$\mathbf{F}_m(\mathbf{P}) = \sum_{i_1 \ldots i_m} f_m(p_{i_1}, \ldots, p_{i_m}).$$  6.5

Consider for simplicity the case $m = 2$. First, for an integer $n$ the fragmentation invariance yields $\mathbf{F}_2(np) = \mathbf{F}_2(p, p, \ldots, p)$, where the argument on the r.h.s. is a state with $n$ equal particles. From this and 6.5 one obtains $f_2(np, np) = n^2 f_2(p, p)$. From this and a similar restriction with $k$ instead of $n$ one obtains a similar restriction with a rational $r = n/k$ instead of integer $n$. The condition that the observables should be insensitive to adding soft particles results in continuity with respect to energies whence follows a similar scaling for any real $r$. Then $f_2(p, p) = E^2 f_2(\hat{p}, \hat{p})$. In a similar way one obtains $f_2(p, nq) = n f_2(p, q)$ for any integer $n$, etc. — until the desired form of energy dependence is obtained. Imposing a simplest requirement of continuity with respect to the angular variables (= minimal sensitivity to almost collinear fragmentations, as motivated by the analysis of Sec. 2.5) results in the restriction of continuity of the angular function.

### Fragmentation invariance of *C*-correlators                                      6.6

The combinatorial structure of *C*-correlators can be studied most easily with the help of the following representation in terms of functional derivatives:

$$\text{Eq. 5.18} = \frac{1}{m!} \mathcal{D}^m \int d\hat{q}_1 \ldots d\hat{q}_m \, f_m(\hat{q}_1, \ldots, \hat{q}_m) \, \eta(\hat{q}_1) \ldots \eta(\hat{q}_m),$$  6.7

where

$$\mathcal{D} = \sum_{i=1}^{N_{\text{part}}} E_i \frac{\delta}{\delta \eta(\hat{p}_i)} .$$  6.8

Now all the dependence on the particle content of the state is localized within $\mathcal{D}$. In particular, the fragmentation invariance of 5.18 and 6.2 (with respect to exact fragmentations 4.25) is an obvious consequence of the linearity of $\mathcal{D}$ in particles' energies.

### *C*-continuity vs. IR safety                                      6.9

An interesting general issue is how *C*-continuity relates to the IR safety, the notion familiar from the QCD theory of hadron jets [2]. The issue of IR safety is usually discussed in the context of perturbative QCD where final states consist of a finite number of partons (quarks and gluons). Let $\mathbf{F}(\mathbf{P})$ be an arbitrary function on final states. Define its component functions $F_n$ by restricting $\mathbf{F}$ to states with finite numbers of particles, as follows:

$$\mathbf{F}(\mathbf{P}) = F_n(\boldsymbol{p}_1, \ldots, \boldsymbol{p}_n),$$  6.10

where for simplicity each of the $n$ particles of the final state is represented by its 3-momentum.

Since the multiparticle state does not change if the particles are permuted, each $F_n$ must be symmetric in its arguments. An observable that can be measured by calorimeters has to be [2], as a minimum, fragmentation invariant, i.e. invariant with respect to exact collinear fragmentations, whence [51]:

$$F_n(\boldsymbol{p}_1,...,\boldsymbol{p}_n) = F_{n+1}(\boldsymbol{p}_1,...,(1-z)\boldsymbol{p}_n, z\boldsymbol{p}_n), \quad 0 \le z \le 1.\qquad 6.11$$

This condition of fragmentation invariance is, of course, well-known from the studies of IR safety in the context of QCD [2], [51]. In particular, if a QCD observable is to be IR finite for a multiparticle (quark and gluon) final state order by order in perturbation theory, then in addition to fragmentation invariance, the functions $F_n$ should satisfy some minimum regularity requirements [51]. It is sufficient that each $F_n$ has bounded (not necessarily continuous) derivatives with respect to its arguments:

$$|f(x) - f(y)| \le M|x - y|,\qquad 6.12$$

for any sufficiently close pair $x$, $y$. Fragmentation invariance plus the above regularity amount to IR safety in the usual sense.

However, such conditions impose no restriction on how fast the functions $F_n$ vary as $n \to \infty$. In particular, there is no limit on how fast *the derivatives* of $F_n$ grow as $n \to \infty$. In the notations of 6.12, this means that $M$ may depend on $n$, and $M$ may grow arbitrarily fast as $n \to \infty$.

## A continuous and perturbatively IR safe function that is not *C*-continuous    6.13

Here is an example of a fragmentation invariant function on final states whose component functions vary arbitrarily fast. Consider the following function:

$$\mathbf{F}_{\mathrm{bad}}(\mathbf{P}) = \sum_{m=2}^{\infty} \mathbf{F}_m(\mathbf{P}),\qquad 6.14$$

with $\mathbf{F}_m$ defined by 5.18, and

$$f_m(\hat{\boldsymbol{p}}_1,...,\hat{\boldsymbol{p}}_m) = \prod_{1 \le i < j \le m} \left( \min[1, \theta_{ij}/z_m] \right),\qquad 6.15$$

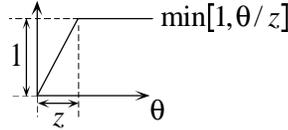

where $\theta_{ij}$ is the angle between $\hat{\boldsymbol{p}}_i$ and $\hat{\boldsymbol{p}}_j$ and $z_m > 0$. The function 6.15 nullifies when any pair of its arguments is collinear so that the series in 6.14 is truncated, therefore well-defined, and satisfies 6.12, in each order or perturbation theory. It follows that it is perturbatively IR safe.

To see that it is not necessarily *C*-continuous, consider a one-particle final state $\mathbf{P} = \{E, \hat{\boldsymbol{p}}\}$. One can see that $\mathbf{F}_{\mathrm{bad}}(\mathbf{P}) = 0$. Now fix an infinite sequence of states $\mathbf{P}_n$ such that each of them (i) has the total energy exactly equal to $E$; (ii) consists of $n$ pairwise non-collinear particles with directions localized within a cone of angular radius $n^{-1}$ around $\hat{\boldsymbol{p}}$. Then $\mathbf{P}_n \to \mathbf{P}$ in the sense of *C*-continuity. However, the values $\mathbf{F}_{\mathrm{bad}}(\mathbf{P}_n)$ are expressed in terms of $F_m$, $m \le n$, and choosing $z_m \to 0$ for $m \to \infty$ fast enough, one can ensure that the values $\mathbf{F}_{\mathrm{bad}}(\mathbf{P}_n)$ do not converge to $\mathbf{F}_{\mathrm{bad}}(\mathbf{P}) = 0$.

Although one is not likely to encounter an infinite sum of *C*-correlators like 6.14 in applications, the moral of the example is that manipulations involving limiting procedures may violate *C*-continuity. This may be the case e.g. with maximization/minimization with respect to a parameter which affects the rate of variation of the functions describing angular dependences (cf. the dependence on $z_m$ in 6.15).

## Aspects of theoretical calculations                                             6.16

The two general issues to be considered here are standard perturbative calculations and the problem of power corrections. One should also distinguish two types of observables: the basic class of relatively simple $C$-correlators (Sec. 5.1; the most important example of $C$-correlators are the jet-number discriminators, Sec. 8) and the complex observables such as spectral discriminators (Sec. 10).

## Analytical calculations of $C$-correlators                                      6.17

Consider theoretical calculations of such observables as the jet-number discriminators of Sec. 8. In general, if one deals with hadronic processes and non-$4\pi$ detector geometries, the perturbative calculations should follow the standard scheme: first one evaluates the matrix elements, and then one performs integrations (via Monte Carlo) over the phase space. However, the highly regular analytical structure of jet-number discriminators opens a possibility of their direct analytical evaluation — at least for the kinematically simplest but fundamental annihilation process $e^+e^- \to$ hadrons.

Indeed, the expressions of the jet-number discriminators $\mathbf{J}_m$, Eq. 8.8, differ from the case of total cross sections only by the weights that have to be inserted into the phase space integrals. The weights are built of scalar products (cf. 13.4), and their form is such that calculations of a large number of diagrams is greatly simplified: due to masslessness, $2p_i p_j = (p_i + p_j)^2$, and if partons $p_i$ and $p_j$ resulted from a decay of another virtual parton, the denominator of the propagator of the latter is canceled. There are still powers of energies in the denominator (the factors $(p_i P)^{-1}$ in 13.4) but their dependence on partons' momenta is linear which in some cases simplifies calculations. Of course, there are also diagrams in which the above cancellation does not occur, but nevertheless the situation here is rather less difficult than in the case of other shape observables that can only be handled numerically.

Another nice feature here is that since there are no phase space cutoffs involved, the standard renormalization group equations are sufficient to perform resummation of logarithms — neither Sudakov resummations are necessary, nor factorization theorems much different from those for total cross sections [51], [55].

It is interesting to compare the case of jet-number discriminators $\langle \mathbf{J}_m \rangle$ with the case of total cross section $\sigma_{\text{tot}}(e^+e^- \to \text{hadrons})$ which is nothing but $\langle \mathbf{J}_2 \rangle$ (up to a numerical coefficient). For this quantity, analytical calculations were pushed through next-next-to-leading order [33]. That feat was made possible by the calculational methods [34], [35], [36] that reduced the calculation to a rather mechanical feeding of the corresponding diagrams into a computer [56]. Unfortunately, the methods of [34], [35], [36] are based in an essential way on the possibility to Wick-rotate the unitarity diagrams in the case of $\sigma_{\text{tot}}(e^+e^- \to \text{hadrons}) \sim \langle \mathbf{J}_2 \rangle$ into Euclidean region. But that is impossible (at least in a straightforward manner) for $\langle \mathbf{J}_m \rangle$. $m > 2$, so the algorithms of [34], [35], [36] cannot be employed. Therefore, although the cancellations mentioned above actually make calculation of quite a few diagrams for $\langle \mathbf{J}_m \rangle$. $m > 2$, easier than for $\langle \mathbf{J}_2 \rangle \sim \sigma_{\text{tot}}(e^+e^- \to \text{hadrons})$, nevertheless one should not expect the NNL-order calculations to be doable in an entirely analytical fashion. But a combination of analytical and numerical techniques (cf. e.g. [57]) may work although the calculations remain very hard.

Note that for the purposes of precision measurement of $\overline{\alpha}_s$, it should probably be amply sufficient to have NNL corrections to $\langle \mathbf{J}_3 \rangle$ (i.e. three terms in the QCD expansion in $\overline{\alpha}_s$). The NL corrections for $\langle \mathbf{J}_4 \rangle$ are of about the same calculational complexity while the leading term for $\langle \mathbf{J}_5 \rangle$ should actually be simpler. Notice that the QCD expansion for $\langle \mathbf{J}_m \rangle$, $m > 2$ in the case of $e^+e^- \to$ hadrons starts at $O(\overline{\alpha}_s^{m-2})$; the values of $\langle \mathbf{J}_4 \rangle$ and $\langle \mathbf{J}_5 \rangle$ can be used as a check for the rate of convergence of perturbation series.

Power corrections                                                                              6.18

As was discussed in [37], power corrections in the physics of jets seem to be numerically important (although, strictly speaking, it remains to be seen whether they will have the same numerical significance for, say, the jet-number discriminators). On the other hand, a theoretical study of hadronization effects reduces simply to (i) computing logarithmic corrections (via higher order perturbative contributions as discussed above) and (ii) studying and estimating power corrections to $\langle \mathbf{J}_m \rangle$. Our formalism offers an option for a theoretical study of power corrections.

Recall the case of $e^+e^- \to$ hadrons where one has $\langle \mathbf{J}_2 \rangle \propto \sigma_{\text{tot}}$, and where the structure of $\sigma_{\text{tot}}$ including power corrections is well known. In this case $\sigma_{\text{tot}}$ can be connected via a dispersion relation with the vacuum average of a chronological product of two currents, so that its asymptotic behavior is connected to the Wilson operator product expansion [58]. Then power corrections emerge as vacuum averages of local operators of the Wilson expansion (the so-called vacuum condensates) with perturbatively calculable coefficients. This was used for phenomenological purposes in [59].

From a theoretical viewpoint, it was demonstrated in [60], [61] that such vacuum condensates are directly related to soft singularities of the expansion of the correlator in quark masses. Moreover, the structure of condensates can be determined explicitly even within perturbation theory [60]. The vacuum condensates are perturbatively uncalculable but they may be studied e.g. using the methods of lattice QCD. On the other hand, the coefficients with which the condensates enter the expression for the cross section, and which contain all the dependence on the energy of the process, are perturbatively calculable (cf. e.g. the two-loop calculations of [63]).

A similar procedure can be used to study the structure of $\langle \mathbf{J}_m \rangle$ for $m > 2$: One would start with the corresponding perturbative expressions (a sum over Feynman diagrams with massive quarks etc.), and perform the expansion in powers and logarithms of quark masses. The non-analytic contributions associated with IR singularities (soft and collinear) would be organized into some kind of operators (cf. the formalism and results of [61]).

Note that the presence of phase space weights in the expressions for jet-number discriminators does not allow one to perform the Wick rotation and reduce the integrals to Euclidean ones in the usual way. Therefore, one would need a non-Euclidean extension of the method of asymptotic operation used in [61]. Such an extension is feasible [62], [64].

## Computing *C*-continuous observables from data          7

In the preceding section we discussed theoretical calculability of *C*-correlators. In this section we consider the problem of their optimal computation from data.[i] The point is that although the defining formula 5.18 is simple, the volume of arithmetic involved may be rather large. For instance, for a final state that lit up 200 calorimeter modules, computation of the 5-th jet-number discriminator $\mathbf{J}_5$ (Eq. 8.8) involves adding ~3×10$^9$ terms, each containing 5 energy and 10 angular factors. Therefore, a straightforward approach may be unacceptable. However, one should bear in mind the following:

• Such large numbers of particles in the final state are more typical for future colliders such as LHC, and by then computing power will become many times cheaper.

• The quality and amount of information from *C*-observables is higher than in the case of the conventional processing. The resulting safety margin leaves room for approximations.

• The very regular analytic form of *C*-correlators 5.18 allows a number of optimizations.

In what follows we concentrate on purely analytical optimizations and ignore programming tricks (e.g. parallelization). Also, an ideal optimization method may depend, in general, on whether the event one deals with is "fuzzy" or has needle-like jets, etc., so that a sophisticated computational scheme may depend on a concrete data sample. Therefore, only some typical options are described below.

### Understanding the problem          7.1

We are going to investigate optimal ways to compute the value $\mathbf{F}(\mathbf{P})$ of a *C*-correlator 5.18 on one given event $\mathbf{P}$. The event is represented by a data record of the form 4.18, and each calorimeter module is treated as a particle. The correctness of this is ensured by the *C*-continuity by definition of the latter.

### Sources of optimizations          7.2

In the case of *C*-observables there are three (groups of) properties that can be made use of for optimization:

1) The *C*-continuity (recall that it determined the energy dependence of *C*-correlators 5.18) which comes in two flavors (cf. Sec. 4.24):

1a) Stability (i.e. continuous variation) with respect to almost collinear fragmentations. The small parameter that can be utilized here is a small angle between the fragments.

1b) Stability with respect to adding soft particles, i.e. analyticity in particles' energies. The small parameter here may be taken to be the total energy fraction of soft particles.

2) Information on a concrete form of the angular dependence, i.e. on the function $f_m(\hat{\boldsymbol{p}}_1, \ldots \hat{\boldsymbol{p}}_m)$.

3) Information on the structure of the event $\mathbf{P}$.

The properties 2) and 3), we only touch upon. The main focus below is on the *C*-continuity.

### Criterion of optimization          7.3

Unlike the conventional paradigm based on jet finding algorithms where observables are defined in terms of the output of such algorithms, in our theory observables are defined directly in terms of unprocessed events (final states). Therefore, the issue of what an observable should be is cleanly separated from how it is to be computed. The first advantage of such a separation is that once the observable is explicitly defined, one is free to use any tricks in order to compute it (including jet algorithms). The second advantage is that one now has a clear and unambiguous criterion to choose among optimization tricks: The criterion is the numeric quality of the resulting approximation.

---

[i] There is no need to consider general *C*-continuous observables here because they are built from *C*-correlators.

## Soft particles                                                         7.4

We begin by considering the property 1b) from Sec. 7.2. We will first show how the analyticity of $C$-correlators at zero particles' energies can be exploited to obtain expansions in energies of soft particles. Then we consider an important special case — the problem of estimating effects of undetected particles (the so-called missing energy; Sec. 7.14). The very fact that the latter problem can be meaningfully approached in the framework of the new formalism is a clear demonstration of its analytical superiority over the conventional scheme.

### Expanding in energies of soft particles                               7.5

The expression 5.18 allows one to perform Taylor expansion in energies of particles. Divide all particles of the event into two groups, "soft" and "hard":

$$\mathbf{P} = \mathbf{P}_{soft} \oplus \mathbf{P}_{hard} \ . \tag{7.6}$$

Then one can expand 5.18 as follows (e.g. using 6.7):

$$\mathbf{F}(\mathbf{P}) = \mathbf{F}(\mathbf{P}_{hard}) + \mathbf{F}^{(1)}(\mathbf{P}_{hard};\mathbf{P}_{soft}) + \mathbf{F}^{(2)}(\mathbf{P}_{hard};\mathbf{P}_{soft}) + \dots, \tag{7.7}$$

where $\mathbf{F}(\mathbf{P}_{hard})$ is the value of the observable $\mathbf{F}$ computed on the state $\mathbf{P}_{hard}$ that consists of only "hard" particles, and

$$\mathbf{F}^{(1)}(\mathbf{P}_{hard};\mathbf{P}_{soft}) = m \sum_{i_1,\dots,i_{m-1}\in\mathbf{P}_{hard}} E_{i_1}\dots E_{i_{m-1}} \sum_{i_m\in\mathbf{P}_{soft}} E_{i_m} f_m(\hat{\boldsymbol{p}}_{i_1},\dots,\hat{\boldsymbol{p}}_{i_m}), \tag{7.8}$$

$$\mathbf{F}^{(2)}(\mathbf{P}_{hard};\mathbf{P}_{soft}) = \frac{m(m-1)}{2} \sum_{i_1,\dots,i_{m-2}\in\mathbf{P}_{hard}} E_{i_1}\dots E_{i_{m-2}} \sum_{i_{m-1},i_m\in\mathbf{P}_{soft}} E_{i_{m-1}} E_{i_m} f_m(\hat{\boldsymbol{p}}_{i_1},\dots,\hat{\boldsymbol{p}}_{i_m}). \tag{7.9}$$

To see the computational savings, assume the numbers of hard and soft particles are equal. Then computing just the first term in 7.7, i.e. $\mathbf{F}(\mathbf{P}_{hard})$, involves $2^m$ times fewer terms than the complete calculation. Taking into account the first correction doubles the number of terms, which still means $2^{m-1}$ times fewer terms than the complete calculation.

### Estimating the error                                                   7.10

As a simple example, consider the error involved in retaining only the first term $\mathbf{F}(\mathbf{P}_{hard})$ on the r.h.s. of 7.7. The error can be estimated from the second term given by 7.8. There is a spectrum of possibilities for writing a bound for it. The following inequality seems to be adequate:

$$\left|\mathbf{F}^{(1)}(\mathbf{P}_{hard};\mathbf{P}_{soft})\right| \le m\,\varepsilon_{miss}\,\mathbf{E}(\mathbf{P}_{hard})\mathbf{F}^{[1]}(\mathbf{P}_{hard}), \tag{7.11}$$

where the $C$-correlator $\mathbf{E}$ is defined in 5.7,

$$\varepsilon_{miss} = \frac{\mathbf{E}(\mathbf{P}_{soft})}{\mathbf{E}(\mathbf{P}_{hard})} \ , \tag{7.12}$$

and $\mathbf{F}^{[1]}(\mathbf{P}_{hard})$ is the value on the hard subsystem $\mathbf{P}_{hard}$ of the following $C$-correlator:

$$\mathbf{F}^{[1]}(\mathbf{P}) = \sum_{j_1,\dots,j_{m-1}\in\mathbf{P}} E_{j_1}\dots E_{j_{m-1}} f^{[1]}_{m-1}(\hat{\boldsymbol{p}}_{i_1},\dots,\hat{\boldsymbol{p}}_{i_{m-1}}),$$

$$f^{[1]}_{m-1}(\hat{\boldsymbol{p}}_1,\dots,\hat{\boldsymbol{p}}_{m-1}) = \sup_{\hat{\boldsymbol{p}}} f_m(\hat{\boldsymbol{p}}_1,\dots,\hat{\boldsymbol{p}}_{m-1},\hat{\boldsymbol{p}}). \tag{7.13}$$

It is assumed here that the angular function $f_m$ is non-negative; cf. Sec. 5.15.

Similar estimates can be obtained for higher terms.

## Errors due to missing energy　　　　　　　　　　　　　　　　　　　　7.14

The factor $m$ in 7.11 indicates that the higher $C$-correlators are increasingly more sensitive to the errors of such an approximation. In particular, they are increasingly more sensitive to errors due to undetected particles.

To estimate the error due to missing energy, i.e. the particles that escaped undetected, one simply uses 7.11, treating all missing particles as "soft" irrespective of their actual energies. (The estimate 7.11 does not assume that the individual particles are soft in any sense.) The missing energy enters the r.h.s. of 7.11 only via ε. All other quantities are directly measurable.

## Missing energy for jet-number discriminators　　　　　　　　　　　　7.15

The above estimates did not employ any concrete information on the angular function $f_m$. Let us show how such information can be made use of. As an example, we will estimate the effect of missing energy for the case of the jet-number discriminators 8.8. (Their concrete physical meaning is of no importance here.) We are going to obtain estimates of a type somewhat different from 7.11 for the *defect* defined as follows:

$$\operatorname{def} \mathbf{J}_m(\mathbf{P}) \stackrel{\text{def}}{=} \mathbf{J}_m(\mathbf{P}) - \mathbf{J}_m(\mathbf{P}_{\text{hard}}) > 0. \qquad\qquad 7.16$$

It is important here that all terms in the expression for $\mathbf{J}_m(\mathbf{P})$ 8.8 are non-negative.

The reasoning below will be in terms of energy fractions rather than absolute energies (cf. the normalization in 8.8).

Assume that the total energy fraction of missing particles $\varepsilon_{\text{soft}}$ (it is always less than $\varepsilon_{\text{miss}}$, Eq. 7.12) is distributed uniformly over the unit sphere and their total number is large. Then an estimate can be obtained by replacing the summation over the soft particles by an integration over the unit sphere. For definiteness, consider the 3rd jet-number discriminator $\mathbf{J}_3(\mathbf{P})$. The resulting expression is as follows:

$$\operatorname{def} \mathbf{J}_3(\mathbf{P}) = \tfrac{27}{4}\,\varepsilon_{\text{soft}} \sum_{i<j} E_i E_j\, \Delta_{ij} \times (1 + \tfrac{1}{3}\hat{\boldsymbol{p}}_i\hat{\boldsymbol{p}}_j) + O(\varepsilon_{\text{soft}}^2), \qquad 7.17$$

where the summation is restricted to "hard" particles. The factor in parentheses is between 2/3 and 4/3. If one simply estimates it by 1 (numerical experimentation may suggest a different value), one obtains:

$$\operatorname{def} \mathbf{J}_3(\mathbf{P}) \approx 3.4\,\varepsilon_{\text{soft}} + O(\varepsilon_{\text{soft}}^2). \qquad\qquad 7.18$$

Similarly,

$$\operatorname{def} \mathbf{J}_4(\mathbf{P}) \approx 5.3\ \varepsilon_{\text{soft}}\ \mathbf{J}_3(\mathbf{P}_{\text{hard}}) + O(\varepsilon_{\text{soft}}^2). \qquad\qquad 7.19$$

One could use the expressions thus obtained as corrections to the numbers computed from data in order to reduce the effects of missing energy.

## Preclustering　　　　　　　　　　　　　　　　　　　　　　　　　7.20

Let us turn to the property 1a). It can be directly utilized to derive an important approximation that consists in recombining (groups of) particles of a given final state into one provided this affects the observables to be computed only within specified errors. We call this *preclustering*, and it bears a resemblance to the conventional jet algorithms. But there are also differences which it is important to elucidate.

The preclustering is simply a computational approximation trick aimed at reducing the amount of arithmetic needed to compute $C$-continuous observables by exploiting the property of $C$-continuity shared by *all* $C$-continuous observables.[i] Its numerical effect, however, depends on the form of a concrete observable; for some observables it works better than for others. Our preclustering has a mathematically well-defined purpose and involves a parameter (see below) that directly controls the resulting approximation errors. Therefore, all arbitrariness is to be judged against the well-defined criterion of whether or not one can achieve the desired precision for specific observables using available computer resources.

The small parameter which the preclustering exploits is the small angle between almost collinear fragments. However, the specific range of what constitutes "small" angles will be seen to depend on the energies of the particles. This should be no surprise because the energy dependence of $C$-correlators is fixed and known (cf. 5.18).

Because we will be considering here $C$-correlators that are uniform functions of energies, we can take all particles' energies $E_i$ to be energy fractions rather than absolute energies:

$$E_i \leftarrow E_i \, / \, E_{\text{tot}}. \qquad\qquad 7.21$$

Such an assumption entails no loss of generality.

## On masslessness of pseudoparticles                                          7.22

Before we turn to formulas, the following subtlety should be emphasized. The preclustering replaces one energy flow (as defined in our formalism) with another object of exactly the same type. Therefore, the pseudoparticles that emerge in our case (and that are analogous to the protojets of the conventional algorithms) are also to be formally interpreted as massless. But one should not try to assign a profound physical meaning to this fact because we are simply dealing with a computational trick here.

On the other hand, in the conventional algorithms protojets may emerge with non-zero masses. Incorporating non-zero pseudoparticles' masses within our formalism would be equivalent to constructing approximations for an observable not in terms of the same observable — which is the case that we are considering — but in terms of a different one. This may be an interesting option, but we only concentrate on the optimizations that do not require a detailed knowledge of the structure of observables. In that sense, our formulas are universal.

One also notices that the recombination is defined unambiguously only for infinitesimal angles between particles (cf. below). The energy and momentum conservation of the conventional algorithms can be regarded as ways to extend the recombination criteria to finite angular separations while preserving maximum information about the event.

## Estimating errors induced by recombinations                                 7.23

Let us begin by considering the simplest $C$-correlators that are linear in energies. This is meaningful because the general $C$-correlators 5.18 can be obtained from these using algebraic combinations and appropriate limiting procedures (Sec. 5.22). We will see that the criterion we are going to derive remains essentially the same in the general case.

Consider the following correlator:

$$\mathbf{F}(\mathbf{P}) = \sum\nolimits_i E_i \, f(\hat{\boldsymbol{p}}_i). \qquad\qquad 7.24$$

On a state $\mathbf{P}' = p \oplus \mathbf{P}$ with one particle, $p$, singled out, the correlator becomes

$$\mathbf{F}(p \oplus \mathbf{P}) = E \, f(\hat{\boldsymbol{p}}) + \sum\nolimits_i E_i \, f(\hat{\boldsymbol{p}}_i). \qquad\qquad 7.25$$

---

[i] In fact, experimentalists routinely use a similar preclustering of the raw data except that jet-algorithm-based observables are computed in the end instead of $C$-continuous observables, and using criteria different from the optimal one explained below in this section. I thank D. A. Stewart, Jr. for explaining this to me.

Replace $p$ with two particles $p_a$ and $p_b$:

$$\mathbf{F}(p_a \oplus p_b \oplus \mathbf{P}) = E_a\, f(\hat{\boldsymbol{p}}_a) + E_b\, f(\hat{\boldsymbol{p}}_b) + \sum_i E_i\, f(\hat{\boldsymbol{p}}_i)\,. \qquad 7.26$$

To compare 7.26 and 7.25, consider their difference:

$$\mathbf{F}(p_a \oplus p_b \oplus \mathbf{P}) - \mathbf{F}(p \oplus \mathbf{P}) = E_a\, f(\hat{\boldsymbol{p}}_a) + E_b\, f(\hat{\boldsymbol{p}}_b) - E\, f(\hat{\boldsymbol{p}})\,. \qquad 7.27$$

At this point, it is convenient to formally extend the function $f$ to non-unit 3-vectors by imposing the condition

$$f(\lambda \hat{\boldsymbol{q}}) = f(\hat{\boldsymbol{q}})\,, \qquad \text{for any } \hat{\boldsymbol{q}}\,. \qquad 7.28$$

Assuming that the angles between each of the fragments and the initial particle $p$ are small, we are going to Taylor-expand the expression 7.27 in $\hat{\boldsymbol{p}}_a$ and $\hat{\boldsymbol{p}}_b$ around $\hat{\boldsymbol{p}}$, using the following formula:

$$f(\hat{\boldsymbol{p}}_\#) = f(\hat{\boldsymbol{p}}) + \Delta\hat{\boldsymbol{p}}_\#\, f'(\hat{\boldsymbol{p}}) + \tfrac{1}{2}(\Delta\hat{\boldsymbol{p}}_\#)^2\, f''(\hat{\boldsymbol{p}}) + \dots. \qquad 7.29$$

(Because one deals with vector arguments here, the products are tensorial.) We obtain:

$$\begin{aligned} &(E_a + E_b - E)\, f(\hat{\boldsymbol{p}}) \\ &+ (E_a\Delta\hat{\boldsymbol{p}}_a + E_b\Delta\hat{\boldsymbol{p}}_b)\, f'(\hat{\boldsymbol{p}}) \\ &+ \big[E_a(\Delta\hat{\boldsymbol{p}}_a)^2 + E_b(\Delta\hat{\boldsymbol{p}}_b)^2\big]\tfrac{1}{2}\, f''(\hat{\boldsymbol{p}}) + \dots \end{aligned} \qquad 7.30$$

Requiring that the first two lines nullify, we obtain:

$$E_a + E_b = E\,, \qquad 7.31$$

$$E\,\hat{\boldsymbol{p}} = E_a\hat{\boldsymbol{p}}_a + E_b\hat{\boldsymbol{p}}_b\,. \qquad 7.32$$

The latter relation cannot be satisfied by three unit vectors — but that is of little import here: First, the relation is only meant to be exact for infinitesimal angles, and its extension to finite angles involves an arbitrariness that cannot be avoided in principle. Second, the trick of defining $f$ for non-unit vectors, Eq. 7.28, offers a simple variant of such an extension, and it allows one to simply enforce a correct normalization whenever necessary.

The remainder is bounded by the following expression:

$$\tfrac{1}{2}\big[E_a|\Delta\hat{\boldsymbol{p}}_a|^2 + E_b|\Delta\hat{\boldsymbol{p}}_b|^2\big] \times M_f\,, \qquad 7.33$$

where $M_f \geq 0$ involves a maximal value of 2nd order derivatives of $f$ (we will not need its precise form). Notice that

$$\Delta\hat{\boldsymbol{p}}_a = (\hat{\boldsymbol{p}}_a - \hat{\boldsymbol{p}}_b)\frac{E_b}{E}\,, \quad \Delta\hat{\boldsymbol{p}}_b = -(\hat{\boldsymbol{p}}_a - \hat{\boldsymbol{p}}_b)\frac{E_a}{E}\,. \qquad 7.34$$

Finally:

$$|\mathbf{F}(p_a \oplus p_b \oplus \mathbf{P}) - \mathbf{F}(p \oplus \mathbf{P})| \leq \frac{E_a E_b}{E} \times \tfrac{1}{2}|\hat{\boldsymbol{p}}_a - \hat{\boldsymbol{p}}_b|^2 \times M_f\,. \qquad 7.35$$

## Modifications for general $C$-correlators       7.36

The above reasoning is extended in a straightforward manner to the case of the general $C$-correlators 5.18. (This is done easiest using the representation 6.7.) One obtains

$$|\mathbf{F}(p_a \oplus p_b \oplus \mathbf{P}) - \mathbf{F}(p \oplus \mathbf{P})| \leq \frac{E_a E_b}{E} \times \tfrac{1}{2}|\hat{\boldsymbol{p}}_a - \hat{\boldsymbol{p}}_b|^2 \times \sum_{k=0}^{m-1} E^k M_f^{(k)}\,, \qquad 7.37$$

where $M_f^{(k)}$ are constants that depend on second order derivatives of $f$. Because $E < 1$, this reduces to 7.35 (with a different $M_f$). Therefore, the form of dependence on energies and angle in 7.35 is completely general — as was to be expected in view of the construction of multilinear observables from linear ones described in Sec. 5.8.

## The optimal preclustering prescription                                                7.38

Now suppose we wish to simplify computation of some $C$-continuous observables via reducing the number of particles by combining a pair of particles, $E_a, \hat{\boldsymbol{p}}_a$ and $E_b, \hat{\boldsymbol{p}}_b$, into one, $E, \hat{\boldsymbol{p}}$. From 7.35 one sees that the criterion to combine two particles into one is

$$\frac{E_a E_b}{E} \times \Delta_{ab} < y_{\text{cut}} , \qquad\qquad\qquad 7.39$$

where

$$\Delta_{ab} \stackrel{\text{def}}{=} \tfrac{1}{2} |\hat{\boldsymbol{p}}_a - \hat{\boldsymbol{p}}_b|^2 = 1 - \cos\theta_{ab}, \qquad\qquad\qquad 7.40$$

and the parameter $y_{\text{cut}}$ effectively controls the error induced thereby. The actual error differs from $y_{\text{cut}}$ by a factor which depends on the observable and is best determined via numerical experiments.

The energy of the new particle is determined from 7.31. Its direction is found from 7.32 (with the normalization enforced as described after 7.32).

Recall that one deals with energy fractions in 7.39 (cf. Eq. 7.21).

## Comments on the definition of $\Delta_{ab}$                                            7.41

The optimal preclustering criterion 7.39 has been rephrased, as compared with the underlying inequality 7.35, in terms of the angular separation $\Delta_{ab}$ defined in Sec. 4.5. Eq. 7.40 involves an extension to large $\theta_{ab}$ which in principle is arbitrary, whereas the bound 7.35 was derived only for infinitesimal $\theta_{ab}$. The considerations behind the definition 7.40 were discussed in Sec. 4.5.

It should be clearly understood that such an extension, theoretically speaking, can be chosen arbitrarily as far as preclustering is concerned. Indeed, whereas in the conventional approach the radius of jets (or $y_{\text{cut}}$ which effectively controls it) is physically significant (because e.g. reconstructed masses of multijet substates — therefore, masses of new particles — depend on it), in our formalism the preclustering is only an approximation trick (e.g. the spectral discriminators described in Secs. 10–11 will have a bump due to a new particle at the right value of invariant mass irrespective of whether or not the preclustering is used). Therefore, if the arbitrariness in the definition of the angular separation turns out to be numerically important for the chosen $y_{\text{cut}}$, one should conclude that the approximation errors due to the preclustering are out of control and $y_{\text{cut}}$ should be reduced.

Another point one should have in view is that the vector norm on the r.h.s. of 7.35 can be any vector norm, not necessarily the Euclidean one. This means that any alternative $\Delta'_{ab}$ such that

$$c_1 \Delta_{ab} < \Delta'_{ab} < c_2 \Delta_{ab}, \quad \text{for some } c_{1,2} > 0, \q\qquad\qquad\qquad 7.42$$

is allowed. The preclustering criterion 7.39 (and also the criteria discussed in Sec. 7.47) based on such $\Delta'_{ab}$ instead of $\Delta_{ab}$ is equivalent to the one based on $\Delta_{ab}$ but may result, e.g., in a faster code (cf. Sec. 13.13).

In general, the overriding consideration for the choice of the angular separation used in the preclustering algorithms is computational simplicity.

### Comparison with the conventional algorithms 7.43

It is interesting to treat the above "optimal" algorithm as a conventional jet algorithm because one can use the criterion 7.39 in a Luclus-type iterative recombination scheme [14]. (Note, however, that a better way is to use the $n \to 1$ variant of preclustering described in Sec. 7.47.) Various recombination algorithms currently in use were discussed in detail in [30]. Our criterion 7.39 directly compares with the JADE criterion [28]:

$$E_a E_b \times \Delta_{ab} < y_{\text{cut}} \,, \qquad\qquad 7.44$$

and the Geneva criterion [30]:

$$\frac{E_a E_b}{E^2} \times \Delta_{ab} < y_{\text{cut}} \,. \qquad\qquad 7.45$$

Recall that the energy squared in the denominator of 7.45 was introduced [30] to avoid problems with soft particles encountered by the JADE algorithm; its role is to provide a repulsion that ensures, in particular, that no "lumps" (i.e. spurious jets consisting of widely separated soft particles) are produced. Since there was no a priory criterion to choose the form of such a factor, the squared energy of the pair was taken to match the dimensionality of the numerator. Note that we use energy fractions so the l.h.s. of the optimal criterion 7.39 is made dimensionless by $E_{\text{tot}}$ in the denominator rather than an additional $E$ as in the Geneva criterion 7.45.

Note also that the Luclus criterion (eq. (38) of [14]) is equivalent to $E_a E_b E^{-1} \times \sqrt{\Delta_{ab}} < y_{\text{cut}}$ at small angles but involves absolute (unnormalized) energies.

From the point of view of the errors induced in the "physical information" (i.e. the collection of all $C$-correlators; cf. Sec. 5.21), neither Geneva nor Luclus criteria err because they overestimate the errors ($E^{-2} > E^{-1}$ for $E < 1$, and $\sqrt{\Delta_{ab}} > \Delta_{ab}$ for small angles, respectively). But for the same reason, they are not optimal: both are overcautions with certain pairs of particles.

For two soft widely separated particles (cf. sec. 2 of [30]) our criterion results in a cutoff of the form

$$\min(E_1, E_2) \sim y_{\text{cut}} \,, \qquad\qquad 7.46$$

which resembles the Durham algorithm [29] except that the latter involves squares of energy fractions.

Because $C$-continuous observables are IR finite, our algorithm does not give rise to IR divergences if used for jet counting in a conventional way. This can be verified directly (cf. the reasoning in sec. 2 of [30]).

### Preclustering $n \to 1$ 7.47

Because $C$-continuity allows fragmentations into any number of particles, it is natural to extend the above preclustering to more than two particles. The modifications of the reasoning of Sec. 7.23 are as follows. One compares the values of $C$-correlators on two multiparticle states, $p \oplus \mathbf{P}$ and $(\oplus_a p_a) \oplus \mathbf{P}$. One forms their difference and Taylor-expands it in angular variables. One finds that the leading terms cancel if

$$E = \sum_a E_a \,, \quad \hat{\boldsymbol{p}} \propto \sum_a E_a \hat{\boldsymbol{p}}_a \,. \qquad\qquad 7.48$$

The remainder is bounded by an expression of the form 7.33. From 7.48 one finds:

$$\Delta \hat{\boldsymbol{p}}_a = \hat{\boldsymbol{p}}_a - \hat{\boldsymbol{p}} = \frac{1}{E} \sum_{b \neq a} E_b (\hat{\boldsymbol{p}}_a - \hat{\boldsymbol{p}}_b) \,, \qquad\qquad 7.49$$

$$\tfrac{1}{2} |\Delta \hat{\boldsymbol{p}}_a|^2 \leq \frac{(E - E_a)^2}{E^2} \times \max_b \Delta_{ab} \,. \qquad\qquad 7.50$$

One finds:

$$|\mathbf{F}((\oplus_a p_a) \oplus \mathbf{P}) - \mathbf{F}(p \oplus \mathbf{P})| \leq \sum_a \frac{E_a (E - E_a)^2}{E^2} \max_b \Delta_{ab} \times M_f.$$　　7.51

The corresponding clustering criterion is as follows:

$$\sum_a \frac{E_a (E - E_a)^2}{E^2} \max_b \Delta_{ab} \leq y_{\text{cut}}.$$　　7.52

This is more precise than the following simpler but less optimal version described in [19], [20]:

$$\max_{a,b} \Delta_{ab} \times \sum_a \frac{E_a (E - E_a)^2}{E^2} \leq y_{\text{cut}}.$$　　7.53

Note that $M_f$ in 7.51 is the same as in 7.35. Therefore, the numerical value of the cutoff parameter here is to be taken the same as in the case of two particles. (For two particles, Eqs. 7.52 and 7.53 coincide with 7.39.)

An even simpler and cruder version is as follows:

$$\max_{a,b} \Delta_{ab} \times E \leq y_{\text{cut}}.$$　　7.54

This is because $E - E_a \leq E$.

The advantage of the simpler version 7.54 is that it is easy to accumulate the particles to be clustered iteratively: One may, e.g., begin with an energetic particle and proceed by adding other particles — rejecting particles that are too energetic — until the threshold is achieved. In the end, one may revert to the more precise formula 7.53 or 7.52, which may be needed to take into account accumulation of errors correctly (cf. the discussion in Sec. 7.55 below).

The simplicity of 7.54 reminds one of the cone-type algorithms traditionally used in hadron collisions (for a review see [10]). In particular, the quantity $\max_{a,b} \Delta_{ab}$ corresponds to the angular diameter of the jet (also notice the presence of an interesting energy factor $E$ which is the energy of the protojet normalized by the total energy of the event). However, in our case one worries neither about overlapping cones (as with cone-type algorithms) nor about irregularity of jets (as in the case of recombination algorithms) because the errors induced by preclustering are under control by analytical means, so the issue of jet shapes is irrelevant.

I emphasize that there can be no restriction on the geometric form of the cone spanned by the particles being clustered as long as the above analytical criteria are met. For instance, two particles that are too energetic to be recombined into one can, nevertheless, attract all softer particles around them even from behind each other and even if the angular distances of the latter from the former are much larger than the distance between the energetic particles. The only criterion of clustering is the numerical quality of the resulting approximation.

Lastly, it might also be possible to develop variants of preclustering with more than one resulting pseudoparticles (cf. the $3 \to 2$ scheme of [65]). Since elimination of a particle in such a scheme would take into account more information about the final state, one might be able to achieve a better quality of the resulting approximation (e.g. eliminate or reduce the quadratic terms in 7.30).

## Accumulation of errors 7.55

Errors induced by several (say, $n$) instances of preclustering — no matter how many particles are recombined in each, and irrespective of whether or not the particles affected are pseudoparticles[i] — are accumulated. For example, if all $n$ instances of preclustering have been done so that the error induced by each is $\sim y_{\text{cut}}$, then the cumulative error might be estimated by $\sim \sqrt{n} \times y_{\text{cut}}$. (This issue should be further studied via numerical experimenting.)

But whatever law governs the accumulation of errors, the very fact of their accumulation implies that an actual preclustering algorithm should, perhaps, somehow take the fact into account, e.g. start with a preset allowance $Y > 0$ which is decreased after each instance of preclustering until it reaches zero. How much of $Y$ is used up in successive instances of preclustering (whose corresponding $y_{\text{cut}}$ may all be different), is up to the programmer's ingenuity. Note also that $Y$ may be chosen differently for different events (the harder the event for computations, the larger $Y$ one may be willing to allocate to it, and vice versa). Optimal criteria for this depend on concrete samples of events.

Note that the values of $y_{\text{cut}}$ (and $Y$) used in the preclustering should typically be smaller than in the case of conventional algorithms.

Finally, it should be noted that we have only derived a criterion for an optimal preclustering. A complete algorithm should efficiently find the clusters of particles to combine into a protojet. The simplest variant is to use the code for a usual recombination ($2 \rightarrow 1$) algorithm and replace the criterion by the above optimal one. One could also use more sophisticated schemes similar to [40]. In general, different implementations may not be identical and may have to be fine-tuned for a particular application.

---

[i] Note that the $n \rightarrow 1$ recombinations may eliminate the need in iterative preclustering (i.e. such in which pseudoparticles are further clustered along with the unclustered particles from the initial multiparticle state).

## Applications to measuring the "number of jets"        8

There are two main uses for the quantity known as the "number of jets" of an event. It can be used either to compute relative fractions of events with a given number of jets — the observables called $n$-jet fractions — or as a tag to select events when searching for new particles. In this section we derive a special sequence of $C$-correlators — the jet-number discriminators [15] — that can be used similarly. Their properties will be studied in Sec. 9 while an example of their uses as building blocks in more complex observables can be found in Sec. 13.32. Since jet counting depends on the kinematics of the concrete reaction, here we consider the simplest case of annihilation $e^+e^- \to$ hadrons with a full $4\pi$ detector. Modifications for hadronic collisions are discussed in Sec. 13.

### "Jet counting" with $C$-correlators        8.1

### Understanding the problem        8.2

We would like to construct $C$-continuous observables that quantify the qualitative feature of final states known as the number of jets. A jet, qualitatively, is a spray of particles radiated within a small solid angle (in a given reference frame) and carrying a substantial fraction of the total energy of the event. The conventional approach consists in a straightforward formalization of this description which results in a use of cuts and an integer-valued observable (the "number of jets").

First of all, it should be emphasized that the linguistic restrictions of the $C$-algebra (which, by construction, reflect the limitations of multimodule calorimetric detectors) do not allow one, as a matter of principle, to write down an expression for the "number of jets" because the latter is an integer number and the only integer-valued $C$-continuous observables are constants on the entire collection of final states. This, however, is in perfect agreement with the fact that the "number of jets" is an ambiguous notion for a non-negligible number of final states. On the other hand, one should draw a distinction between a qualitative feature of the physical phenomena one studies and its numerical expression in terms of an underlying theory. Such an expression need not be a straightforward definition of what one feels one sees but it must conform to the kinematical ("linguistic") requirements of the theory. For instance, the "position" of an electron is desribed by a vector in Hilbert space — a notion not quite exactly intuitive but precise. The liberties one allows oneself to take with kinematical/"linguistic" restrictions are inversely proportional to the precision one aims to attain.

It proved possible to construct a sequence[i] of $C$-correlators of a special form (the jet-number discriminators [15]) that quantify the "number of jets" without actually identifying individual jets. The purpose of the detailed derivation presented below is to present concrete motivations for every element of their construction in order to show that it is essentially unique. (In the simplest and cleanest situation of $e^+e^-$ annihilation into hadrons the qualifier "essentially" seems to be superfluous.)

### Defining multijet states in terms of $C$-continuity        8.3

The following assumption deserves to be explicated: An "$m$-jet state" is one that is "similar" to a state which contains exactly $m$ energetic particles with large angular separation. One cannot fail to see that the crucial point here is what "similar" means, exactly. It is natural to express it in mathematical terms using the $C$-convergence: at a qualitative level, two states are "similar" if they are indistinguishable by calorimetric detectors with poor enough energy and angular resolutions.

Unfortunately, the $C$-convergence cannot be expressed in terms of a useful distance function. Therefore, there is no single numerical criterion to measure the above "similarity". But fortunately, the wis-dom of general topology [46] tells us that the most general option to discriminate the elements with a particular property in spaces with unusual convergences is to employ continuous[i] functions that take special values (e.g. 0) on those elements. The number of the functions used in such a comparison and their form depend on the specifics of the particular problem as well as practical expediency. A natural thought would then be to construct a function that takes, say, the value 0 on $m$-jet states, and is appreciably different from zero on all states that cannot be described as "similar" to $m$-jet states.

---

[i] The sequence might be thought of as an infinitely dimensional vector. Recall that some physical quantities are described by 4-vectors, some by 6-component antisymmetric 4-tensors, etc. The "number of jets" happens to be correctly described by an infinite sequence of scalar continuous components.

Unfortunately, the notion of "$m$-jet state" is ill-defined (if one wishes to avoid cuts, as we do). But fortunately, a little further thought reveals that the definition of $m$-jet states does have an unambiguous aspect, namely, that a state consisting of exactly $m$ particles cannot have more than $m$ jets. The latter property ("cannot have more than $m$ jets") is shared by all states consisting of no more than $m$ particles, and the collection of all such states is defined unambiguously. Any state that is sufficiently close to this collection (in the sense of $C$-convergence) shares the same property — it cannot have more than $m$ jets (within the precision of "sufficiently").

Therefore, the problem of "jet counting" has transformed into that of constructing $C$-continuous functions that take a special value (zero) only on all multiparticle states with less than a certain number of particles.

Going back for a moment to the conventional jet counting, one can see that instead of one integer-valued function ("number of jets") one could use a sequence of step functions: each such function should take the value 0 on all states with less than a certain number of jets, and the value 1 on all other states. A sequence of such functions ($m = 1,2,\ldots$) would do the job of jet counting just fine. The functions we are going to construct can be regarded as $C$-continuous weights that replace such step functions in accordance with the philosophy of Sec. 2.5.

Among all $C$-continuous functions, it is natural first to try the basic ones — i.e. the $C$-correlators 5.18 — as candidates for that role.

Explicit formulas for jet-number discriminators                                    8.4

Consider a $C$-correlator 5.18 that is exactly 0 on any state with less than $m$ particles. Then $f_m(\hat{p},\hat{p},\ldots) = 0$. Then $f_m(\hat{p}_1,\hat{p}_2,\ldots)$ should contain a nullifying factor (cf. Theorem 119 in [45] which states that if a sufficiently smooth function $f(\nu)$ is zero on a manifold described by the equation $\varphi(\nu) = 0$ then $f(\nu) = \varphi(\nu) f'(\nu)$ where $f'(\nu)$ is also a smooth function). We choose the nullifying factor to be $\Delta_{12} = 1 - \cos\theta_{12}$, an object that has already occurred in the study of optimal preclustering, Eq. 7.34.

The nullifying factor $1 - \cos\theta_{ij}$ is a rotationally invariant function that is analytic everywhere on the unit sphere and is simply connected to scalar products, which makes it a perfect choice. It also seems to be the *only* reasonable choice — apart from raising it to some power, a complication so substantial[ii] that one should have very good reasons to justify it.

Symmetry requires a similar factor for each pair of arguments of $f$:

$$f_m(\hat{\pmb{p}}_1,\ldots,\hat{\pmb{p}}_m) = \prod_{1 \leq i < j \leq m} \Delta_{ij} \times \widetilde{f}_m(\hat{\pmb{p}}_1,\ldots,\hat{\pmb{p}}_m),$$                    8.5

where

$$\Delta_{ij} = 1 - \cos\theta_{ij} = 1 - \hat{\pmb{p}}_i \hat{\pmb{p}}_j.$$                    8.6

In what follows it will be important that $\Delta_{ij} \equiv \Delta(\theta_{ij})$ is non-negative, monotonic and smoothly interpolates between $\Delta(0) = 0$ and $\Delta(\pi) < +\infty$.

---

[i] with respect to the concrete convergence defined for the elements of that space.

[ii] from the point of view of analytical calculations. This is because of a combinatorial blow up of the number of terms at intermediate stages; cf. the analytical calculation of normalizations in Sec. 8.12.

To fix $\widetilde{f}_m$, it is natural to check out the simplest option first, so set

$$\widetilde{f}_m = 1 \,. \qquad\qquad\qquad 8.7$$

Since the perceived jet structure is independent of the energy scale (which is reflected in the fact that the corresponding amplitudes scale up to logarithmic scaling violating corrections), it is natural to make the observable dimensionless by introducing a factor $E_{\text{tot}}^{-m}$, which is equivalent to using energy fractions instead of absolute energies.

One arrives at the following expressions for the jet-number discriminators [15]:

$$\mathbf{J}_m(\mathbf{P}) = N_m\, E_{\text{tot}}^{-m} \sum_{i_1 < \ldots < i_m} E_{i_1} \ldots E_{i_m}\; j_m(\hat{\mathbf{p}}_{i_1}, \ldots, \hat{\mathbf{p}}_{i_m}), \qquad j_m(\hat{\mathbf{p}}_1, \ldots, \hat{\mathbf{p}}_m) = \prod_{1 \le i < j \le m} \Delta_{ij}\,. \qquad 8.8$$

The summation here runs over all selections of $m$ different particles from all the particles of the final state $\mathbf{P}$. The terms with equal angular arguments are absent here as compared with 5.18 because they are nullified by the $\Delta$'s.

The normalization $N_m$ in 8.8 can be chosen so that the discriminators always vary between 0 and 1 (see Sec. 8.12):

$$0 \le \mathbf{J}_m(\mathbf{P}) \le 1, \quad \text{for all } m \text{ and } \mathbf{P}. \qquad\qquad 8.9$$

For $m = 1, 2$ the discriminators are trivial:

$$\mathbf{J}_1(\mathbf{P}) = N_1\, E_{\text{tot}}^{-1} \sum_i E_i \equiv 1, \qquad\qquad 8.10$$

$$\mathbf{J}_2(\mathbf{P}) = N_2\, E_{\text{tot}}^{-2} \sum_{i<j} E_i E_j (1 - \cos\theta_{ij}) \equiv \frac{s}{E_{\text{tot}}^2}, \qquad\qquad 8.11$$

where $s$ is the total invariant mass of the event. In the center of mass reference frame, $\mathbf{J}_2(\mathbf{P}) \equiv 1$.

## Normalization                                                                                            8.12

The jet-number discriminators 8.8 are always bounded:

$$\mathbf{J}_m(\mathbf{P}) \le N_m\, E_{\text{tot}}^{-m} \left( \sum_{i_1 < \ldots < i_m} E_{i_1} \ldots E_{i_m} \right) \sup_{\hat{\mathbf{p}}_1, \ldots, \hat{\mathbf{p}}_m} j_m(\hat{\mathbf{p}}_1, \ldots, \hat{\mathbf{p}}_m) \le \text{const}. \qquad 8.13$$

In general, $N_m$ can only be found numerically because they depend on the geometry of the experimental setup (e.g. whether or not one deals with a full $4\pi$ detector, etc.). For the case of $e^+e^- \to$ hadrons and full $4\pi$ geometry, numeric experiments show that the maximal value is reached on the configuration $\mathbf{P}_\infty^{\text{sym}}$ with the energy uniformly spread over the unit sphere (as well as on a few highly symmetric configurations; cf. Sec. 8.17). It can be regarded as a limit of final states consisting of $N$ particles with equal energy $E_i = N^{-1}$ uniformly distributed over the unit sphere so that $\sum_i \to \frac{N}{4\pi} \int_{S_2} \mathrm{d}\hat{\mathbf{p}}_i$ as $N \to \infty$. Then one can choose $N_m$ from the condition

$$\mathbf{J}_m(\mathbf{P}_\infty^{\text{sym}}) = 1\,. \qquad\qquad 8.14$$

The angular integrals are done using the following formula:

$$\frac{1}{4\pi} \int_{S_2} \mathrm{d}\hat{\mathbf{p}}\; (\hat{\mathbf{p}}\,\hat{\mathbf{q}}_1) \times \ldots \times (\hat{\mathbf{p}}\,\hat{\mathbf{q}}_{2k}) = \frac{1}{3 \cdot 5 \cdot \ldots \cdot (1 + 2k)} \sum_\pi (\hat{\mathbf{q}}_{\pi_1} \hat{\mathbf{q}}_{\pi_2}) \times \ldots \times (\hat{\mathbf{q}}_{\pi_{2k-1}} \hat{\mathbf{q}}_{\pi_{2k}}), \qquad 8.15$$

where the summation runs over all non-equivalent decompositions of $2k$ objects into $k$ pairs (decompositions differing by the order of pairs or the order of objects in each pair are treated as

equivalent). One finds:

| $m$ | 1 | 2 | 3 | 4 | 5 | 6 | |
|-----|---|---|----|----|------|--------|---|
| $N_m$ | 1 | 2 | $\frac{27}{4}$ | 36 | $\frac{9375}{32}$ | $\frac{455625}{128}$ | … |

          8.16

Note that a very large number of terms is generated here at intermediate stages of analytical calculations.[i,ii]

## Some special values         8.17

To illustrate how the discriminators approach their maximum ($= 1$), here are the values of $\mathbf{J}_m\!\left(\mathbf{P}_n^{\text{sym}}\right)$ on some highly symmetric configurations of particles with equal energies ( $n = 4$ corresponds to vertices of a tetrahedron inscribed into the unit sphere; $n = 6$ corresponds to an octahedron; the case $n = \infty$ is defined in 8.12):

|  | $\mathbf{J}_1$ | $\mathbf{J}_2$ | $\mathbf{J}_3$ | $\mathbf{J}_4$ | … |
|---|---|---|---|---|---|
| $\mathbf{P}_1^{\text{sym}}$ | 1 | 0 | 0 | 0 | 0 |
| $\mathbf{P}_2^{\text{sym}}$ | 1 | 1 | 0 | 0 | 0 |
| $\mathbf{P}_3^{\text{sym}}$ | 1 | 1 | $\frac{27}{32} \approx 0.84$ | 0 | 0 |
| $\mathbf{P}_4^{\text{sym}}$ | 1 | 1 | 1 | $\frac{64}{81} \approx 0.79$ | 0 |
| $\mathbf{P}_6^{\text{sym}}$ | 1 | 1 | 1 | 1 | … |
| … | … | … | … | … | … |
| $\mathbf{P}_\infty^{\text{sym}}$ | 1 | 1 | 1 | 1 | 1 |

        8.18

(Recall that, up to normalizations, $\mathbf{J}_1$ and $\mathbf{J}_2$ are the total energy and the invariant mass of the state, respectively. Actually, $\mathbf{J}_1(\mathbf{P}) \equiv 1$ in the spherically symmetric case, and the corresponding column is included for completeness.) The symmetric states have the largest angular distances between the particles. The table shows that the uniformity of the angular spread together with the uniformity of energy distribution are more important factors for achieving the maximal value than the total number of particles.

Note that the state $\mathbf{P}_\infty^{\text{sym}}$ must be considered as having an infinite rather than zero number of jets. Note also the following experimental fact: all global maxima of the jet-number discriminators (i.e. the states on which at least one $\mathbf{J}_m$ besides $\mathbf{J}_1$ takes the value 1) correspond to the following eight configurations: the five Platonic solids (the tetrahedron $\mathbf{P}_4^{\text{sym}}$, the octahedron $\mathbf{P}_6^{\text{sym}}$, the cube $\mathbf{P}_8^{\text{sym}}$, the icosahedron $\mathbf{P}_{12}^{\text{sym}}$, the dodecahedron $\mathbf{P}_{20}^{\text{sym}}$; see e.g. [67]) as well as the sphere $\mathbf{P}_\infty^{\text{sym}}$ and the two configurations that can be regarded as degenerate regular polyhedra, namely, the 'dipole' $\mathbf{P}_2^{\text{sym}}$ and the 'dihedron' $\mathbf{P}_3^{\text{sym}}$. A failure to include the latter three degenerate configurations into the analysis might explain the limited ( $\sim 5\%$ ) precision of an earlier purely classical attempt [68] to find a kinematical explanation of the well-known profound significance of the Platonic solids.

---

[i] I failed to find a way to evaluate $N_7$ with my copy of FORM-2 [64].

[ii] I thank B. B. Levtchenko for some numerical checks of the normalizations.

## Understanding the jet-number discriminators 9

For definiteness, we consider here the standard jet-number discriminators $\mathbf{J}_m$ exactly as defined in Sec. 8.4 for the $4\pi$ geometry parametrized in terms of two angles (the unit sphere; modifications to the cylindrical geometry of hadron-hadron collisions will be discussed in Sec. 13). But we allow the total 3-momentum of the states to be non-zero. The qualitative features observed in the examples remain valid in more general cases.

### Qualitative behavior of $\mathbf{J}_m$ 9.1

We will first explain how to obtain crude estimates for $\mathbf{J}_m$, and then discuss in detail various features of their qualitative behavior. Examples for states with a few particles are presented in Sec. 9.6.

### Estimating $\mathbf{J}_m$ via a crude clustering 9.2

The jet-number discriminators — as any other $C$-correlators — are fragmentation invariant (Sec. 6.6). Therefore to obtain a rough estimate for the values of the discriminators on a given final state it is sufficient to replace each jet with one particle carrying the same total energy and going in the direction of the jet and drop soft particles. (A more refined version of this procedure can be used to optimize computation of $C$-correlators from data; see Sec. 7.)

### A typical picture of values 9.3

The values of $\mathbf{J}_m$ for a typical final state $\mathbf{P}$ with 3 jets are roughly as follows :

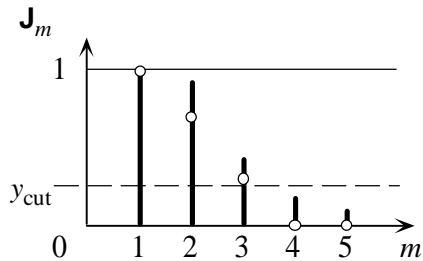

9.4

The white circles represent the values of discriminators for the 3-particle state $\mathbf{P}_0$ obtained by recombining particles from each jet into one, as described in Sec. 9.2. $\mathbf{P}_0$ can be interpreted as the parton state prior to hadronization; then $\mathbf{P}$ is the corresponding hadronic state. As a rule, $\mathbf{J}_m(\mathbf{P}) > \mathbf{J}_m(\mathbf{P}_0)$ as a result of fragmentation. The $C$-continuity ensures that the closer (in the sense of $C$-convergence) the final hadronic state to the underlying parton state, the less the upward drift of $\mathbf{J}_m$ during fragmentation.

In particular, $\mathbf{J}_m \sim 0$ for $m$ larger than the perceived number of jets; the non-zero values (the tail at large $m$) are due to fragmentation (almost collinear radiation from hard partons as well as "drops of glue" between jets).

Note, however, that since the maximum is reached on some highly symmetric configurations (cf. 8.17), fragmentation of the latter results in a downward (rather than upward) shift of the values of $\mathbf{J}_m$. But such cases are exceptional.

### Understanding the decrease at large $m$ [i] 9.5

The monotonic decrease at large $m$ is explained as follows. Let $M$ and $\Theta$ be, respectively, the perceived number of distinct jets in a state and their average (small) angular width. The state may also contain a "soft" component that consists of particles with small total energy $\varepsilon_{\text{soft}}$ spread between the jets. Then one can see from the explicit expression 8.8 that for $m > M$, each term in the sum for $\mathbf{J}_m$ is suppressed by additional powers of $\Theta^2$ and $\varepsilon_{\text{soft}}$ (cf. the examples below). The larger $m$, the larger the number of such factors, which explains the decrease.

______________________

[i] I thank B. Straub for this question.

Numerical experiments show that the decrease of $\mathbf{J}_m$ is a universal feature even for $M \lesssim m$.[i,ii]

## Simple examples        9.6

As explained above, up to a small upward shift and a non-zero tail at large $m$, the values of $\mathbf{J}_m$ are roughly the same as for the states with particles clustered into the corresponding jets. Thus, it is useful to consider states with just a few particles, each representing a jet, which we do below. One-jet states being too simple to be interesting, we begin directly with two jets.

## Two pure jets        9.7

Here one deals with two hard widely separated particles:

$$\mathbf{P}_{2h} = \{E_1, \hat{\boldsymbol{p}}_1; E_2, \hat{\boldsymbol{p}}_2\}, \quad E_{1,2} >> 0, \quad \Delta_{12} >> 0.        9.8$$

Our usual notation is $\Delta_{ij} = 1 - \cos\theta_{ij} = 1 - \hat{\boldsymbol{p}}_i\hat{\boldsymbol{p}}_j$. Due to the normalization of $\mathbf{J}_m$ all energies are to be compared with $E_{\text{tot}}$, the total energy of all particles of the event. One has:

$$\mathbf{J}_2(\mathbf{P}_{2h}) = 4\frac{E_1 E_2}{E_{\text{tot}}^2}\Delta_{12}, \quad \mathbf{J}_3(\mathbf{P}_{2h}) = \mathbf{J}_4(\mathbf{P}_{2h}) = \ldots = 0.        9.9$$

One can see that the conditions 9.8 that ensure that we deal with two jets are equivalent to one condition, namely, $\mathbf{J}_2(\mathbf{P}_{2h}) >> 0$. Vice versa, it is sufficient to require that

$$\mathbf{J}_2(\mathbf{P}_{2h}) < y_{\text{cut}}        9.10$$

with a small $y_{\text{cut}}$ to obtain a state with one jet. (For a comparison of our formalism and conventional jet algorithms see Sec. 7. Here we only note that the resemblance of the above to the Geneva criterion, Eq. 7.45, is due to the fact that there are just two particles so that $E_{\text{tot}} \equiv E_1 + E_2$ in this simple case.)

## Adding one soft particle        9.11

Here one deals with a state

$$\mathbf{P}_{2h+1s} = \mathbf{P}_{2h} \oplus \{\varepsilon_3, \hat{\boldsymbol{p}}_3\}, \quad \varepsilon_3 << E_{1,2},        9.12$$

that can be shown as follows:

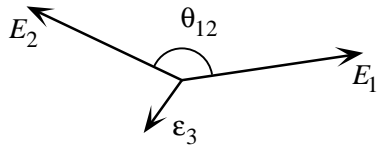

9.13

One has ($E_{\text{hard}} \equiv E_1 + E_2$):

---

[i] I thank B. B. Levchenko for numerical checks of this property.
[ii] The relationship between jet clustering algorithms and event shape measures, including the monotonicity, is discussed in [65].

$$\mathbf{J}_2(\mathbf{P}_{2h+1s}) = \mathbf{J}_2(\mathbf{P}_{2h})\frac{E_{\text{hard}}^2}{E_{\text{tot}}^2} + 2\varepsilon_3(E_1\Delta_{13} + E_2\Delta_{23}),$$      9.14

$$\mathbf{J}_3(\mathbf{P}_{2h+1s}) = \tfrac{27}{8}\mathbf{J}_2(\mathbf{P}_{2h})\times\frac{E_{\text{hard}}^2}{E_{\text{tot}}^2}\times\frac{\varepsilon_3\Delta_{13}\Delta_{23}}{E_{\text{tot}}},$$      9.15

$$\mathbf{J}_m(\mathbf{P}_{2h+1s}) \equiv 0, \quad m \ge 4.$$      9.16

Eq. 9.14 illustrates the fact that one can take into account soft particles via an expansion in their energy (cf. Eq. 9.21 and Sec. 7.4). Eq. 9.15 shows that if the soft particle becomes almost collinear to any of the hard ones then the restriction on its energy is relaxed.

For completeness, here is the expression of $\mathbf{J}_3$ on a state with exactly three hard particles:

$$\mathbf{J}_3(\mathbf{P}_{3h}) = \tfrac{27}{4}E_{\text{tot}}^{-3}\ E_1E_2E_3\ \Delta_{12}\Delta_{13}\Delta_{23}\ .$$      9.17

## 2 hard + 2 soft particles      9.18

As a further example, add another soft particle:

$$\mathbf{P}_{2h+2s} = \mathbf{P}_{2h} \oplus \{\varepsilon_3, \hat{\boldsymbol{p}}_3; \varepsilon_4, \hat{\boldsymbol{p}}_4\}, \quad \varepsilon_{3,4} << E_{1,2}.$$      9.19

This can be depicted as follows:

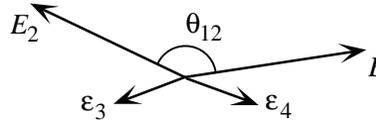

     9.20

Consider again the most interesting case of the third discriminator:

$$\mathbf{J}_3(\mathbf{P}_{2h+2s}) = \tfrac{27}{8}\mathbf{J}_2(\mathbf{P}_{2h})\times\frac{E_{\text{hard}}^2}{E_{\text{tot}}^2}\times\frac{\varepsilon_3\Delta_{13}\Delta_{23} + \varepsilon_4\Delta_{14}\Delta_{24}}{E_{\text{tot}}} + O(\varepsilon^2).$$      9.21

This expression should be considered in view of what was said in Sec. 9.2. In particular, we wish to see whether the pair of widely separated soft particles could imitate a jet by pushing the value of $\mathbf{J}_3(\mathbf{P}_{2h+2s})$ well above 0. Comparing 9.21 with 9.15 and taking into account the fact that $\Delta_{ij} \le \text{const}$, one can see that this does not occur as long as the total energy fraction of the soft particles remains bounded by a small value.

## $\mathbf{J}_m$ never see spurious jets      9.22

This issue was discussed in the literature [31], [30] in connection with the fact that some recombination algorithms (e.g. JADE) suffer from the problem of counting "lumps" instead of jets. A lump is a state consisting of soft particles with large relative angles but erroneously identified as a jet. It is interesting to consider whether the problem of lumps reemerges in the context of the jet-number discriminators (and $C$-correlators in general). If this were so, then the values of $C$-continuous observables would be affected by "lumps" in a numerically significant way.

Recall that the energy dependence of $C$-correlators in general — and $\mathbf{J}_m$ in particular — is analytic, which means that an expansion in powers of soft energies is always possible. Moreover, it is always possible (cf. the examples 9.14, 9.15, and 9.21) to obtain estimates for the terms that are linear in energies of the soft particles (as well as for higher terms) by a power of $\varepsilon_{\text{soft}}$ (the total energy of soft particles) times a factor that is independent of the geometry of the state. We see that the problem of lumps in the context of $C$-correlators reduces to that of determining a correct measure of "softness". The latter turns out to be given by the total energy fraction of the soft particles: The values of $C$-correlators are guaranteed (by their $C$-continuity) to remain within a given small error interval if the total energy fraction of the soft particles that are added to (or eliminated from) the state, is bounded by the corresponding small value $\varepsilon$.

## Width of jets[i]                                                                                       9.23

The problem of the width of jets in the conventional approach arises due to the fact that observables such as mass spectra of multijet substates depend on the control parameters of jet finding algorithms which (parameters) correspond — directly or indirectly — to the angular width of jets. Since in the present formalism all observables are redefined in such a way that the intermediate representation in terms of jets is avoided, the issue does not emerge in such an interpretation. Nevertheless, as a further example of expressive power of the $C$-algebra, it is interesting to find a $C$-continuous measure for what could be called "width of jets".

The preceding examples motivate the use of ratios of jet-number discriminators for that purpose. Indeed, consider the ratios

$$\mathbf{W}_m(\mathbf{P}) = \frac{\mathbf{J}_{m+1}(\mathbf{P})}{\mathbf{J}_m(\mathbf{P})} \ . \qquad\qquad 9.24$$

From the explicit expressions 8.8 and the examples above one can see that if there are $M \leq m$ distinct jets in the event then the above ratios are suppressed by powers of angular distances squared between particles in the same jet, and/or by powers of energy fractions of soft particles. The latter fact together with the dependence of the ratios on the global geometry of the event, does not allow one to interpret them straightforwardly in terms of average width of jets.[ii] _Dynamical width_, therefore, is a better name for such observables. Their concrete form is chosen solely from the consideration of simplicity (one might e.g. consider taking square root of the r.h.s.).

The averaging over events is done after evaluating the ratios, which ensures that $\langle \mathbf{W}_m \rangle$ carry new information as compared with the jet-number discriminators $\langle \mathbf{J}_m \rangle$. In particular, the sensitivity of $\langle \mathbf{W}_m \rangle$ to hadronization is different.

## $\mathbf{J}_m$ vs. jet counting                                                                         9.25

The conventional jet counting assigns an integer number of jets to each event and classifies the events accordingly. Fix a multiparticle state $\mathbf{P}$ and consider any jet counting algorithm $A$ that yields an integer number of jets $N_A(y_{\text{cut}};\mathbf{P})$ for each $y_{\text{cut}}$; $N_A(y_{\text{cut}};\mathbf{P})$ should decrease monotonically as $y_{\text{cut}} \to 0$:

---

[i] I thank E. Kushnirenko for suggesting this problem.

[ii] Note that there is actually no point in trying to reproduce some arbitrary — however visual — definition of "jet width". Human eye was created by Nature for purposes other then studying multiparticle systems. "Physical meaning" is not the same as semiclassical visualization. The correct apparatus of "vision" in high-energy jet physics is $C$-correlators. This also answers an objection sometimes raised that a jet-number discriminator takes different values on different states that "apparently" have the same number of jets. Just because one can "see" that the states have the same "number of jets", can hardly be regarded as a truly physical reason to count them with the same weight in an observable that is supposed to be sensitive to "jettiness".

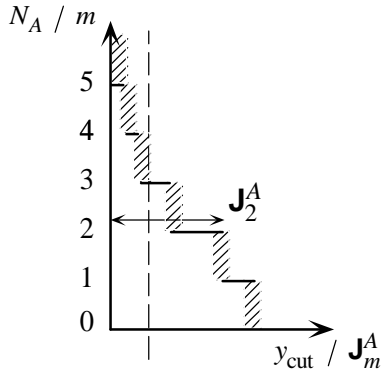

9.26

(The hashed areas correspond to experimental and theoretical uncertainties.) One sees from the figure that one could, in theory, restore a sequence of jet-number discriminators $\mathbf{J}_m^A(\mathbf{P})$ similar to $\mathbf{J}_m(\mathbf{P})$. Therefore, the information contents of $\mathbf{J}_m^A(\mathbf{P})$ and $N_A(y_{\text{cut}};\mathbf{P})$ for one event are essentially equivalent. The difference emerges when one performs an averaging over many events and takes into account errors (Sec. 9.28).

It is, of course, hardly possible to find simple expressions for $\mathbf{J}_m^A(\mathbf{P})$ for the popular algorithms. Our $\mathbf{J}_m(\mathbf{P})$ are singled out by the transparency of their analytical structure.

## Counting jets with $\mathbf{J}_m$          9.27

If the physical contents of the conventional jet counting and the discriminators are equivalent, then it should be possible to use the latter for jet counting of a conventional kind. Let us discuss this briefly. Actually, it is interesting to consider a jet-counting-type procedure not for its own sake but rather in connection with the composite observables discussed in Sec. 13 as an auxiliary tool for their approximate computation.

Recall that the state $\mathbf{P}$ contains exactly $M$ particles (up to exactly collinear fragmentations) *if and only if* $\mathbf{J}_m(\mathbf{P}) \equiv 0$ for all $m > M$ but not for $m \leq M$. This follows from the construction of the discriminators $\mathbf{J}_m$. Therefore, $\mathbf{J}_m(\mathbf{P}) \gg 0$ indicates that there are no less than $m$ jets in $\mathbf{P}$, while $\mathbf{J}_m(\mathbf{P}) \sim 0$ implies that there are less than $m$ jets in $\mathbf{P}$.

Due to fragmentation one has to introduce a small "resolution parameter" $y_{\text{cut}} > 0$. Then the state $\mathbf{P}$ is said to have $M$ jets if $\mathbf{J}_m(\mathbf{P}) < y_{\text{cut}}$ for $m = M + 1$ but not for $m \leq M$ (cf. Fig. 9.4). The jet counting now proceeds in a sieve-like manner[i]: one first computes $\mathbf{J}_2(\mathbf{P})$ (in the center of mass frame one starts with $\mathbf{J}_3(\mathbf{P})$ because $\mathbf{J}_2(\mathbf{P}) \equiv 1$). If $\mathbf{J}_2(\mathbf{P}) < y_{\text{cut}}$ then $\mathbf{P}$ contains just one jet. Otherwise one computes $\mathbf{J}_3(\mathbf{P})$. If $\mathbf{J}_3(\mathbf{P}) < y_{\text{cut}}$ then one deals with two jets… etc.

In general there is no reason why the cut should be the same for all $m$, so one would (and probably should) use a sequence of cuts $y_{\text{cut},m}$ — i.e. a separate parameter for each $m$.

## Instability of the conventional jet counting          9.28

Fig. 9.26 is an illustration of the difference with respect to experimental data errors and the unknown higher order corrections between the conventional jet counting, on the one hand, and the jet-number discriminators, on the other hand. (From the discussion in Sec. 9.25 it is clear that our conclusions will be valid for any jet counting algorithm of a conventional type.) Suppose the errors are purely statistical and distributed with a given variance. In the case of jet-number discriminators, if the statistics is increased, the statistical error of the results goes to zero.

---

[i] Following the custom of giving jet counting algorithms geographic names (Geneva, Durham), the procedure being described was called 'Moscow sieve' in [16].

The conventional jet counting corresponds to putting events into different bins depending on their jet number. Due to errors, some events would be assigned a wrong number of jets and go into a wrong bin. Therefore, each bin would have a certain fraction of events from other bins, depending on the width of the error intervals only — increasing the statistics cannot help. Thus, a statistical error is transformed into a systematic one. Although such a transformation occurs for any non-linear function of a random variable, the analysis of Sec. 2.5 shows that the effect is expected to be alleviated in the case of *C*-correlators as compared with the discontinuous mappings of conventional algorithms.[i]

Note that changing the jet resolution parameter $y_{\text{cut}}$ does not help one to get rid of such a smearing.

Lastly, the conventional jet counting is particularly sensitive to the errors at large *m* / small jet resolutions. One source of such errors is the so-called Sudakov radiation of almost collinear partons. The above analysis suggests that the importance of Sudakov effects within the framework of the conventional jet counting is an artifact due to the instability of the latter.

## $\mathbf{J}_m$ vs. *n*-jet fractions                                                                 9.29

Recall the interpretation of continuous observables as weights measuring the content of a particular physical feature in final states (Sec. 2.6). Also recall that $\mathbf{J}_m$ is a continuous analogue of a step function that would take the value 0 on final states with less than *m* jets and 1 otherwise. It follows that the average value of $\mathbf{J}_m$ over all events is naturally interpreted as the "weight" of the $\geq m$-jet component in the entire ensemble of final states. Then the average value $\langle \mathbf{J}_m \rangle$ is a natural replacement for $\sigma^{m\,\text{jets}} + \sigma^{m+1\,\text{jets}} + \dots$. The quantity $\sum_m \langle \mathbf{J}_m \rangle$ is then naturally interpreted as the average multiplicity of jets.

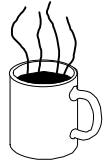

Take a Break....

---

[i] It was noticed by F. Dydak that for jet discriminators, the energy factors in each term in the sum are independent measurements from independent detector modules, and their errors are also independent. As a result, there is no systematic shift due to nonlinearity with respect to individual energy measurements. The property is also true for the *C*-correlators from which the spectral discriminators are built (see below).

## Applications to mass measurements.
## Simplest spectral discriminator                                      10

In the preceding sections we have constructed a correct numerical description of the feature of multi-hadron final states usually referred to as the "number of jets". The description involves only continuous shape observables from the $C$-algebra (the sequence of jet-number discriminators $\mathbf{J}_m$, $m = 1,2,3…$) while identification of individual jets is avoided. Yet we have seen that the physical content of the new observables remains equivalent to that of the conventional $n$-jet fractions. What did change was the mathematical expression of that physical content.

Now we are going to perform a similar transformation for the other important class of applications of jet algorithms, namely, searches for new particles based on studying invariant masses of multijet sub-states that resulted from the decays of those particles. We are going to reexpress in the correct language of $C$-algebra the observables such as invariant mass distributions of $n$-jet substates. Needless to say, identification of individual jets will be avoided.

In this section we define and study in detail spectral discriminators that accomplish that task in the toy model of a particle decaying into one jet and, as an immediate extension, in the more realistic case of decays into 1 jet + "muon" (Sec. 10.30; by "muon" we mean any particle that can be reliably identified, using calorimeters or not, including its energy and direction). A generalization to the case of multi-jet substates and the corresponding computational procedures will be described, respectively, in Secs. 11 and 12.

### 1-jet spectral discriminator                                        10.1

Let us begin with a very simple situation where one studies masses of just one-jet substates. Much of the reasoning will remain the same in the general case.

Suppose there is a particle $X$ that decays into one jet. Then the invariant mass of the particles consti-tuting the jet is equal to the mass of $X$. For simplicity, assume first that $X$ is the *only* particle produced in the collision. Then the final state consists of just one jet from $X$. How can one determine the mass of $X$ using the tools we have at our disposal?

We have seen (Sec. 5.25) that to select a substate from a final state within our formalism one uses the so-called *filters* — continuous functions interpolating between 0 and 1 in different parts of the unit sphere. For each filter $\Phi$ one obtains a substate $\Phi \circ \mathbf{P}$, and evaluates its invariant mass $\mathbf{S}(\Phi \circ \mathbf{P})$, Eq. 5.28. (It is convenient to say in such a case that one uses a *mass detector* described by the filter $\Phi$. Since one can analyze other properties of the substate selected by the filter, one can talk about $E_T$-detectors, $\mathbf{J}_3$-detectors etc.)

### Elementary mass detector                                            10.2

In order to be able to select a jet substate, consider a mass detector centered at the point $\hat{\boldsymbol{q}}$ with angu-lar size $O(R)$ (measured in units of the angular separation $\Delta$; cf. Secs. 4.5 and 7.41). The corresponding filter is defined as follows:

$$\Phi_{\hat{\boldsymbol{q}};R}(\hat{\boldsymbol{p}}) = \Phi(\Delta_{qp} / 2R),$$                10.3

where $\Delta_{qp}$ is the angular separation between the directions $\hat{\boldsymbol{q}}$ and $\hat{\boldsymbol{p}}$,

$$
\begin{aligned}
\Phi(d) &= 1 && \text{if } d \leq 1, \\
&= 0 && \text{if } d \geq 2, \\
&= 2 - d && \text{otherwise,}
\end{aligned}
$$                10.4

and the factor 2 is introduced into the r.h.s. of 10.3 to make $R$ vary between 0 and 1.

The continuous linear interpolation between 1 and 0 in 10.4 regulates the cut in accordance with the philosophy of Sec. 2.15. The concrete form is chosen from considerations of simplicity. Note that for $R = 1$ the filter is identically equal to 1: $\Phi_{\hat{q};R=1}(\hat{p}) \equiv 1$.

One can, of course, determine the mass of $X$ by simply *looking* at an event, *seeing* what the position and size of the jet are (i.e. $\hat{q}$ and $R$), and putting the mass detector described by 10.3 over the jet, after which one obtains the invariant mass. But such a procedure involves a step that cannot be expressed in the language of $C$-observables; namely, there is no recipe to choose one position and size for the filter from the family 10.3.

Anyhow, *choosing one* such filter involves a *comparison between all* filters. Therefore, consider *all* filters 10.3, and for *each* position $\hat{q}$ and size $R$, compute the invariant mass according to 5.28:

$$S(\hat{q}, R) \equiv \mathbf{S}(\Phi_{\hat{q},R} \circ \mathbf{P}) \geq 0 \,. \qquad \text{10.5}$$

Let us examine the properties of this expression as a function of $\hat{q}$ and $R$ in order to determine what sort of information it can yield.

For all mass detectors that cover the jet completely, $S(\hat{q}, R)$ is the same and equal to the jet mass. This can only occur for $R$ larger than the jet's size. The larger $R$, the larger the number of the positions $\hat{q}$ for which the measured mass is equal to that of the jet. On the other hand, for $R$ smaller than the jet's size, the detector and the jet overlap only partially, and $S(\hat{q}, R)$ yields values that are spread below the jet mass in a more or less continuous fashion.

### Definition                                                                                    10.6

We are thus naturally led to consider the distribution of the values of $S(\hat{q}, R)$ over the real half axis parametrized by a variable, say, $s$. A straightforward way to do so is to consider the following expression:

$$\rho_{1\,\text{jet}}(s; R) \overset{\text{def}}{=} Z^{-1} \int d\hat{q} \; \delta(s - S(\hat{q}, R)), \qquad \text{10.7}$$

which we call *one-jet spectral discriminator*.

The dependence on $R$ carries important information about the width of the jet, so it would be a good idea to leave it as a free parameter — one can always integrate over it later. (Notice certain parallels between the above logic and the cone-type jet algorithms. For a more detailed comparison see Sec. 11.25.)

It is convenient to choose the normalization in 10.7 so that

$$\int_0^{+\infty} ds \, \rho_{1\,\text{jet}}(s; R) = 1. \qquad \text{10.8}$$

In the spherically symmetric case $Z = 4\pi$.

### Simple properties of $\rho_{1\,\text{jet}}(s; R)$                                              10.9

For brevity, we will be writing $\rho(s; R) = \rho_{1\,\text{jet}}(s; R)$ in what follows.

### Small mass detector ($R \to 0$)                                                               10.10

Infinitesimal mass detectors measure infinitesimal mass so that all spectral weight is near the origin:

$$\rho(s; R) \underset{R \to 0}{\to} \delta(s). \qquad \text{10.11}$$

## Large mass detector ($R \to 1$)                          10.12

When the mass detector covers the entire sphere — which corresponds to $R \to 1$ — all the spectral weight is concentrated at the point $s = S_{\text{tot}}$, the total invariant mass of the state:

$$\rho(s;R) \underset{R \to 1}{\longrightarrow} \delta(s - S_{\text{tot}}).$$             10.13

## Flow of spectral weight                          10.14

Obviously, the regions of small and large $R$ are physically uninformative. But as $R$ varies from 0 to 1, the spectral discriminator undergoes an evolution that can be described as a flow of spectral weight from $s = 0$ to $s = S_{\text{tot}}$. Indeed, the normalization of the spectral discriminator 10.8 does not depend on $R$, and each position of the mass detector contributes an infinitesimal unit of spectral weight, the total number of such units being fixed and independent of $R$. As $R$ increases a little, so does, in general, the mass measured by the mass detector in the same position. This means that the infinitesimal weight from that position is shifted a little towards higher $s$. Ultimately, all the spectral weight concentrated initially at $s = 0$ is collected at the point $s = S_{\text{tot}}$. But what happens in between?

We will see that localized clusters of particles in the event manifest themselves as 'barriers' for the flow of spectral weight that take a form of isolated $\delta$-functions ($\delta$-spikes) in spectral discriminators, but first we consider the opposite extreme case.

## Uniform distribution of energy over the sphere                          10.15

In this degenerate case the spectral flow is simple. In any position, the mass detector measures the same invariant mass, so that all spectral density is localized at the same value $s = S_R$ which is a mono-tonic function of $R$. The spectral flow in this case is visualized in the following figure:

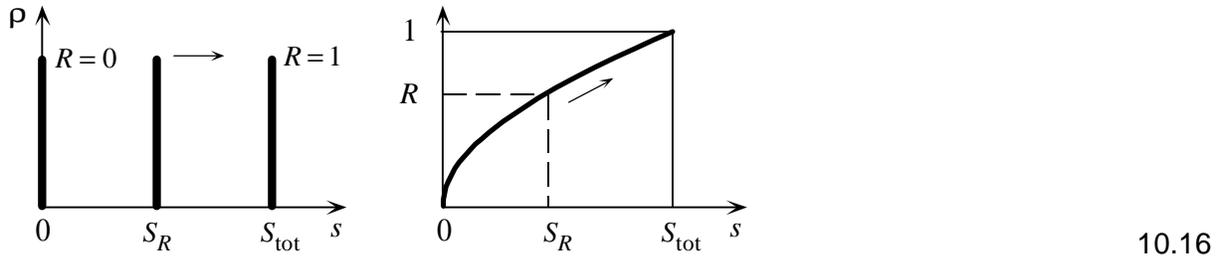

10.16

The left figure shows three positions of the $\delta$-spike corresponding to three values of $R$; the height of the $\delta$-spike does not change as $R$ varies. The right figure shows the trajectory of the $\delta$-spike on the $s$–$R$ plane.

On the other hand, for a non-uniform energy distribution different positions of the mass detector measure different values of $s$ for a fixed $R$. Therefore, the spectral weight is spread over $s$ for a fixed $R$ (a horizontal spread along each line $R = \text{const}$ in the right figure in 10.16). In what follows we study the qualitative features of spectral flow with emphasis on the case of "extremely non-uniform" energy distribution, i.e. when the final state contains a few clear well-separated jets.

## Spectral density is localized within $0 \le s \le s_R^{\max} \le S_{\text{tot}}$                          10.17

This is obvious. Moreover, the boundary value $s_R^{\max}$ monotonically increases with $R$.

## Zero-mass $\delta$-spike for $0 < R < 1$                          10.18

This is a simple illustration of some of the features encountered in less trivial cases.

If $R$ is small enough, there will be many positions $\hat{q}$ of the mass detector where it will measure zero

mass — because the mass detector will be covering just one or no particles. Denote as $\Omega_0^R$ the collection of all such positions, and let $|\Omega_0^R|$ be its normalized surface:

$$|\Omega_0^R| \equiv \frac{1}{4\pi} \int_{\Omega_0^R} d\hat{q}.$$  10.19

If $|\Omega_0^R| > 0$, the discriminator $\rho(s;R)$ contains a contribution of the form $|\Omega_0^R| \delta(s)$. It is clear that the coefficient decreases,

$$|\Omega_0^R| \underset{R \to 1}{\to} 0,$$  10.20

monotonically for all $R$, starting from the initial value 1.

## Jets and δ-spikes                                               10.21

### Events with a single jet                                        10.22

Suppose the event contains just one jet of particles. If all particles are strictly collinear then mass detectors always measure zero mass, and $\rho(s;R) \equiv \delta(s)$. Therefore suppose there is a non-zero angular spread of particles of the jet. Its invariant mass $S_j$ is the total invariant mass of the state $S_{\text{tot}}$. If the mass detector is not large enough to cover the entire jet then the mass of the latter will be underestimated in all positions of the mass detector, and the spread of $\rho(s;R)$ does not extend far enough to the right to reach the boundary $s = S_j$.

But it comes closer as $R$ increases and reaches $s = S_j$ when $R$ reaches a critical value $R_j^-$ (which roughly corresponds to the angular size of the jet). At $R = R_j^-$ there is just one position of the mass detector when it measures the correct mass.

Once $R$ passed the critical value, there will be many positions around the jet (the wider the mass detector, the larger the number of such positions) the correct mass is measured. Let $\Omega_j^R$ be the part of the unit sphere consisting of all positions of the mass detector in which it measures $S_j$. Then for $R < R_j^-$ the spectral discriminator contains an isolated δ-function,

$$|\Omega_j^R| \delta(s - S_j).$$  10.23

The weight $|\Omega_j^R|$ (= normalized surface of $\Omega_j^R$) grows monotonically from 0 to 1 as $R$ varies from $R_j^-$ to 1 while the position of the spike on the $s$-axis does not change and no spectral weight moves above it.

Evolution of the spectral discriminator in this case as a function of $R$ can be visualized as follows:

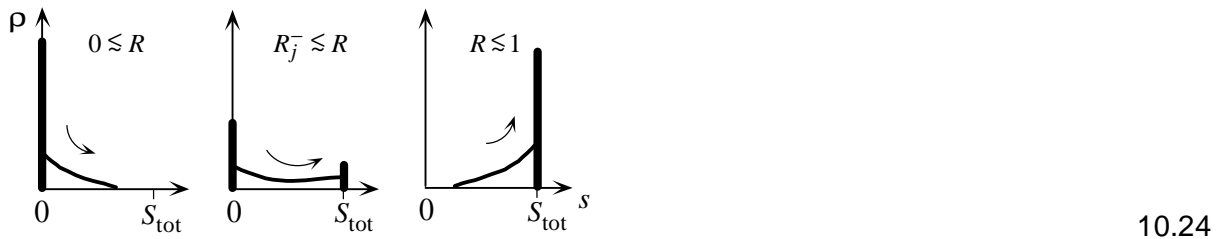

10.24

The fat vertical lines represent the δ-spikes. The shape of the continuous component depends on the distribution of energy within the jet.

**Events with several well-separated jets**                    10.25

Let the index $j$ label the jets of the event, so that $S_j$ are their masses. The situation is similar to the single jet case in the sense that a contribution of the form 10.23 will show up for $R > R_j^-$. However, since there are other jets in the event, then starting from some value $R_j^*$ (dependent upon the distance from the jet $j$ to other jets) there will be *fewer* positions in which the mass detector covers only the jet $j$, and there will also be an increasing number of positions in which the mass detector feels particles from other jets and measures masses larger than $S_j$. This means that for $R_j \gtrsim R_j^*$ the spectral weight accumulated at $s = S_j$ will start to spill over to the right from that point. Finally, there will be another critical value $R_j^+$ such that for $R > R_j^+$ the mass detector is too large to cover only the jet $j$ — then the $\delta$-spike at $s = S_j$ completely disappears and all its spectral weight is pushed above its position:

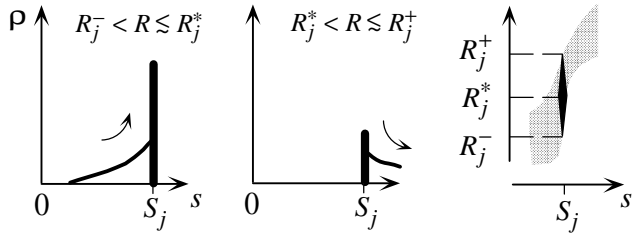

10.26

The two left figures show two points in the evolution of the spectral discriminator. The right figure shows the spectral flow as a density on the $s$–$R$ plane. The vertical fat line represents the $\delta$-spike; its variable width corresponds to $|\Omega_j^R|$, the size of the coefficient of the $\delta$-function.

If there occurs an overlap of a few $\delta$-spikes, the picture will be more chaotic. This may often happen in the 1-jet case because all jets may be expected to have roughly the same size determined by the dynamics of hadronization — but not in the case of $\geq 2$-jet substates whose invariant masses are scattered over a much larger interval.

**$\delta$-spikes corresponding to groups of jets**                    10.27

As the size $R$ of the mass detector is increased, it may cover more than one jet. Such a cluster of jets will also show up as a $\delta$-spike in $\rho(s; R)$ — but for larger $R$ than for individual jets.

It is not difficult to realize that choosing a better shape for the mass detector (e.g. a 'sum' of two elementary detectors Eq. 11.15; cf. Sec. 11.14) would allow one to focus much better on clusters of jets — and therefore on particles that decay into jets. This is exactly the idea behind the definition of more complex spectral discriminators in Sec. 11.13.

**Spurious $\delta$-spikes**                    10.28

In certain exceptional ideal situations (e.g. a wide jet with energy uniformly distributed over the jet cone) the mass detector for smaller $R$ may have a freedom of movement within the jet while measuring the same mass (which is a fraction of the jet mass $S_j$ and approaches the latter as $R \to R_j^-$ from below). Then there will be a spurious $\delta$-spike. Such exact spikes, however, are practically improbable and should rather be expected to materialize as an enhancement advancing towards $S_j$ from the left for $R \lesssim R_j^-$.

Minijets will be represented by minispikes for small values of $R$.

### Summary. Evolution of a δ-spike          10.29

We have seen that the presence of a well-separated jet in the final state is manifested in the spectral discriminator as a δ-functional contribution (δ-spike) with a specific dependence on $R$. This dependence can be visualized as a flow of spectral weight as $R$ changes. The flow is in the direction of larger $s$. The jet corresponds to a barrier for the spectral flow at $s = S_j$.

The relevant parameters are the jet's invariant mass $S_j$ and the three values $R_j^- \leq R_j^* \leq R_j^+$ (cf. Fig. 10.26). As $R$ approaches $R_j^-$ from below, the continuous component of the spectral density extends to the right and approaches $s = S_j$. When $R$ passes $R_j^-$, there develops a δ-spike at $s = S_j$, and the spike increases as $R$ increases from $R_j^-$ to $R_j^*$ — the spectral weight accumulates at the barrier. When $R$ exceeds $R_j^*$, the accumulated spectral density spills over the barrier and begins to spread continuously to the right of the barrier while the δ-spike at $s = S_j$ shrinks. At last when $R$ passes $R_j^+$, the δ-spike at the barrier disappears and is replaced by a distribution above $s = S_j$.

The characteristic form of the δ-spike (shown as a density in the rightmost figure 10.26) is expected to persist in the more general cases of spectral discriminators for multijet substates where its position will correspond to invariant masses of multijet substates.

### Realistic example: $X \to$ jet + "muon"          10.30

This case is actually very similar to the simplest 1-jet spectral discriminator. Suppose the particle $X$ decays into one jet and a muon μ. (By "muon" we mean any particle that can be reliably indentified, using calorimeters or not, including its energy and direction.) Let $E_\mu, \hat{p}_\mu$ be the muon's energy and direction. The muon is treated as a non-calorimetric "external parameter" (recall the remarks in Sec. 4.8). It is sufficient to make only one change in the formulas of Sec. 10.1, namely, for each filter one should "measure" the total invariant mass of the hadronic substate selected by the filter (as usual, the hadrons' energies $E_i$ enter all expressions with the weights $\Phi_{\hat{q};R}(\hat{p}_i)$) *and* the muon — irrespective of whether the muon's direction is covered by the filter or not:

$$S(\hat{q}, R) = 2 \sum_{1 \leq i < j \leq N} \Phi_{\hat{q};R}(\hat{p}_i) \Phi_{\hat{q};R}(\hat{p}_j) \times E_i E_j (1 - \hat{p}_i \hat{p}_j)$$

$$+ 2 E_\mu \sum_{1 \leq i \leq N} \Phi_{\hat{q};R}(\hat{p}_i) \times E_i (1 - \hat{p}_\mu \hat{p}_i).          10.31$$

Then one defines:

$$\rho_{1\,\text{jet}+\mu}(s; R) \overset{\text{def}}{=} Z^{-1} \int d\hat{q} \; \delta(s - S(\hat{q}, R)).          10.32$$

Because the structure of this expression is similar to 10.7, the only difference from the simple 1-jet case[i] is that the position of δ-spikes is shifted to higher $s$. The characteristic shape of the density distribution in the rightmost figure 10.26 is not affected.

### Smearing and accumulation of δ-spikes          10.33

In actuality, the δ-spikes from $X$'s decays will get smeared into a more or less wide bumps for various reasons (e.g. averaging over all final states and non-zero width of the particles that decay into jets; overlaps of δ-spikes; the numerical procedures described in Sec. 12 that require one to deal with spe-cially chosen continuous replacements for exact spectral discriminators). But apart from non-zero width, the qualitative behavior of the smeared δ-spikes is expected, due to $C$-continuity, to retain a resemblance to the right figure in 10.26 as long as there remains a resemblance to the ideal final states for which the behavior of δ-spikes was established.

---

[i] apart from the unphysical region of very small $R$ which is left for an interested reader to investigate as an exercise.

Suppose now the sample of events one studies contains both QCD background jets and jets from the decay of $X$. (The reasoning is the same whether one considers the decay into just one jet or jet + "muon".) Then averaging of the discriminator over all events results in an accumulation of δ-spikes from $X$ at about the same value of $s$ while the background jets are spread in a more or less uniform fashion over a wide interval of $s$. Therefore, the signal from $X$ will appear as a bump against a continuous background. To enhance the signal, one may integrate over $R$ but then the potentially important information would be lost (the two tails in the right figure in 10.26). Whether or not the characteristic $S$-shaped density distribution on the $s-R$ plane will be preserved after averaging, depends on how clean and well-isolated the jets from $X$ are, etc.

## General spectral discriminators                                11

In Sec. 10 we have studied the simplest spectral discriminator for 1-jet substates, $\rho(s;R) \equiv \rho_{1\,\text{jet}}(s;R)$. We have found that, for a single event, it may contain isolated $\delta$-functions ($\delta$-spikes). Each such $\delta$-spike corresponds to an isolated jet — or a cluster of jets if the $\delta$-spike occurs for larger $R$. We have also observed that modifying the configuration of the mass detectors employed should allow one to focus better on multijet substates. The purpose of this section is to discuss such modifications in more detail. First in Sec. 11.1 we consider a general definition of spectral discriminators which is a direct extension from the 1-jet case. Then in Sec. 11.13 we present a simple concrete version for $n$-jet substates in the spherically symmetric case of $e^+e^- \to$ hadrons. Modifications for hadronic collisions will be considered in Sec. 13.

### Mass detectors and spectral discriminators                   11.1

One considers a family of filters $\Phi_\gamma$ describing a family of substates $\Phi_\gamma \circ \mathbf{P}$ (the continuous parameter $\gamma$ describes the configuration — size and shape — of the filters; in general, $\gamma$ contains several scalar components, see examples below). Each filter $\Phi_\gamma$ is thought of as corresponding to an ideal _mass detector_ that measures the invariant mass of the substate it selects, $\mathbf{S}(\Phi_\gamma \circ \mathbf{P})$; cf. Eq. 5.28.[i] Then one constructs, following the pattern of Sec. 5.35, a differential observable with respect to the invariant masses:

$$\rho(s;\mathbf{P}) = Z^{-1} \int d\gamma \; \delta\big(s - \mathbf{S}(\Phi_\gamma \circ \mathbf{P})\big). \qquad\qquad 11.2$$

This is a continuous sum of $\delta$-functions and does not, in general, reduce to an ordinary function (cf. Sec. 11.11).

Observables 11.2 will be referred to as _spectral discriminators_. Their normalization can be defined so that

$$\int_0^{+\infty} ds \; \rho(s;\mathbf{P}) = 1. \qquad\qquad 11.3$$

### Changing the spectral variable                               11.4

If one chooses to work with true masses $M = \sqrt{s}$ instead of $s$, the $\delta$-function in 11.2 is transformed as follows:

$$\delta\big(s - \mathbf{S}(\Phi_\gamma \circ \mathbf{P})\big) = (2M)^{-1}\delta\big(M - \sqrt{\mathbf{S}(\Phi_\gamma \circ \mathbf{P})}\big). \qquad\qquad 11.5$$

On the other hand, spectral discriminators are measures with respect to $s$ and one should take into account a Jacobian:

$$\rho(s;R)ds = \big[2M\rho(M^2;R)\big]dM. \qquad\qquad 11.6$$

The net effect is that the extra factors cancel out:

$$\rho(M;\mathbf{P}) = Z^{-1} \int d\gamma \; \delta\big(M - \sqrt{\mathbf{S}(\Phi_\gamma \circ \mathbf{P})}\big). \qquad\qquad 11.7$$

One may wish to make a non-linear change of the variable so as to focus better on a particular range of $s$. This is because the algorithms described in Sec. 12 use a uniform discretization of $s$ and a non-uniform discretization can be best achieved via a non-linear change of variable.

---

[i] Note that one can study characteristics of multijet substates other than masses in a completely similar manner.

## Continuous component of $\rho(s; \mathbf{P})$      11.8

Recall that $\gamma$ in 11.2 is a continuous parameter that describes the shape (configuration) of the filter $\Phi_\gamma(\hat{\mathbf{p}})$. In general, $\gamma$ has several scalar components, and runs over a multidimensional manifold $\Gamma$ (e.g. a direct product of unit spheres; cf. 11.14). To study the structure of 11.2 consider the equation

$$s = \mathbf{S}(\mathbf{P}; \Phi_\gamma) \qquad\qquad 11.9$$

with given fixed $s$ and $\mathbf{P}$, and $\gamma$ the unknown.

In a generic case, its solutions with respect to $\gamma$ form a hypersurface $\Gamma_s$ in $\Gamma$. For such $s$, $\rho(s; \mathbf{P})$ is an ordinary function, and its values can be represented in the following form:

$$\rho(s; \mathbf{P}) \sim \int_{\Gamma_s} d\sigma \left| \nabla_\gamma \mathbf{S}(\mathbf{P}; \Phi_\gamma) \right|^{-1}, \qquad\qquad 11.10$$

where $d\sigma$ is the infinitesimal element of the hypersurface $\Gamma_s$ (in natural units), and $\nabla_\gamma$ is the corresponding gradient operator. The formula means that the larger $\Gamma_s$ and the slower the variation of $\mathbf{S}(\mathbf{P}; \Phi_\gamma)$ on $\gamma \in \Gamma_s$, the larger $\rho(s; \mathbf{P})$.

## δ-functional components of $\rho(s; \mathbf{P})$      11.11

If, on the other hand, the values of $\gamma$ that are solutions of the equation 11.9 for $s = s'$ constitute a part $\Gamma'$ of the manifold $\Gamma$ with a non-zero hypervolume

$$|\Gamma'| \equiv \int_{\Gamma' \subset \Gamma} d\gamma > 0, \qquad\qquad 11.12$$

then $\rho(s; \mathbf{P})$ contains a contribution $|\Gamma'| \delta(s - s')$. Such a situation occurs, e.g., if for some value $\gamma = \gamma_0$ the mass detector described by $\Phi_{\gamma_0}$ completely covers a cluster of particles (i.e. $\Phi_{\gamma_0}(\hat{\mathbf{p}}_j) = 1$ for each particle of the group), and for small variations of $\gamma$ in all directions around $\gamma_0$ the cluster remains covered so that the measured mass $\mathbf{S}(\mathbf{P}; \Phi_\gamma)$ stays the same.

There is little one can say a priori about the continuous component — its shape depends on the details of shape and geometry of the final state. But about δ-spikes, a qualitative information can be obtained by purely analytical means — exactly as was done in Sec. 10: *All the conclusions about δ-spikes and their evolution made there remain valid in the general case.* In particular, the characteristic shape of the density on the $s$-$R$ plane (Fig. 10.26) persists in the general case (cf. below Sec. 11.24).

## Multijet spectral discriminators      11.13

## Composite mass detectors      11.14

A *composite mass detector* can be regarded as consisting of two or more elementary mass detectors described by 10.3. For instance, consider two elementary mass detectors of the same size $R$, centered at $\hat{\mathbf{q}}_1$ and $\hat{\mathbf{q}}_2$, respectively. The two functions are $\Phi_{\hat{\mathbf{q}}_1; R}(\hat{\mathbf{p}})$ and $\Phi_{\hat{\mathbf{q}}_2; R}(\hat{\mathbf{p}})$. We combine them as follows:

$$\Phi_{\hat{\mathbf{q}}_1, \hat{\mathbf{q}}_2; R}(\hat{\mathbf{p}}) = \Phi_{\hat{\mathbf{q}}_1; R}(\hat{\mathbf{p}}) \oplus \Phi_{\hat{\mathbf{q}}_2; R}(\hat{\mathbf{p}}). \qquad\qquad 11.15$$

The operation $\oplus$ takes two or more numbers as operands, and its exact form is not important[i] as long as it has the following properties:

(i)      continuity (the result is a continuous function);

(ii)      it should give the result from the interval [0,1] if the operands are from that interval;

(iii)      commutativity (the result is independent of the order of operands).

---

[i] To insist on a "physical" interpretation here would be inappropriate.

The simplest choice is:

$$a \oplus b = \max(a, b).$$ 11.16

One can combine more than two mass detectors, e.g.:

$$\Phi_{\hat{q}_1, \hat{q}_2, \hat{q}_3; R}(\hat{p}) = \Phi_{\hat{q}_1; R}(\hat{p}) \oplus \Phi_{\hat{q}_2; R}(\hat{p}) \oplus \Phi_{\hat{q}_3; R}(\hat{p}),$$ 11.17

with the result independent of the order.

So, a composite mass detector is characterized by:

(i)    $n$, the number of its constituent elementary detectors; (heuristically, this corresponds to the number of jets in the substates the detector probes)

(ii)    their positions (*configuration*) $\hat{q}_1, ..., \hat{q}_n$. This is $\gamma$ in terms of Sec. 11.1 and the manifold $\Gamma$ is the direct product of $n$ unit spheres; it can be parametrized e.g. by $2n$ angles;

(iii)   their size $R^{\text{i}}$ (it is not included into "configuration" because we want to keep it a free parameter);

(iv)   the filter:

$$\Phi^{(n)}_{\hat{q}_1, ..., \hat{q}_n; R}(\hat{p}) = \Phi_{\hat{q}_1; R}(\hat{p}) \oplus \cdots \oplus \Phi_{\hat{q}_n; R}(\hat{p}).$$ 11.18

## Example. Spectral discriminator for 2-jet substates    11.19

The following filter selects two clusters of particles:

$$\Phi(\hat{p}) = \Phi^{(2)}_{\hat{q}_1, \hat{q}_2; R}(\hat{p}).$$ 11.20

Evaluate

$$S(\hat{q}_1, \hat{q}_2; R) \equiv \mathbf{S}(\Phi \circ \mathbf{P})$$ 11.21

according to the definition 5.28. This is the invariant mass of the pair of clusters selected by $\Phi$. The simplest 2-jet spectral discriminator is defined as follows (cf. 10.7 and 11.2):

$$\rho_{2\,\text{jets}}(s; R) \stackrel{\text{def}}{=} Z^{-1} \int d\hat{q}_1 \int d\hat{q}_2 \, \delta(s - S(\hat{q}_1, \hat{q}_2; R)).$$ 11.22

The normalization is such that

$$\int_0^{+\infty} ds \, \rho_{2\,\text{jets}}(s; R) = 1.$$ 11.23

In the spherically symmetric case $Z = (4\pi)^2$.

Generalizations to 3-, 4-…jet spectral discriminators as well as to the cases such as 2 jets + "muon" are straightforward.

## Qualitative behavior of $\rho_{2\,\text{jet}}(s; R)$    11.24

In this case the mass detector is a 'sum' of two elementary 'modules', and its configuration is described by the pair $\hat{q}_1, \hat{q}_2$ (four angles). An infinitesimal unit of spectral weight is now associated with each such configuration.

A $\delta$-spike occurs whenever the final state contains a pair of well isolated jets, and the size of elementary detectors is large enough to cover them; then the two modules can be moved independently around the corresponding jets without changing the measured mass, which corresponds to the situation of Sec. 11.11. The reasoning of Sec. 10.21 is repeated almost verbatim, and one concludes that the evolu-tion of the resulting $\delta$-spike as $R$ changes follows the same qualitative pattern as in the 1-jet case (Sec. 10.29). In particular, the $\delta$-spike on the $s-R$ plot has a similar form with a characteristic vertical line ($\delta$-functional component) and two wider tails (cf. the right figure in 10.26). Of course, $S_j$ is now the invariant mass of the pair of jets.

---

[i] We only consider the case when all elementary mass detectors constituting a composite one have the same size. The complication of different sizes would have to be well justified.

If the final state has more than two jets, then $\rho_{2\,\text{jet}}(s;R)$ exhibits a $\delta$-spike for each pair of jets; e.g. for a state with four jets, one would have 6 $\delta$-spikes. They will be positioned on the $s$-axis at the invariant masses of the corresponding 2-jet substates.

The remarks of Secs. 10.33 concerning smearing of $\delta$-spikes in realistic situations and searches for particles decaying into a certain number of jets, remain valid in the most general case.

A new point here as compared with the 1-jet case is as follows. Suppose the final state contains a pair of jets that are not well separated from each other. Nevertheless, $\rho_{2\,\text{jet}}(s;R)$ would exhibit a $\delta$-spike at $s$ equal to the invariant mass of the pair. The contributions to the $\delta$-spike come from those configurations of the mass detector when the two elementary modules partially overlap while covering both jets. The problem of separating the two jets does not arise here.

## Comparison with the conventional practice                                    11.25

The spectral distribution from, say, $\rho_{2\,\text{jet}}(s;R)$ after averaging over all final states is similar to mass distributions for 2-jet substates obtained within the conventional approach. But there are also important differences.

The comparison is easiest with the cone-type algorithms. In that case, the conventional approach consists, essentially, in choosing *one* position for an elementary mass detector per each jet, for any given $R$. After that is done, one value of invariant mass is computed per each 2-jet substate. Denote as $S_n(R)$ the resulting values of invariant mass, where $n$ numerates the 2-jet substates. One then computes the distribution of the masses for the entire sample of events. This is exactly the same as if one replaced our $\rho_{2\,\text{jet}}(s;R)$ with the following expression:

$$\rho_{2\,\text{jet}}^{\text{conv}}(s;R) = Z^{-1} \sum_n \delta(s - S_n(R)),$$                   11.26

where $n$ enumerates the 2-jet substates as determined by the jet algorithm used. (The normalization factor $Z$ here is not the same as in Eq. 11.22.) Performing an averaging of $\rho_{2\,\text{jet}}^{\text{conv}}$ over all events is exactly equivalent to computing the invariant mass distribution of 2-jet substates. (The fact that we deal with invariant masses squared is not important; cf. Sec. 11.4.)

Comparing Eqs. 11.26 and 11.22 allows one to see exactly what is the difference between the conventional mass distributions and our spectral discriminators.

Graphically, using the conventional $\rho_{2\,\text{jet}}^{\text{conv}}$ instead of our $\rho_{2\,\text{jet}}$ is equivalent to replacing the density distribution in the right figure of 10.26 with a line parametrized by

$$s = S_n(R).$$                                                        11.27

Now if jets are well separated, the curves 11.27 would have vertical segments that pass exactly over (parts of) the $\delta$-spikes of the exact $\rho_{2\,\text{jet}}(s;R)$ (provided energy and momentum conservation is respected when jets' 4-momenta are determined). But even then the corresponding $\delta$-functions enter into the r.h.s. of 11.26 with equal coefficients, which is not the case with the exact $\rho_{2\,\text{jet}}(s;R)$. In other words, even in the best case important information about the event is lost. Moreover, in the conventional approach the coefficients of $\delta$-functions on the r.h.s. of 11.26 are independent of $R$. So if the curve 11.27 has no vertical segments due to ill-defined jets it is hard to choose one value for the value of the mass. On the other hand, the $\delta$-spikes (and the resulting bumps) of $\rho_{2\,\text{jet}}(s;R)$ have a variable height even for one event, which information can be used to better pinpoint the $X$'s mass.

But the worst problem is encountered when the jets are not well-defined and the $\delta$-spikes are smeared and/or overlap[i]. Then a conventional algorithm attempts to represent a continuous distribution with a few $\delta$-functions; graphically, the smeared $\delta$-spike of the right figure 10.26 is replaced with a curve. If the continuous distribution has narrow peaks then such a representation may be meaningful. But if the peaks are not narrow, then whatever method one uses to position the $\delta$-functions, one cannot do that stably. For instance, if one attempts to determine a weighted average point for the position of the curve for each $R$, then the result depends on how one separates the bump for this pair of jets from other similar bumps, which involves an ambiguity. Alternatively, if one attempts to draw the line along the maximum of the bump, such a procedure is unstable against measurement errors (cf. the discussion in Sec. 2.3, esp. Fig. 2.4). In either case, there occur "ambiguities" resulting in an enhancement of errors in the determination of the mass of the particle one searches for.

On the other hand, when one computes a spectral discriminator, more information about each event is involved: One computes a spectral weight distribution from each event which depends in a well-defined $C$-continuous manner on the final state, and one fits the entire spectral weight distribution against theoretical predictions. At no point in data processing does one have to make unstable choices, and no instabilities occur.

## Discussion                                                                                     11.28

(i)     The pure QCD background contribution to spectral discriminators reflects the QCD dynamics of production of high-mass virtual partons — in fact, the structure of QCD matrix elements. Therefore, measuring spectral discriminators (perhaps, modified by appropriate weights; cf. Sec. 13.30) may be used as tests of QCD.

(ii)    It is sufficient to perform the processing for the values of $R$ only within an interval of typical widths of jets resulting from the decay of $X$. An integration over an appropriate range of $R$ may make the bump more prominent against the continuous background, especially its maximum.

(iii)   The overall picture ($\delta$-spikes and their evolution on the $s$-$R$ plane) in more complex cases (e.g. in the case of 3-jet substates, or mixed cases like 2 jets + "muon" etc.) will be similar to what was described in Sec. 10.29.

(iv)   Recall the example of Sec. 10.30 (jet + "muon"). It does not matter what are, say, the experimental cuts for the muon: one may simply sum contributions from all events — whatever the energies of the muon. The position of the bump relative to $s$ should normally not be affected much by this.

(v)    Similarly, if the geometry of the entire detector installation is such that it covers only part of the sphere around the collision point, then the integrations over the unit sphere would have to be correspondingly restricted. Of course, theoretical predictions would have to take such things into account.

(vi)   One does not have to determine the exact number of jets in each event: all events are processed, in principle, in the same way[ii]. Nor does one have to identify jets. Even the events with $X$, in which individual jets are hard or impossible to resolve in a conventional way, will contribute their share to the bump at $s = S_X$ in the spectral discriminator.

(vii)  From the point of view of theoretical predictions, higher order corrections may prove important for a precise description of the characteristic tails in the right figure 10.26. It may well turn out that to extract the most from spectral discriminators one would need to include such corrections into theoretical predictions. Therefore, whatever discrepancies may remain between theoretical predictions and experimental curves, they are either due to experimental data errors or to unknown theoretical higher order corrections — but never to "ambiguities" of jet definition.

---

[i] We are talking about one event here; the smearing due to averaging over many events is a separate issue.

[ii] A realistic code may involve optimizations such that different events would be treated differently; cf. Sec. 7.

## Practical computation of spectral discriminators 12

This section summarizes an algorithm to compute a spectral discriminator from data. The scheme described is insensitive to which concrete spectral discriminator one deals with (except that the integration over configurations of 'mass detectors' depends on how many — and what type of — parameters each configuration is characterized by). It also generalizes to other differential observables (including higher differential observables considered in Sec. 13.34) in a straightforward manner. The two realistic examples are jet + "muon" (Sec. 10.30) and two jets (Sec. 11.19). The exact discriminator is denoted as $\rho(s; R)$.

Note that a good part of the described scheme (the notion of a continuous regularization; the use of linear splines; the folding trick) might be of interest whenever one computes a differential cross section etc. from a random sample of events with more or less significant data errors (in addition to the limitation of finite statistics). This is because the scheme systematically takes into account the arguments of Sec. 2.5 so that all cuts are regularized and linear splines are used instead of the more conventional bin-type algorithms etc. Whether or not such a "politically correct" scheme would lead to a noticeable improvement of results depends on many factors (e.g. the size of statistical data errors and the desired precision of resultss). However, since additional computing costs required are minor one may consider using the continuous splines-based scenario instead of the bin-type algorithms as a matter of routine.

In the case of spectral discriminators there are — in addition to data errors — also approximation errors due to discretized integrations and an approximate representation of $\delta$-spikes for each event. So here one has an additional motivation for the use of splines instead of bin-type algorithms.

### The data 12.1

One deals with a finite sample of events. Each event is represented by a finite list of physical "particles" which in practice are the physical calorimeter modules lit up for that event. The length of the list depends on the event. Each "particle" then is a non-negative energy $E_i$ and a unit 3D vector (direction) $\hat{p}_i$; the latter may be represented e.g. by three Cartesian coordinates or by two angles $\theta_i, \varphi_i$.

In the jet + "muon" case one also has the muon's energy $E_\mu$ and direction $\hat{p}_\mu$.

### The grid of masses 12.2

One has to choose the interval of invariant masses $s$ to work with (concerning other parametrizations see Sec. 11.4). Theoretically, it is $[0, S_{\max}]$ — from zero to the maximal invariant mass of the events in the sample.

One might be tempted to truncate it at both ends. But that would affect the results of the folding procedure described below in Sec. 12.11. The folding trick involves a smearing between adjacent values of $s$, so if a truncation were introduced, then the validity interval of the results would get narrower after each folding, which may be undesirable.

Anyhow, the interval of $s$ should be divided into many equal[i] subintervals. It does not seem to make sense to choose the length of a subinterval to be much less than the error in computation of the invariant masses of filtered substates.[ii] However, memory permitting (CPU resources are not affected), one may choose it to be, say, half that — the folding trick of Sec. 12.11 allows one to double it any number of times.[iii]

In what follows, $s_i$, $i = 0, \dots N$, denotes the boundaries of subintervals; $s_0 = 0$ and $s_N = S_{\max}$. Also,

---

[i] Non-uniformity can be introduced via a change of the mass variable; cf. Sec. 11.4. One might wish to do this to focus better on a particular mass range.

[ii] Remember that there are not only data errors but also errors due to discretization of integrations over unit spheres etc.

[iii] If the number of subintervals $N$ is chosen as a power of 2 then the foldings can be done all the way down to one subinterval. In general, if $N$ is proportional to $2^t$, then $t$ foldings are possible.

$$\Delta s \equiv s_{i+1} - s_i \qquad\qquad 12.3$$

is independent of $i$.

## The grid of $R$    12.4

The maximal theoretically possible interval for $R$ is [0,1] (Sec. 10.2). However, it does not seem to make sense to go beyond $R = \frac{1}{2}$ (which corresponds to the elementary mass detector covering half the sphere; even that value may be too large). Because there is no interaction in the algorithm between different values of $R$, one can choose any values to work with and employ optimizations that allow simultaneous computation for several values of $R$.

Note, however, that one would like to determine the value(s) of $R$ for which the bump corresponding to the particle one searches for is the narrowest. This should roughly correspond to the average angular size of the resulting jets.

## Discretization of integrations    12.5

The definition of spectral discriminators involves an integration over all configurations of mass detectors. Each such configuration is characterized by positions of the corresponding elementary mass detectors. In practice, one would choose a finite number of positions $\hat{q}_i$ and distribute them over the unit sphere uniformly[i] to form a dense grid. Then configurations of mass detectors can be labeled by a discrete index (say, $\gamma$).

Denote $S_\gamma \equiv \mathbf{S}(\Phi_{\gamma,R} \circ \mathbf{P})$, the mass measured by the $\gamma$-th mass detector. Then

$$\rho(s;R) \approx Z^{-1} \sum_\gamma \delta(s - S_\gamma), \qquad Z = \sum_\gamma 1, \qquad\qquad 12.6$$

where the normalization factor $Z$ ensures validity of 10.8.

It should be remembered that even in the continuum limit for the integration in 10.7, the exact spectral discriminator $\rho$ need not be a continuous function and may contain $\delta$-functional contributions (Sec. 11.8).

If one defines the discriminator with special weights (cf. Sec. 13.32), those weights have to be inserted into the summands of 12.6.

## General scenario. Regularization    12.7

Numerical work with objects such as spectral discriminators that involve $\delta$-functional contributions is based on the use of a *regularization* (Sec. 15.29). One computes a sequence of arrays $r_N(s_i;R)$ from the available data, and then chooses (empirically) an optimal value $N = \tilde{N}$ which is neither too large nor too small. It should not be too large so that the stochastic irregularities are sufficiently smeared, and it should not be too small so that the signal from the particle one searches for is seen. Whether such a balance can be achieved depends on the size of the event sample and the precision of integrations. (Recall that *any* pair of jets in a final state — in the case of 2-jet spectral discriminator; any one jet in the jet + muon case; etc. — contributes a bump to the spectral discriminator. But only the bumps that correspond to a particle decaying into jets are added up instead of being smeared away after averaging over many events.)

Practically, in view of the simplicity of the folding trick (Sec. 12.11) one computes from data one ar-ray $r_N(s_i;R)$ for a large enough $N$ and then finds an optimal $\tilde{N}$ using foldings while, say, watching the results on computer screen.

---

[i] Uniformity is, strictly speaking, not necessary and is assumed for simplicity. With a non-uniform distribution, one should introduce appropriate weights in all the formulas (e.g. via systematic mappings of coordinates). This may be used as an optimization trick; for instance, one could take more points around the peaks of the energy flow due to energetic jets — similarly to the adaptive integration routines.

It should be emphasized that each element of the array $r_N(s_i;R)$ is a correct observable in its own right; given infinite statistics and computer resources, it could be computed with infinite precision; it accumulates contributions from all events as is usual with any quantum mechanical observable.

## Computation of $r_N(s_i;R)$ for one event                                12.8

The event is represented by the data described in 12.1.

One runs a loop over all configurations of the mass detector (each configuration is labeled by $\gamma$). At each step one runs a loop over the chosen values of $R$. At each step one computes $\Phi(\hat{p}_i) \equiv \Phi_{\gamma;R}(\hat{p}_i)$ and then the invariant mass $S \equiv \mathbf{S}(\Phi \circ \mathbf{P})$. The obtained value $S$ is used to add a contribution to $r_N(s_i;R)$.

## Adding a contribution to $r_N(s_i;R)$                                      12.9

Each configuration of mass detector adds a unit of spectral weight to the spectral discriminator; the contribution is located at $s = S$ computed previously for each $R$. In terms of $r_N(s_i;R)$, one does the following.

One determines $i$ such that $s_i \leq S \leq s_{i+1}$, and redefines

$$r_N(s_i;R) \leftarrow r_N(s_i;R) + \frac{s_{i+1} - S}{\Delta s},$$

$$r_N(s_{i+1};R) \leftarrow r_N(s_{i+1};R) + \frac{S - s_i}{\Delta s}.\qquad\qquad\textbf{12.10}$$

At this point all the loops are CONTINUE'd.

In the end one divides $r_N(s_i;R)$ by $Z$, the total number of points used to discretize integration(s) over the unit sphere(s) (cf. 12.6).

## The folding trick                                                         12.11

This is based on the observation that linear splines for the number of subdivisions $N/2$ can be obtained from the splines for the number of subdivision $N$ in the following simple way:

$$h_N(s - s_i) = \tfrac{1}{4} h_{2N}(s - s_{2i-1}) + \tfrac{1}{2} h_{2N}(s - s_{2i}) + \tfrac{1}{4} h_{2N}(s - s_{2i+1}).\qquad\textbf{12.12}$$

The formula for the array $r_N(s_i;R)$ is a replica from 12.12:

$$r_N(s_i;R) = \tfrac{1}{4} r_{2N}(s_{2i-1};R) + \tfrac{1}{2} r_{2N}(s_{2i};R) + \tfrac{1}{4} r_{2N}(s_{2i+1};R).\qquad\textbf{12.13}$$

It is implied that whenever an index goes beyond its boundaries, the value returned is zero.

Note that the array on the l.h.s. has (almost) twice as few elements as the array on the r.h.s., and the corresponding values $s_i$ are twice as far apart.

The folding trick effectively halves $N$ without complete recalculations, and without making any further approximations.

A concrete implementation of the algorithm need not follow the above description in every detail.

## Advanced options        13

There are several ways in which the formalism can be extended. We have so far been considering the simplest kinematical context of the c.m.s. annihilation $e^+e^- \to$ hadrons. Now we turn to modifications needed to adapt the formalism to the case of hadronic reactions. As was discussed in [26], there are two points of view on jet counting in the case of hadronic initial states. One is to modify the algorithms developed for $e^+e^- \to$ hadrons in a straightforward manner [69]. The other is to emphasize the specifics of hadronic reactions, most notably, the invariance with respect to boosts in the direction of the colliding beams and the inclusive nature of the corresponding jet-related observables [26]. Two simple modifications in the spirit of the first point of view are discussed in Sec. 13.1. Accommodating the ideas of the second approach requires more involved and extensive modifications of the formalism, as discussed in Sec. 13.7. Lastly, in Sec. 13.29 we discuss the options available for constructing more complex $C$-continuous observables that use additional dynamical information that may be available about the reaction one studied in order to enhance the signal. The examples we consider are motivated by the top search experiments at FERMILAB [12], [13].

## Modifications for hadronic reactions, recombination-style        13.1

### Special reference frames for DIS        13.2

In the case of deeply inelastic lepton-nucleon scattering it may be desirable to define jets in special reference frames [50]. To achieve this, we first rewrite particles' energies and the factors $\Delta_{ij}$ defined in 8.6 in a covariant form. Define $P = (1, \mathbf{0})$ and

$$p_i = (E_i, \boldsymbol{p}_i), \quad \boldsymbol{p}_i = E_i \hat{\boldsymbol{p}}_i, \quad p_i^2 = 0. \qquad 13.3$$

Then

$$E_i = p_i P, \quad \Delta_{ij} = (p_i p_j)(p_i P)^{-1}(p_j P)^{-1}. \qquad 13.4$$

Now to make e.g. the jet-number discriminators (Sec. 8) "count jets" in any other reference frame, it is sufficient to choose a 4-vector such that $P^2 = 1$ and its rest frame is the desired reference frame. Other observables (such as spectral discriminators of Sec. 11) are modified similarly.

### Suppressing "forward jets"        13.5

In the case of hadrons in the initial state one may wish to modify the jet-number discriminators 8.8 to suppress contributions from spectator partons (forward jets). Such modifications are easily done by analogy with how the recombination algorithms are modified in such cases (cf. [50]). For instance, it may be sufficient to introduce into $j_m$ (see 8.8) the factor

$$\Delta_i = 1 - \cos^2 \theta_i \qquad 13.6$$

per each particle, where $\theta_i$ is the angle between the particle's direction and the beam axis.

## Modifications for hadron-hadron reactions, rapidity-inclusive-style        13.7

### Using rapidity: cylinder replaces unit sphere        13.8

Recall that in the spherically symmetric case of $e^+e^- \to$ hadrons we represented particles' directions $\hat{\boldsymbol{p}}$ by unit 3-vectors, i.e. points of the unit sphere parametrized e.g. with $\theta$ and $\varphi$. In hadron-hadron collisions it proves advantageous to replace $\theta$ with pseudorapidity $\eta = \ln \cot \frac{\theta}{2}$ (cf. [10], [26]), and to work with transverse components of particles' 3-momenta. This is equivalent to representing the directions $\hat{\boldsymbol{p}}$ in terms of $\eta$ and a unit 2-vector[i] $\hat{\boldsymbol{p}}_T$ orthogonal to the beam axis:

$$\hat{\boldsymbol{p}} \Leftrightarrow (\hat{\boldsymbol{p}}_T, \eta) \Leftrightarrow (\varphi, \eta).$$          13.9

This means that, instead of the unit sphere, one now deals with a cylinder, a direct product of a unit circle (the collection of all $\hat{\boldsymbol{p}}_T$) and a real axis corresponding to $\eta$. This is equivalent to taking out the two polar points of the sphere and stretching it to $\pm\infty$.

$\hat{\boldsymbol{p}}_T$ can also be referred to as _transverse direction_. Note that

$$\hat{\boldsymbol{p}}_{T,a} \hat{\boldsymbol{p}}_{T,b} = \cos\varphi_{ab} \equiv \cos(\varphi_a - \varphi_b).$$          13.10

On the other hand, to avoid confusion it is reasonable to retain the interpretation of the scalar product of two directions as cosine of the angle between them:

$$\hat{\boldsymbol{p}}_a \hat{\boldsymbol{p}}_b = \cos\theta_{ab}\ .$$          13.11

The natural integration measure is modified as follows. If $\Omega$ is a part of the cylinder 13.9 then Eq. 4.3 is still valid but "surface" is interpreted differently; in place of 4.4 one now has

$$\int d\hat{\boldsymbol{p}}\ \psi(\hat{\boldsymbol{p}}) = \int_{S_1} d\hat{\boldsymbol{p}}_T \int_{-\infty}^{+\infty} d\eta\ \psi(\hat{\boldsymbol{p}}_T, \eta) = \int_0^{2\pi} d\varphi \int_{-\infty}^{+\infty} d\eta\ \psi(\varphi, \eta).$$          13.12

## Defining angular separation          13.13

The definition of angular separation for the spherically symmetric case was based on the considerations discussed in Sec. 4.5. For the cylindrical geometry and small angular distances one has, instead of Eq. 4.7,

$$\Delta_{ab} \approx \tfrac{1}{2}|\hat{\boldsymbol{p}}_a - \hat{\boldsymbol{p}}_b|^2 = \tfrac{1}{2}\left[|\hat{\boldsymbol{p}}_{T,a} - \hat{\boldsymbol{p}}_{T,b}|^2 + \eta_{ab}^2\right] = (1 - \cos\varphi_{ab}) + \tfrac{1}{2}\,\eta_{ab}^2,$$          13.14

where $\eta_{ab} = \eta_a - \eta_b$. This definition retains a maximal resemblance both to the spherically symmetric version, Eq. 4.7, and, for small $\varphi_{ab}$, to the conventional measure of angular distance used in cone-type jet definitions:

$$\tfrac{1}{2}R_{ab}^2 = \tfrac{1}{2}\left[\varphi_{ab}^2 + \eta_{ab}^2\right].$$          13.15

Note, however, that the Euclidean form of 13.15 (and 13.14 for small $\varphi_{ab}$) is completely ad hoc. It may have had some sense in the context of conventional algorithms where it might have been expected that jets should rather have smooth regular shapes (but even then the issue of the form of jets is rather murky [26]). As was remarked in Sec. 7.41, in the context of the new formalism the preclustering algorithm is expected to be mostly used for smaller angular resolutions so that the issue of the protojet shapes is simply irrelevant as long as the induced errors are kept sufficiently small (cf. Sec. 7.41). On the other hand, the considerations of computational efficiency may induce one to use a simpler definition, e.g.

$$\Delta_{ab} = \left[\max(|\varphi_{ab}|, |\eta_{ab}|)\right]^2.$$          13.16

There are many possible variants, and the choice may depend e.g. on the specific form of representation of the information about angular positions of detector cells of a concrete detector installation.

---

[i] Instead of which one could use the angle $\varphi$. I prefer the coordinateless vector notation as a more natural one.

The theoretical considerations that fixed the definition of angular separation in the spherically symmetric case of $e^+e^- \to$ hadrons (ease of analytical study of the jet-number discriminators) seem less important in the hadronic case because analytical calculations are hardly possible here anyway. However, should one wish to take this into account, one may opt for the following definition which corresponds to one of the angular distances discussed in [69]:

$$\Delta_{ab} = \cosh \eta_{ab} - \cos \varphi_{ab}.$$  13.17

This expression is said to conform to the structure of eikonal factors in QCD matrix elements [69].

Concerning the arbitrariness of the definition of angular separation, the following should be understood. The choice of angular separation is of little consequence for the preclustering of Sec. 7 (because sensitivity to a particular choice means that the chosen value of $y_{\text{cut}}$ is too large; cf. Sec. 7.41) and the choice may vary depending on a particular implementation of the algorithm. In the definition of the jet-number discriminators, however, it may be desirable to adhere to one definition from the very beginning. Eventually, some form of an "accord" may perhaps be needed to fix the definition of $\Delta_{ab}$ (perhaps, a suitable topic for SNOWMASS'96). Because considerations other than computational simplicity do not seem to carry sufficient weight in the case of hadron-hadron collisions, the definition 13.16 appears to be a strong favourite.

The reasoning below is independent of the particular choice of the angular separation $\Delta_{ab}$.

## Transverse energy　　　13.18

Another modification in the case of hadronic reactions is that, instead of the total energy of the $i$-th particle $E_i$ it is considered more natural in this context to use its _transverse energy_ [26],

$$E_{i,\text{T}} \stackrel{\text{def}}{=} E_i |\sin \theta_i|.$$  13.19

The ordinary energy should be replaced with the transverse energy,

$$E_i \leftarrow E_{i,\text{T}},$$  13.20

in all the formulas and interpretations of the preceding sections. Thus, the "energy flow" is now an abstract measure on the cylinder $\hat{\boldsymbol{p}}_{\text{T}} \times \eta$ whose values are transvers energy, etc. Practically, all one has to do is to reinterpret Eqs. 4.20, 4.21, etc. accordingly.

## $C$-correlators　　　13.21

The general formula for $C$-correlators 5.18 remains true with the above reinterpretations. However, since the two poles of the unit sphere have been removed, the required continuity of the angular functions ($f_m$ in 5.18; remember that it is now expressed in terms of $\eta$ instead of $\theta$) does not now concern the "points" $\eta = \pm \infty$ (formerly $\theta = 0, \pi$). This effect is seen e.g. in the formula for the $C$-correlator that computes invariant mass, Eq. 5.14, which now becomes

$$S_{\text{tot}} = \mathbf{S}(\mathbf{P}) = \sum_{ij} E_{i,\text{T}} E_{j,\text{T}} \frac{(1 - \hat{\boldsymbol{p}}_i \hat{\boldsymbol{p}}_j)}{|\sin \theta_i \sin \theta_j|},$$  13.22

with the angular function that is singular as $\eta = \pm \infty$. However, the total energy of the event is bounded by a constant so that the behavior of the "transverse energy flow" $E_{i,\text{T}}$ at $\eta \to \pm \infty$ suppresses the singularities of the angular functions ($E_{i,\text{T}}$ depends on $\eta$ via $i$). Such singularities are, therefore, spurious.

### Modifications of the optimal preclustering    13.23

The reasoning of Sec. 7 remains valid with the reinterpretation of Eqs. 13.9, 13.20 and any of the definitions of the angular separation $\Delta_{ab}$ discussed in Sec. 13.13. It should be emphasized that for smaller $y_{cut}$ one should opt for the computationally simplest among all equivalent definitions.

### Modifications of the jet-number discriminators $J_m$    13.24

The formula 8.8 and the reasoning that led to it remain valid in the new context provided one uses a new definition for the angular separation $\Delta_{ab}$ and all energies are transverse energies. Since the hadronic case is kinematically more complex, it is no longer possible to present analytical formulas for, e.g., normalizations of $J_m$, which will have to be determined numerically for each choice of $\Delta_{ab}$. Other properties (such as monotonic decrease for large $m$ etc.; see Sec. 9) are retained independently of this choice.

### Modifications of spectral discriminators    13.25

All one has to do is the following:

- Reinterpret the angular separation $\Delta_{ab}$ in the definition of the elementary filter Eq. 10.3 according to Sec. 13.13.

- Recompute the normalization factors for all spectral discriminators, Eqs. 10.7, 10.32, 11.22.

- Restrict each integration over $\hat{q}$ (position of an elementary mass detector) to the interval of pseudorapidity corresponding to the actual experimental data.

The key qualitative features of spectral discriminators — $\delta$-spikes, their evolution etc. — are independent of the particular kinematic context.

### Jet distributions with respect to $E_T$    13.26

As was clarified in [26] there are physical arguments in the case of hadron-hadron collisions to prefer inclusive observables such as 1-jet inclusive differential cross section $d\sigma/dE_T$ (used e.g. for precision determination of $\overline{\alpha}_S$ [70]). Such an observable can be translated into the language of $C$-algebra in a most natural manner. In fact, the relevant construction is a simple analogue of spectral discriminators. Define:

$$\mathbf{E}_T(\mathbf{P}) \stackrel{\text{def}}{=} \sum_{i=1}^{N_{\text{part}}} E_{i,T} \; .    \qquad 13.27$$

Consider the 1-jet spectral discriminator 10.7 with kinematic modifications described above, and use the observable 13.27 in place of the invariant mass. One obtains:

$$\rho_{1\,\text{jet}}(E_T;R) \stackrel{\text{def}}{=} \int d\hat{q} \; \delta(E_T - \mathbf{E}_T(\hat{q},R)).    \qquad 13.28$$

The meaning of this expression is similar to spectral discriminators already considered: Each clean jet in the final state gives rise to a $\delta$-spike in $\rho_{1\,\text{jet}}(E_T;R)$ for an appropriate range of $R$. Averaging over all events yields a continuous distribution that is analogous to $d\sigma/dE_T$ in the high-$E_T$ region. The analogy would be rather complete if all $\delta$-spikes were of the same height. This, however, is not so, and the $\delta$-spikes have different height depending on the structure of the surrounding event. This means that there is some dynamical information blended into $\rho_{1\,\text{jet}}(E_T;R)$ as compared with the simple counting of events in the case of $d\sigma/dE_T$. But this can hardly be considered a drawback — an advantage rather — given the intrinsically natural manner of how it is accomplished.

## Further options for $C$-continuous observables    13.29

Even within the restrictions of $C$-algebra there still is a considerable freedom of choice to allow one to take into account the dynamical information that may be available for a particular experimental situation (e.g. from a preliminary study of the data using conventional methods). The general options outlined below should help one with that.

## Generalized differential $C$-continuous observables    13.30

In practice it is convenient to have in view a rather wide class of $C$-continuous observables of the following form:

$$\mathbf{F}(\mathbf{P})\int d\gamma\,\lambda(\gamma)\,\mathbf{G}(\Psi_\gamma\circ\mathbf{P})\,\delta\big(s-\mathbf{H}(\Phi_\gamma\circ\mathbf{P})\big), \qquad 13.31$$

where $\mathbf{F}$, $\mathbf{G}$ and $\mathbf{H}$ are, typically, $C$-correlators (other $C$-continuous observables are also allowed), $\Phi$ and $\Psi$ are two families of filters parametrized by $\gamma$ which describes the geometry of the filters — in fact, the structure of substates the observable probes (cf. Sec. 11.13), $\mathbf{H}$ describes the physical feature with respect to which the distribution is studied (e.g. the invariant mass of the substate or its total transverse energy $E_\mathrm{T}$), while $\mathbf{F}$, $\mathbf{G}$ and $\lambda$ may be used to enhance the signal (an example is given in Sec. 13.32).

The $C$-continuity of the construct 13.31 follows from the results of Sec. 5.

## Example: using additional dynamical information to enhance signal    13.32

Consider a search for particle $X$ that decays into two jets, for which one uses 2-jet spectral discriminator 11.22 expecting to see a bump signaling a virtual presence of $X$. Let us show how additional dynamical information may allow one to enhance the signal/background ratio by modifying the definition 11.22 along the line of 13.31.

For instance, suppose the jets resulting from the decay are expected to be accompanied by no less than, say, 3 jets. Then the signal can be enhanced by multiplying each contribution to the spectral discriminator (the integrand of 11.22) by $\mathbf{J}_3(\overline{\Phi}_{\hat{q}_1,\hat{q}_2;R}\circ\mathbf{P})$ ($=\mathbf{G}$ in 13.31) where

$$\overline{\Phi}_{\hat{q}_1,\hat{q}_2;R}(\hat{p}_i)\equiv 1-\Phi_{\hat{q}_1,\hat{q}_2;R}(\hat{p}_i)\ , \qquad 13.33$$

and $\Phi_{\hat{q}_1,\hat{q}_2;R}$ is the same filter that describes the mass detector used in the definition of the spectral discriminator (cf. 11.20). The point is that the filter $\overline{\Phi}_{\hat{q}_1,\hat{q}_2;R}$ defines a substate that is complementary to the one being tested by $\Phi_{\hat{q}_1,\hat{q}_2;R}$. If the complementary substate has less than 3 jets then $\mathbf{J}_3(\overline{\Phi}_{\hat{q}_1,\hat{q}_2;R}\circ\mathbf{P})$ suppresses the contribution (the properties of the observables $\mathbf{J}_m$ are described in Secs. 8 and 9). This would affect the normalization 11.23 but that is of little consequence as long as the desired suppression is effected. Note that such a factor can be raised, without spoiling the $C$-continuity, to any positive power which may make the suppression sharper (perhaps, at the cost of decreased stability with respect to data errors and statistical fluctuations).

Similarly, suppose one expects that 2-jet events cannot contain the particle one searches for (i.e. $X$ is always accompanied by at least another jet). Then it is natural to modify the expression of spectral discriminator by simply multiplying the contribution to it from each event $\mathbf{P}$ by (a power of) the ordinary 3-jet discriminator $\mathbf{J}_3(\mathbf{P})$ ($=\mathbf{F}$ in Eq. 13.31).

One could also use information about expected angular distribution of the jets in the final state to introduce angular weights ($=\lambda$ in Eq. 13.31).

The above scheme is generalized in a straightforward way to the case of, say, particles that decay into 3-jets and are mostly produced in 6-jet events (which corresponds to the top search at the FERMILAB

Tevatron in the purely hadronic channel [25]).

## Higher differential *C*-continuous observables 13.34

Similarly to differential cross sections, one can define differential *C*-continuous observables with respect to more than one parameter. For this, it is sufficient to insert additional δ-functions into the integrand of 13.31.

As an example, consider the Lego plots in Fig. 4 of [13]. Their (simplified) *C*-continuous analogue would be as follows. One considers a 3-jet spectral discriminator and inserts an additional δ-function,

$$\rho_{3j-2j}\left(s_{3j}, s_{2j}; R\right) = \int d\hat{q}_1 \, d\hat{q}_2 \, d\hat{q}_3 \, \delta\big(s_{3j} - S'(\hat{q}_1, \hat{q}_2, \hat{q}_3; R)\big) \, \delta\big(s_{2j} - S''(\hat{q}_1, \hat{q}_2, \hat{q}_3; R)\big), \qquad 13.35$$

where

$$S'(\hat{q}_1, \hat{q}_2, \hat{q}_3; R) = \mathbf{S}\big(\Phi_{\hat{q}_1, \hat{q}_2, \hat{q}_3; R} \circ \mathbf{P}\big), \qquad S''(\hat{q}_1, \hat{q}_2, \hat{q}_3; R) = \min_{\substack{a,b=1,2,3 \\ a \neq b}} \mathbf{S}\big(\Phi_{\hat{q}_a, \hat{q}_b; R} \circ \mathbf{P}\big). \qquad 13.36$$

The function $S''$ is the minimal invariant mass among pairs of jets from the three jets selected by the filter in $S'$.

Transition to $m_{nj} = s_{nj}^{1/2}$ can be performed as explained in Sec. 11.4.

The above examples demonstrate that the expressive power of *C*-algebra is sufficient for practically any application of the precision measurement class.

## Summary and conclusions                                                                                      14

The results of this work can be summed up as follows:

• A systematic analysis allows one to reinterpret the "ambiguities" of the conventional jet finding algorithms as instabilities caused by the intrinsic discontinuity of such algorithms. Such instabilities are inherited in the form of systematic errors by any observable defined on the basis of such algorithms.

• The instabilities are eliminated if observables are chosen to be continuous in an appropriate sense. A proper continuity of observables is determined uniquely by the structure of measurement errors of multimodule calorimetric detectors. It can be described as a stability with respect to "almost collinear" fragmentations that cannot be distinguished by a finite precision calorimetric detector. The fact that real-life calorimetric detectors have finite energy and angular resolutions is a key consideration that has been missing in the discussion of jet-related fragmentation-invariant observables: The resulting continuity (which we call calorimetric continuity, or $C$-continuity) restricts the structure of observables rather severely — yet leaving enough freedom to allow one to describe any jets-related physics in the language of such observables.

• The $C$-continuity ensures that observables possess optimal properties with respect to data errors. On the other hand, a use of continuous weights instead of hard cuts corresponds better to the physical reality of absence of any internal boundaries separating final states with different "numbers of jets" in the continuum of all possible final states.

• A rather wide family of $C$-continuous observables can be constructed within the framework of the so-called $C$-algebra. The $C$-algebra contains:

— A special class of observables (the so-called $C$-correlators) that have a rigid analytical structure; in particular, their dependence on particles' energies is fixed. The most important among them are: (transverse) energy; invariant mass; a sequence $\mathbf{J}_m$, $m = 1, 2, \ldots, \infty$ of the jet-number discriminators (Secs. 8, 9 and 13.24). (Note that the construction of the latter differs slightly for different kinematical situations such as $e^+e^- \to$ hadrons, deeply inelastic lepton-nucleon scattering or hadron-hadron collisions; cf. Sec. 13.)

— A set of rules to construct new $C$-continuous observables from those already available (Secs. 5.22 and 13.29).

— A set of rules to translate the conventional observables based on jet algorithms into the language of $C$-algebra (the filtering to select "multijet substates", Secs. 5.25, 10.2, 11.14; the "spectral" construction to describe distributions of properties — e.g. masses — of "multijet substates", Secs. 10.1 and 11.1).

• The $C$-correlators that form a basis of $C$-algebra have the form of multiparticle correlators and, therefore, fit tightly and naturally into the framework of Quantum Field Theory (they can be rewritten in terms of the energy-momentum tensor [54]). It is fair to say that any property that can be meaningfully studied within QFT (which summarizes a vast body of experimental knowledge) should be expressible in the form of multiparticle correlators. Since there is, really, little reason to expect that QFT may break down at the energies of the current and planned colliders, one concludes that the taking into account of "the kinematical QFT aspect" is the other key ingredient that has been missing from the discussion of jet-related observables. The fact that the $C$-correlators fit naturally into the QFT framework opens a prospect for higher quality theoretical predictions in the physics of jets.

• It should come as no surprise that because only kinematical restrictions went into the condition of $C$-continuity, it proves possible to express *any* physics conventionally studied via jet finding algorithms, in the language of $C$-continuous observables. With such observables, at no stage does one have to identify individual jets. As a result, the problem of instabilities of jet algorithms in its current form simply vanishes.

• In the new formalism, jet algorithms retain the role of an approximation trick for faster computation of such observables from data (preclustering). The jet resolution parameters of conventional algorithms become parameters that control the corresponding approximation errors. In particular, there is one algorithm (Sec. 7.20) that is optimal from the point of view of minimization of such approximation errors. Such an "optimal preclustering" happens to possess features of various conventional jet algorithms; in particular, it allows one to recombine any number of particles into one (or more) protojet(s) using well-defined analytical criteria (Eq. 7.39 and Sec. 7.47).

• The explicit — and rather simple — analytical structure of the observables from $C$-algebra results in a much greater flexibility as regards construction of approximation tricks etc. than in the case of the conventional scheme. One example is the expansions in energies of soft particles (Sec. 7.4; cf. the analysis of errors due to missing energy in Sec. 7.15). Other tricks will undoubtedly be found in the context of conrete applications.

• The advantages of the new formalism — a lesser sensitivity to data errors, absence of instabilities, advantages for theoretical studies including calculations, computational flexibility — should be expected to be least negligible in the applications that can be characterized as precision measurements, i.e. whenever the quality of both experimental numbers and theoretical predictions is important, including low-signal situations.

## Objections                                                                                                  14.1

## Higher computing resources                                                                    14.2

One objection to the new formalism is that $C$-continuous observables require more computer resources to compute them from raw data. That, however, is a purely technical matter: First, the regular analytical structure of new observables allows many analytical and programming optimizations. Second, little additional computer power seems to be needed for computing new observables for the values of control parameters of jet algorithms that (values) are currently employed. As more computing power becomes available (which happens very fast), more precise computations can be done (see below).

## Ambiguities                                                                                              14.3

One may argue that there are still ambiguities to fix in the definitions of various $C$-continuous observables (e.g. the shape of elementary mass detectors in the construction of spectral discriminators in Secs. 10 and 11; the choice of angular separation in the hadron-hadron case Sec. 13.13, etc.). The answer to this objection is that it was not ambiguities per se that were the problem with the conventional jet algorithms, but *their enhancement by instabilities*. Once the cause of the enhancement is eliminated, all one has to worry about is a consistency of the definitions used in theoretical and experimental calculations.[i]

## Psychological objections                                                                          14.4

More difficult to cope with (in a sense) are objections of psychological nature, mostly due to an understandable inertia of thought.

One such objection is that a jet finding algorithm, if used in a consistent manner by theorists and experimentalists, is a well-defined observable. Of course, one *can* call any number computed from data an observable. However, it stands to physical reason that defining a physically correct observable *must* take into account errors of measurement and calculation; a measure of stability against such errors *must* be incorporated into the definition; and one *must* be aware of the consequences of the choices one makes and of the available options. Instability of jet algorithms is a self-inflicted woe, and there is no reason why it should not be eliminated whenever technically possible and physically necessary.

---

[i] Of course, having got rid of the systematic errors due to "ambiguities of jet algorithms" one may discover that what one really needs is more theoretical higher order corrections, or a higher precision of one's calorimeters, or a greater bandwidth of one's data acquisition system — but that would be an entirely different situation.

On the other hand, it is not clear how one can rationally argue against the point of view that the jet pattern, i.e. the number and 4-momenta of jets for a final state (cf. Eq. 1.2), is only an approximate description of that final state. One may or may not be happy with such an approximation. In the former case, one can simply ignore the $C$-algebra[i]. In the latter case, however, one would like to know how to improve upon the approximation. The new formalism offers a systematic answer to this question.

Another psychological difficulty is that some effort is needed in order to learn to think in terms of the new formalism rather than keep interpreting and judging it (sometimes with misleading conclusions) in terms of the old paradigm based on a naive definition of jets, which the new formalism is meant to supersede. The habit of thinking in terms of jets is deeply ingrained, in part due to their highly visual character. Physical theories, however, are not always entirely intuitive, and physical meaning — contrary to a wide-spread implicit belief — is not the same as quasiclassical visualization. Although the jet/parton picture remains a valid and greatly useful approximation, it is only a first approximation to the precise QFT-compatible description based on the $C$-algebra.

## From jet finding algorithms to $C$-algebra                                      14.5

A transition to the systematic use of $C$-continuous observables — in situations where the additional computational resources required by the new formalism would be justified by higher precision requirements — could be performed gradually. Indeed, the conventional data processing consists of two steps: first, a jet finding algorithm computes a jet pattern for each event (Eq. 1.2); second, an observable is computed from jet patterns.

The new formalism allows one to translate any such observable into the language of $C$-continuous functions ($C$-correlators, spectral discriminators, etc.). The difficulty here is that the translation is not a mechanical procedure. But once one has worked through typical examples (Secs. 8–13), it should not be particularly difficult. Because the framework of $C$-algebra incorporates only kinematical restrictions, any truly observable physical feature[ii] can be adequately expressed in the language of $C$-continuous observables.

Since jet algorithms are treated as an approximation trick (= preclustering) within the framework of $C$-algebra, one can start by simply replacing the old observables with new $C$-continuous ones — without changing the jet algorithm[iii] — and reprocess the same events in order to compute the new observables (recall e.g. that only $O(100)$ events were used to discover the top quark [12], [13]). This allows one to test the algorithms with a limited but meaningful sample of events.

Next, since $C$-continuous observables retain their meaning and qualitative behavior even if one deals directly with the original events instead of their jet patterns, one can choose a smaller value for the jet resolution $y_{cut}$ (or the jet cone radius $R$) and repeat the entire computation. This means that fewer distortions would be introduced into the computed observables but more computing power would be needed. Continuing in this fashion one should be able to see how far one is able and willing to go with the available computer resources.

Therefore, although a prospective transition to the use of $C$-algebra for quantitative description of physical phenomena involving hadron jets in situations where the limitations of the conventional approach are too restrictive — although such a transition does represent a change of direction, it does not mean a complete break with the tradition of jet finding algorithms, and can be accomplished in an evolutionary manner.

---

[i] As did the Indiana State House of Representatives in 1887 when it decreed that $\pi = 3$ [70].

[ii] Not to be confused with an expression of such a feature in the language of jet algorithms.

[iii] Eventually, the optimal preclustering will have to be used but the switch can be done any time.

# Appendix. Abstract measures on [0,1]          15

An excellent source on abstract measures with emphasis on the functional point of view is [45]. For numeric handling of measures one only needs to put together a few tricks that are well-known in applied mathematics. In this appendix only measures on the unit interval are discussed. The primary purpose is to explain the computational issues associated with spectral discriminators.

## Continuous functions. Constructive aspects          15.1

The notion of (abstract) measure is derivative from that of continuous function. Consider continuous functions defined in the unit interval $0 \leq s \leq 1$, and the continuity extends to the end points. Define the $C^0$-distance[i] between two such functions:

$$\|f - g\|_{C^0} \stackrel{\text{def}}{=} \sup_{0 \leq s \leq 1} |f(s) - g(s)| \ . \qquad\qquad 15.2$$

We say that $f$ approximates $g$ _in the sense of_ $C^0$ if the $C^0$-distance 15.2 is sufficiently small.

## Linear splines          15.3

Split the interval [0,1] into $N$ equal subintervals; the boundaries between them are the $N+1$ points $s_i = i/N$, $i = 0,1,\ldots N$. Fix $N+1$ numbers $f_i$, $i = 0,1,\ldots,N$. Consider a function $\tilde{f}(s)$ such that it takes the values $f_i$ at the points $s_i$, $\tilde{f}(s_i) = f_i$, and interpolates linearly in between:

$$\tilde{f}(s) = N[f_i(s_{i+1} - s) + f_{i+1}(s - s_i)], \quad s_i \leq s \leq s_{i+1} . \qquad\qquad 15.4$$

We call such functions _linear splines_. The following two facts are important here: (a) Linear splines are perfectly constructive objects: they can be represented as arrays $[0:N]$ of floating point numbers. (b) One can use them to approximately represent any continuous function.

Indeed, suppose $f_i = f(s_i)$. By choosing $N$ large enough, $\tilde{f}(s)$ can be made to represent $f(s)$ with any precision $\varepsilon$ for all $s$ simultaneously  (i.e. in the sense of $C^0$):

$$\left\| f - \tilde{f} \right\|_{C^0} < \varepsilon, \quad \text{or} \quad \left| f(s) - \tilde{f}(s) \right| < \varepsilon \quad \text{for all} \ \ 0 \leq s \leq 1. \qquad\qquad 15.5$$

The linear splines play the same role for continuous functions as the floating point numbers do for real numbers. Note that if $N$ is fixed, different continuous functions will be represented with a different $C^0$-precision.

A convenient representation for 15.4 is as follows. Define:

$$h(s) = \max(0, 1 - |s|), \ h_N(s) = Nh(Ns), \ \lim_{N \to \infty} h_N(s) = \delta(s) . \qquad\qquad 15.6$$

Then

$$\tilde{f}(s) \equiv \frac{1}{N} \sum_{i=0}^{N} h_N(s - s_i) f(s_i) \ . \qquad\qquad 15.7$$

## Measures on [0,1]          15.8

Imagine a subroutine (called, say, $\rho$) that accepts as an argument a variable length array $f[0:N]$ of floating point numbers, and returns one number as a result, $\rho(f)$. Since the array $f$ can be regarded as representing a linear spline, the subroutine defines a function on the space of all linear splines. To dis-tinguish linguistically the ordinary functions and these new functions that have ordinary functions as arguments, the new sort of functions are called *functionals*. Thus we say that ρ is a functional on linear splines.

---

[i] $C^0$ is a standard mathematical notation for things pertaining to continuous functions when the latter adjective is inconvenient as in the case of $C^0$-distance. The superscript 0 indicates that one talks about continuity of the function itself only; $C^1$ would mean that there are continuous first order derivatives, etc. The distance is $C^0$ because if a sequence of continuous functions converges in the sense of $C^0$-distance, then the limit is a function that is automatically continuous (i.e. also $C^0$).

## Examples of functionals                                                          15.9

(1)  ρ(f) = surface under $f(s)g(s)$ where $g(s)$ is an integrable function (the algorithm here involves iterations to perform integration).

(2)  ρ(f) = the value $f(\pi/5)$ (can be computed with arbitrary precision).

(3)  ρ(f) = surface under $\sin f(s)$.

(4)  Arrange all rational numbers from [0,1] into a sequence $r_n$ (e.g. $\frac{1}{2}, \frac{1}{3}, \frac{2}{3}, \frac{1}{4}, \frac{3}{4}$ …). Then

$$\rho(f) = \sum_{n=1,2,\ldots} \frac{1}{n^2} f(r_n) \ .$$                          15.10

The series is convergent because $f$ is bounded. But the convergence is slow enough to ensure that one has to deal with long sequences of rational numbers to attain high precision.

The subroutines (1), (2) and (4) have an additional property: their results are *linear* with respect to their argument ($f$).

Furthermore, take a sequence $f_n$ of linear spline functions that approximates a given continuous function $f$ in the sense of $C^0$. What can one say about the sequence $\rho(f_n)$? In general, nothing. So, one has to make the following …

## Strong assumption                                                                15.11

For any sequence of linear spline functions $f_n$ that converges in the sense of $C^0$ to some continuous function $f$, the sequence of numbers $\rho(f_n)$ should converge in the usual numerical sense, and the result should be the same for any sequence that approximates $f$. (This is true in all the above examples.) Then the subroutine ρ effectively defines a functional on *any* continuous function. In other words, for any continuous function $f$ one can compute $\rho(f)$ to *any* accuracy by choosing a linear spline approximation $\widetilde{f}$ that is sufficiently close to $f$ in the sense of $C^0$, and evaluating $\rho(\widetilde{f})$.

I emphasize that such a subroutine need not give meaningful results if a different type of closeness for its argument $f$ is considered. For instance, if one computes the values of ρ for a partial Fourier series for $f$, then — whatever the precision of computations — the resulting numerical sequence will not, in general, converge. Thus, choosing a wrong type of approximation/convergence (in the case of Fourier series, it is the convergence "in the sense of $L^2$") would render computations meaningless.

## What are the most general *linear* functionals on $C^0$?                         15.12

It is exactly such functionals that are called <u>*measures*</u>.[i] The crucial requirements are (i) that measures are *linear* functionals, and (ii) that they are defined on the *entire* space $C^0$.

A convenient syntactic convention is to write a measure as an integral:

$$\rho(f) = \int_0^1 ds \, \rho(s) \, f(s).$$                                        15.13

This emphasizes the fact that the notion of measure is an extension of the notion of ordinary function (because an integrable function $\rho(s)$ defines a measure according to 15.13). But it may prompt one to handle a measure as one would a continuous function, which may lead to erroneous results.

## Continuity of measures                                                           15.14

Since we are dealing with measures in a numerical way the issues of errors — therefore, continuity — are central. In standard textbooks one usually requires of measures to satisfy certain continuity prop-erties. However, a remarkable recent theorem [73] ensures that measures defined as above automatically possess those properties.[i] More precisely, any measure $\rho$ is automatically continuous in the following sense: If a sequence of continuous functions $f_n$ converges in the sense of $C^0$ then the numbers $\rho(f_n)$ converge in the usual numerical sense.

---

[i] In the main text they are called "abstract measures" to avoid interference with physical measurements.

## Measures as additive functions of subsets                                  15.15

In older textbooks and in all advanced textbooks on the theory of probability where measures are routinely studied (e.g. [49]), one defines measures differently — as additive functions on subsets. For instance, if one takes a subset $S$ of the interval $[0,1]$ to consist of a sequence of non-intersecting intervals, then the total length of $S$ is a measure: the length of a union of two such non-intersecting subsets is a sum of their lengths. After that one embarks upon an agonizing study of which subsets are measurable[ii] etc., most of which has nothing to do with constructive mathematics. The only textbook I know of that discusses both definitions systematically with the functional definition as a primary one, is [45]. The modern definition in terms of linear functionals is better suited to our needs: Our measures emerge directly as sums of $\delta$-functions (cf. 10.7).

## General structure of measures                                              15.16

In general, a measure is a sum of two components: one is an integrable function (this is usually a continuous function with a few integrable singularities — e.g. $s^{-1/2}$); the other is localized on a zero length subset of $[0,1]$ (this is usually a discrete — perhaps infinite — sum of $\delta$-functions (cf. 15.10).

A typical spectral discriminator is a sum of a continuous component and a finite number of $\delta$-functions (Sec. 11.1).

## Convergence of sequences of measures                                       15.17

This is a central issue because we want to perform computations with measures. This means that our expressions/subroutines for measures will be only approximations, and if we want them to be indeed *approximations*, we should understand precisely what convergence means in the case of measures.

From among different types of convergence of measures we need the one that is natural in our problem. We call it *convergence in the sense of measures*.[iii]

Consider a sequence of measures $\rho_n$. Suppose it is such that for any continuous function $f$ their values $\rho_n(f)$ form a convergent sequence of numbers. The limit defines a functional that is automatically linear. Since we require convergence for each continuous function, the resulting functional is defined on each continuous function and is, therefore, a measure automatically. Wright's theorem (see Sec. 15.14 above) guarantees continuity of the resulting measure with respect to variations of its argument (in the sense of $C^0$).

The rates of convergence for different $f$ are *not* correlated. This means that there is, in general, no single number to characterize precision of an approximation in the sense of measures.[iv]

A rule of thumb is, the faster $f$ changes, the slower the convergence. This has an important practical consequence:

---

[i] The theorem is valid if one replaces the notorious Axiom of Choice that allows one to prove "existence" of all sorts of pathological counterexamples, with the so-called Axiom of Determinateness due to Myczelski and Steinhaus (the only readable account I am aware of is [73]) which offers a more adequate formalization of constructive aspects of mathematics.

[ii] The modern answer (the Myczelski-Sverczkowski theorem; cf. [73]) is, *all*. It is valid in the same context as the Wright theorem mentioned above.

[iii] In the context of functional analysis it is a special case of the so-called *-weak topology [42]. In the context of the theory of measure proper it is often called simply weak convergence [47].

[iv] Mathematically, this means that the *-weak topology is in general not metrizable [42]; even when it is (e.g. in the case of normalized measures on a compact set), the corresponding distance function is not too useful.

## Quality of approximations 15.18

From the above one can infer that if a measure $\rho$ is known approximately, then numerical errors of $\rho(f)$ differ for different $f$. In particular, for some $f$ the errors may be judged acceptable, for others, too large. Spectral discriminators are constructed by accumulating statistically data from many events. As the number of events in the data sample is increased, the spectral discriminator obtained is expected to converge to the "true" one. But with a limited statistics, on which $f$ the obtained precision will be (un)acceptable?

There is no concrete numerical answer to this question but a rule of thumb is that the "softer" the shape of $f$, the better the precision. For instance, it is easier to obtain a high precision for the jet-number discriminators $\mathbf{J}_m$ (Sec. 8.8) for smaller $m$.

## Approximate description 15.19

Now we turn to practical issues of how to construct and manipulate approximations for a given measure. First, we have to agree on what is "approximate". In all cases we encounter the meaning is dictated by the problem, and is as follows. A sequence of measures $\rho_n$ converges to a given one, $\rho$ if for any continuous test function $f$ one has convergence $\rho_n(f) \to \rho(f)$ in the usual numerical sense. Note that the rate of convergence depends on $f$. Practically, one chooses a sufficiently rich finite set of test functions $f_i$, and measures closeness of $\rho_n$ to $\rho$ by the differences $|\rho_n(f_i) - \rho(f_i)|$.

## Measures represented by continuous functions 15.20

For measures represented by continuous functions, one may use linear splines. Note that a sequence of continuous functions that converges in the sense if $C^0$, also converges in the sense of measures. Linear splines approximate a continuous function in the sense of $C^0$, therefore, will automatically result in a satisfactory approximation in the sense of measures.

Unfortunately, this obvious method is inapplicable in our case because spectral discriminators emerge as sums of $\delta$-functions — not as continuous functions.

## $\delta$-functions 15.21

The fact is, any measure can be approximated to arbitrary precision in the sense of measures by finite linear combinations of $\delta$-functions (cf. 15.26). A well-known case when this occurs is a Monte Carlo evaluation of an observable 2.7. Then one approximates the expression 2.7 by a finite sum, $\sim \sum_i \mathbf{F}(\mathbf{P}_i)$, which is equivalent to approximating the probability measure $\pi$ with a sum of $\delta$-functions as follows:

$$\pi(\mathbf{P}) \sim \sum_i \delta(\mathbf{P}, \mathbf{P}_i). \qquad 15.22$$

In terms of computer data structures, such a linear combination is just a finite-length list of records, each containing two fields: the coefficient of the $\delta$-function, and its location. For measures on $[0,1]$, the location is just a number from this interval.

## Constructing regular approximations 15.23

The problem is that in our case the locations of $\delta$-functions are irregularly scattered over the unit interval so that visualization and comparison are difficult. Moreover, fluctuations due to errors may be unacceptably large. So, a more regular representation is needed.

Recall that a measure is a collection of its values on continuous functions. Approximate a continuous test function $f$ by a linear spline $\tilde{f}_N$, Eqs. 15.4 and 15.7, where $\tilde{f}_N \to f$ as $N \to \infty$ in exactly the sense we need (i.e. in the sense of $C^0$ that ensures convergence of values of any measure $\rho$). This fact allows one to approximately represent:

$$\int_0^1 ds\, \rho(s)\, f(s) \approx \int_0^1 ds\, \rho(s)\, \tilde{f}_N(s) \equiv \frac{1}{N} \sum_{i=0}^{N} r_N(s_i)\, f(s_i). \qquad 15.24$$

The explicit expression for $r_N$ is as follows:

$$r_N(s) \equiv \int_0^1 ds' \rho(s') h_N(s'-s). \tag{15.25}$$

This is a continuous function of $s$ but only a finite array of its values $r_N(s_i)$ is actually used to represent $\rho$ (cf. 15.24) — for a better approximation one takes a larger $N'$ and the new array $r_{N'}(s_i)$ consists of values of a different continuous function.

The r.h.s. of 15.24 is exactly the value on $f$ of a sum of $\delta$-functions located at $s_i$, with weights equal to $r_N(s_i)$. So, we have obtained the following approximation:

$$\rho(s) \approx \rho_{\tilde{N}}(s) \equiv \frac{1}{N} \sum_{i=0}^{N} r_N(s_i)\delta(s-s_i). \tag{15.26}$$

### $\rho_N$ converges to $\rho$ in the sense of measures    15.27

This is ensured by construction.

An important fact is that whereas $\rho$ was any measure — perhaps, a sum of $\delta$-functions scattered irregularly within $[0,1]$ — the $\delta$-functions of $\rho_N$ are distributed in a regular mann*er, independently of what $\rho$ was. Therefore, the array of numbers $r_N(s_i)$, $i=0,1,\ldots,N$ is a convenient approximate representation for a measure, to be interpreted according to 15.26. Choosing $N$ large enough, one can make the approximation as precise as needed.

### $r_N(s_i)$ are valid observables per se    15.28

An advantage of the above construction is that the numbers $r_N(s_i)$ are simply the values of $\rho$ on test functions $h_N(s-s_i)$ — without any approximations involved. This means that if $\rho$ is a spectral discriminator obtained by collecting statistics from many events, then $r_N(s_i)$ are perfectly valid scalar-valued observables per se. In particular, if $\rho$ is known precisely (e.g. from infinite statistics experiments with no systematic errors etc.) then they are also known precisely. In other words, the "measured" values for $r_N(s_i)$ (i.e. the values computed from a finite sample of data) will converge to the ideal values as the statistics is increased, and can be used to compare experimental data with theory.

However, statistical fluctuations (including those due to data errors) are smaller for smaller $N$ (cf. Sec. 2.21).

### Regularization    15.29

This is a fundamental notion formalized and systematically developed for applied problems by Tikhonov and his school [75]. While any exact spectral discriminator $\rho(s)$ is a continuous function after averaging over all final states its approximate representations $\tilde{\rho}$ that emerge due to a discretized integration over $\gamma$ in 11.2 are rather irregular sums of $\delta$-functions. Although in principle $\tilde{\rho}$ converges to $\rho$ in the sense of measures (when the event sample and the precision of integrations are increased), one would like to obtain a more uniform approximation. This is achieved as follows. For each approximation $\tilde{\rho}$, one constructs the sequence $\tilde{\rho}_N$ as described in Sec. 15.23. ($\tilde{\rho}_N$ are practically represented by the corresponding arrays $\tilde{r}_N(s_i)$.) Then one chooses $N=\tilde{N}$ as large as possible — yet not too large so that the irregularities (due to the stochastic nature of the approximation $\tilde{\rho}$) would not be too manifest. The result $\tilde{\rho}_{\tilde{N}}$ (equivalently, the array $\tilde{r}_{\tilde{N}}$) is the required regularization. As $\tilde{\rho}$ is made more precise (e.g. by increasing the event sample), the optimal $\tilde{N}$ goes to $\infty$, and $\tilde{\rho}_{\tilde{N}}$ (or $\tilde{r}_{\tilde{N}}$) $\to \rho$. Precise mathematical criteria for choosing the optimal $\tilde{N}$ are unknown at the time of this writing so it is best to proceed in an empirical fashion. A convenient tool for that is the folding trick of Sec. 12.11.

## Appendix. General theory of abstract measures　　　16

The definitions and facts listed below constitute a minimum of information on abstract measures presented for completeness' sake. A reader inexperienced in this sort of mathematics may find it difficult to understand it; in that case, one should consult the material of Sec. 15 as well as an expert mathematician. A much more complete source is the excellent advanced textbook [45]; general functional-analytic aspects are treated with elegance in [44].

Note that standard expositions (like [45]) are grappling with difficulties due to an unfortunate axiomatization of infinite constructions (pathologies due to the Axiom Of Choice). These have nothing to do with practical mathematics and are ignored below. (See the remarks and references in Sec. 15.8.)

A)　Consider a finite-dimensional smooth manifold $M$. In our case this can be a Euclidean vector space, a sphere, a cylinder, or a direct product of a finite number of these.

B)　*Test function* on $M$ is a continuous numeric-valued function on $M$ such that it takes non-zero values only within a compact subregion $K$ of $M$; i.e. $K$ is such that it does not stretch to infinity in any direction. If $M$ is a (direct product of) sphere(s) than it has no infinite directions and the restriction does not apply.

C)　*Measure* on $M$ is a *linear* functional defined on *all* test functions. (This definition is sufficient for all practical purposes.) Two generic examples: (i) a linear combination of δ-functions; (ii) an integrable function.

D)　Measures form a *linear space*, i.e. one can take their linear combinations.

E)　Convergence of measures can be defined in several ways. The one we need is the most natural one (sometimes called weak convergence of measures; it is a special case of the *-weak topology in spaces of linear functionals; we call it *convergence in the sense of measures*), and is as follows: A sequence of measures converges if their values on any test function form a sequence that converges in the usual numeric sense.

F)　Such convergence is, in general, *not metrizable*, i.e. cannot be described by a single distance function. This is possible in the case of a compact $M$ (e.g. a sphere) but even then such a distance function is not practically useful.

G)　The linear space of measures is *closed* with respect to the convergence in the sense of measures.

H)　Any measure can be *approximated* (with respect to convergence in the sense of measures) with continuous functions (e.g. multidimensional analogues of linear splines). This can also be done with finite linear combinations of δ-functions. These two facts are the basis of how measures are to be handled in numerical applications.

## References     17